\documentclass[useAMS,usenatbib]{mn2e}
\def\farcs{\hbox{$.\!\!^{\prime\prime}$}}

\usepackage{graphicx}
\usepackage{graphics}
\usepackage{threeparttable}
\usepackage{deluxetable}
\usepackage{lscape}
\usepackage{longtable}
\usepackage{color}
\usepackage{mathtools}
 \usepackage{amssymb}
\let\la=\lesssim            
\let\ga=\gtrsim
\bibpunct{(}{)}{;}{a}{}{,}

\def\farcs{\hbox{$.\!\!^{\prime\prime}$}}
\def\re{\hbox{$R_{\rm e}$}}
\def\recirc{\hbox{$R_{\rm e, circ}$}}

\def\msun{\hbox{$M_{\odot}$}}
\def\ae{\hbox{$a_{\rm e}$}}
\newcommand{\pacsd}{PACS-d}
\newcommand{\mipsd}{MIPS-d}
\newcommand{\mipsu}{MIPS-u}

\newcommand{\sbzk}{$sBzK$}

\newcommand{\pbzk}{$pBzK$}
\newcommand{\pbzks}{$pBzK$s}
\newcommand{\sersic}{S\'ersic}

\title[]{Star formation and quenching among the most massive galaxies  at $ \mathbf{ z\sim 1.7}$}
\author[C. Mancini et al.]
{C.~Mancini$^{1,2}$\thanks{E-mail:chiara.mancini@unipd.it}, 
A.~Renzini$^{1}$, E.~Daddi$^{3}$, G.~Rodighiero$^{2}$, 
S.~Berta$^{4}$, N.~Grogin$^{5}$,
  \newauthor
D.~Kocevski$^{6}$, A.~Koekemoer$^{5}$ 
\\
$^{1}$INAF - Osservatorio Astronomico di Padova, Vicolo dell'Osservatorio, 5, I-35122, Padova, Itlay\\
$^{2}$Dipartimento di Fisica e Astronomia ``Galileo Galilei'', Universit\'a di Padova, Vicolo dell'Osservatorio, 3, I-35122, Padova, Itlay\\
$^{3}$CEA-Saclay, Service d’Astrophysique, F-91191 Gif-sur-Yvette, France\\
$^{4}$Max-Planck-Institut f\"ur Extraterrestrische Physik (MPE), Postfach 1312, 85741 Garching, Germany \\
$^{5}$ Space Telescope Science Institute, 3700 San Martin Drive, Baltimore, MD 21218, USA\\
$^{6}$ Department of Physics and Astronomy, University of Kentucky, Lexington, KY, 40506, USA}
\begin{document}

\date{}

\pagerange{\pageref{firstpage}--\pageref{lastpage}} \pubyear{2014}

\maketitle

\label{firstpage}

\begin{abstract}
We have conducted a detailed object-by-object study of a mass-complete ($M_*\geq10^{11}~M_{\odot}$) sample of 56 galaxies at $1.4\leq z\leq 2$ in the GOODS-South field, showing that an accurate de-blending in MIPS/24um images is essential to properly assign to each galaxy its own star formation rate (SFR), whereas an automatic procedure often fails. This applies especially to galaxies with SFRs below the Main Sequence (MS) value, which may be in their quenching phase. After that, the sample splits  evenly between galaxies forming stars within a factor of 4 of the MS rate ($\sim 45\%$), and sub-MS galaxies with SFRs $\sim 10-1000$ times smaller ($\sim 55\%$). We did not find a well defined class of intermediate, transient objects below the MS, suggesting that the conversion of a massive MS galaxy into a quenched remnant may take a relatively short time ($<1$ Gyr), though a larger sample should be analyzed in the same way to set precise limits on the quenching timescale. X-ray detected AGNs represent a $\sim 30\%$ fraction of the sample, and are found among both star-forming and quenched galaxies. The morphological analysis revealed that $\sim 50\%$ of our massive objects are bulge-dominated, and almost all MS galaxies with a relevant bulge component host an AGN. We also found sub-MS SFRs in many bulge-dominated systems, providing support to the notion that bulge growth, AGN activity and quenching of star formation are closely related to each other. 
   
\end{abstract}


\section{Introduction}
The transition from a star forming regime to  passive evolution, that we call {\it quenching}, is perhaps the 
most important event that may happen to a galaxy in the course of its life. Yet, the physical processes causing quenching are still poorly understood,
making  this  a main frontier  issue in galaxy evolution. We know that the first quenched (massive) galaxies appear at redshifts $\sim 2-3$ \citep[e.g.,][]{2004Natur.430..184C,2009ApJ...691.1879W, 2012ApJ...759L..44G,2013ApJ...775..106C} and that their  fraction and number density increase steadily with cosmic time,
to the point that they come to dominate the stellar mass in galaxies in the local Universe \citep[e.g.,][]{2004ApJ...600..681B}. We also know that the quenched fraction increases steeply with galaxy mass independently of local environment, as well as with local environmental overdensity, independently of their mass. We therefore speak of {\it mass quenching} and {\it environment quenching} as two distinct,  {\it separable} processes \citep{2010ApJ...721..193P,2012ApJ...757....4P}. Yet, the physical nature of such processes is still merely conjectural and it has also been suspected that they could be different manifestation of a same underlying process 
\citep{2014arXiv1402.1172C, 2015ApJ...800...24K}.

The probability (or relative frequency) of mass quenching appears to increase exponentially  with stellar mass ($M_*$), or almost indistinguishably with the star formation rate (SFR), given the tight SFR-$M_*$ correlation that exists for {\it main sequence} (MS) galaxies \citep{2007ApJ...670..156D,2007A&A...468...33E, 2007ApJ...660L..43N,2015A&A...575A..74S}. A variety of radically different options are currently entertained  for the mass quenching process, whereby quenching is either an {\it internal} or an {\it external} process. In one option for the former case  sudden energy/momentum release from star formation and/or AGN (feedback) results in the ejection of all gas from galaxies that then turn passive, the `quasar mode' quenching in current jargon \citep[e.g.,][]{2004ApJ...600..580G}. Powerful AGN jets may also   heat the circumgalactic medium to  high temperature  thus preventing further accretion of cold gas, the so-called `radio mode' AGN feedback \citep{ 2006MNRAS.365...11C}. In another option for an external process, the circumgalactic gas is shock-heated to high temperatures as the mass of the host dark matter halo ($M_{\rm h}$) exceeds a critical threshold (of order of $\sim 10^{12}\,\msun$), and therefore it stops to cool and  flow into the galaxy, thus discontinuing to  feed  star formation \citep[e.g.,][]{ 2003MNRAS.345..349B}. 
Finally, the growth of a central mass concentration (bulge) may {\it quench itself}, with increasing shear (differential rotation)  suppressing the disk instability to form actively star-forming clumps, the so-called gravitational (or morphological) quenching \citep{2009ApJ...707..250M,2014ApJ...785...75G}.
So, we have at least four options for the physical nature of mass quenching. Moreover, the tight correlations  existing between halo mass, stellar mass, bulge mass  and mass of the central black hole  make difficult to disentangle between these options, as all result in a correlation of the quenched fraction with galaxy mass \citep{2012ApJ...757....4P}. In other words, it is not clear whether the pertinent mass in mass quenching is $M_*$ or $M_{\rm h}$, or the mass of the central bulge \citep{2014ApJ...788...11L}, or even the mass of the central black hole.

The frequency of AGN activity is found to increase strongly with galaxy mass, especially at $z\gtrsim 1$ \citep{2005ApJ...633..748R,2007ApJ...670..173D,2008ApJ...672...94F,2009A&A...507.1277B,2012MNRAS.427.3103B,2012ApJ...753L..30M,2012MNRAS.419...95M,2013ApJ...779L..13C,2014ApJ...787...38F,2014ApJ...796....7G,2015A&A...574A..82P,2015ApJ...800L..10R}, which provides circumstantial support for a connection between AGN feedback
and mass quenching. However, finding the smoking gun of AGN quenching proved to be very difficult, as most AGNs are hosted by actively star forming galaxies showing no signs of being quenched. Actually, the mere galaxy-AGN {\it co-evolution} concept argues for both growing and be quenched together. On the other hand, supermassive black holes do not grow in one shot. Rather,
a galaxy may experience many nuclear activity cycles interleaved with inactive phases \citep{2012MNRAS.427.2734N,2014ApJ...782....9H}, with the former ones leading to feedback and driving nuclear outflows, but failing to quench. Hence,  if AGN quenching really works, not every AGN cycle leads to quenching but just a final, fatal one. Later, once star formation is quenched by ejecting the gas, so is the nuclear activity as well. This AGN variability on a variety of timescales clearly makes the connection AGN-quenching difficult to unambiguously recognize \citep{2014ApJ...782....9H}. 

While mass quenching works equally for satellites and centrals, environment quenching appears to be exclusive of satellite galaxies \citep{2012ApJ...757....4P}, but see also \citet{2015ApJ...800...24K}. Also for the physics of environment quenching various options exist. Ram-pressure stripping is one possibility \citep{1972ApJ...176....1G}, perhaps favored by the finding that the quenched fraction of satellites correlates better
with the local overdensity inside groups than with group richness, a proxy for the mass of the host halo \citep{2012ApJ...757....4P}. But other options include strangulation \citep{1980ApJ...237..692L} and harassment \citep{1996Natur.379..613M}.

In the Peng et al. phenomenological model, mass quenching starts promptly at relatively high redshifts and, of course, acts preferentially on massive galaxies. Environment quenching sets in progressively, becomes more important at lower redshifts, following the growth of overdensities and large scale structures.

In this paper we focus on mass quenching, and to do so we deal only with the most massive galaxies, namely those with $M_{\star}\geq10^{11}\,\msun$.
We then explore the relatively narrow redshift range $1.4\leq z\leq 2$, because this corresponds to an epoch when mass quenching has started to work at full steam, while environmental quenching is just about to begin\footnote{For example, in the case of the \citet{2010ApJ...721..193P} phenomenological model at $z\sim 2$ and $M*=10^{11}\,\msun$ the mass quenching rate is over two orders of magnitude higher than the environment quenching rate, see their Figure 13.}. All such massive star forming galaxies must soon start to be quenched, otherwise --keeping to form stars at the {\it main sequence} rate-- would soon result in a dramatic overgrowth of their mass \citep{2009MNRAS.398L..58R,2010ApJ...721..193P}. So, all such galaxies are almost immediate precursors to quenched, passively evolving, red and early-type galaxies. Thus, a mass selection  $M_*\geq 10^{11}\,\msun$ of galaxies at the mentioned redshifts ensures that it will include only galaxies which are either already quenched, or being in the course of quenching, or that will soon (i.e., $\lesssim 10^9$ years) enter the quenching phase.

Such mass and redshift range has been already widely explored by many observational studies using large databases, in particular trying to address the quenching issue \citep[e.g.,][]{2011ApJ...742...96W,2013ApJ...765..104B,2014ApJ...780....1W}. Some of these studies deal with very large numbers of galaxies to ensure the statistically significant samples that are crucial to understand the global trends  of galaxy evolution. Given the sheer size of the samples, automated methods are used to derive a few fundamental parameters for each galaxy, such as SFR, mass, structural parameters etc. As such, not all information present in the data could be used. We then restrict ourselves to work, intensively rather than extensively,  on a quite small sample of galaxies, examining them one by one. 
The purpose is first to unambiguously distinguish between galaxies which are forming stars at the main sequence rate, those which are already quenched, and those which appear to be still star-forming, but at a rate significantly below the main sequence, which may be  caught in the {\it quenching} phase. We also aim  to investigate possible links between galaxy specific star formation rate (sSFR), morphology, AGN activity. The presence of a well known overdensity at $z\simeq 1.61$ in the GOODS-S field \citep{2007ApJ...671.1497C,2009A&A...504..331K,2011ApJ...743...95G} also allows us to check the relative abundance of quenched galaxies, bulge-dominated galaxies, and AGNs within and outside the overdensity.  

The paper is organized as follows. Section~\ref{sec:sample} describes the sample selection and the used catalogs and data-sets. In Sections~\ref{sec:MS} we derive galaxy stellar mass and SFR and define the $M_*-$SFR main sequence at $1.4\leq z \leq 2$, used as reference throughout all the paper. AGN hosts are discussed in Section~\ref{sec:agn}. In Section~\ref{sec:subsamples} galaxies are classified in quenched or star-forming, based on their position relative to the main sequence, multi-wavelength information, colors, and SED-fitting analysis. In Section~\ref{sec:morphology} we perform morphological analysis based on single- and double-component surface brightness fitting, and separately treat some peculiar system. Based on our results, we discuss the role of the AGN activity, and of the environment, in quenching star-formation in Sections ~\ref{sec:agn_q}, and \ref{sec:environment_q}, respectively. Section~\ref{sec:conclusions} presents the conclusions. Throughout the paper we assume a $\Lambda$CDM cosmology with $H_{\rm 0}=70$ km\ s$^{-1}$\ Mpc$^{-1}$, $\Omega_{\rm M}=0.27$, and $\Omega_{\Lambda}=0.73$. All stellar masses and SFR are quoted for a \citet{1955ApJ...121..161S} initial mass function (IMF), and magnitudes are given in the AB photometric system, unless explicitly stated otherwise.

\section[Data and Sample]{Data and Sample Selection}\label{sec:sample}
The relatively small field ($10'\times 16'$) of the Great Observatories Origins Deep Survey-South \citep[GOODS-S,][]{2004ApJ...600L..93G} represents the optimal choice for studying in detail a representative sample of the most massive galaxies at $1.4\leq z\leq 2$. In fact, it provides at the same time  both high-resolution images in the near-IR with WFC3/{\it HST} \citep[CANDELS,][]{2011ApJS..197...35G, 2011ApJS..197...36K}, and the deepest data in the mid-IR \citep[{\it Spitzer}, ][]{2004ApJS..154...25R}, far-IR\citep[{\it Herschel}, PEP/HerMES,][]{2011A&A...533A.119E, 2011A&A...532A..90L}, and X-ray band \citep[{\it Chandra} Deep Field-South, CDF-S,][]{2011ApJS..195...10X}. 
The corresponding database is crucial to study the optical rest-frame morphology, accurately estimate the SFR for the most massive objects (most of which are highly obscured by dust), and to unveil X-ray excess due to the AGN contribution. Last but not least, the large amount of ground-based (VLT) and space-based ({\it HST}, {\it Chandra}, {\it Spitzer}) 
data from the GOODS Treasury Program, and the availability of spectroscopic redshifts from the GOODS and GMASS programs \citep[][]{2005A&A...434...53V,2008A&A...482...21C,2009A&A...494..443P,2013A&A...549A..63K} enable us to characterize the studied galaxies in an optimal way as currently possible. Of course, the resulting sample will be too small for coping with statistics or cosmic variance, but this is not a limitation in our {\it intensive} approach.
 
In this work we focus on the 56 most massive ($M_*\geq 10^{11} M_{\odot}$) galaxies at $1.4 \leq z \leq 2$,  31 of which have spectroscopic redshifts. Two more objects entered the original sample, but they have been excluded from this study, because their WFC3/{\it HST} images are not available, or too noisy to perform surface brightness fitting, due to the proximity of saturated stars. 
The sample was culled from the $K$-selected ($K(\rm Vega)<22$) multi-band catalog of \citet[][hereafter, D07]{2007ApJ...670..156D}, including data from all the available filters in GOODS-S, i.e., {\it HST}/ACS optical, F435W ($B$), F606W ($V$), F775W ($I$), and F850LP($z$), VLT/ISAAC near-IR, $J$, $H$, $K$, and {\it Spitzer}/IRAC, 3.6, 4.5, 5.8, and 8.0~$\mu$m \citep[for more details on the data-sets see ][]{2004ApJ...600L..93G}.  
The optical/near IR photometry was then complemented with the 24~$\mu$m catalog (Daddi et al., in prep.), built as summarized in Section~\ref{sec:mipsdata} from the {\it Spitzer}/MIPS images \citep{2004ApJS..154...25R}. 
Far-IR fluxes were extracted (Daddi et al., in prep.) from the publicly released PACS 70-160~$\mu$m data from Herschel GOODS \citep[H-GOODS, ][]{2011A&A...533A.119E}, and SPIRE 250~$\mu$m data from Herschel Multi-tiered Extragalactic Survey \citep[HerMES, ][]{2010A&A...518L..21O}. 
For objects without spectroscopic information we used photometric redshifts from the public GOODS-MUSIC catalog, which agree well with the spectroscopic ones ($\Delta z/(1+z)\simeq 0.03$) for galaxies at $z<2$ \citep[cf.][]{2006A&A...449..951G,2007A&A...465..393G}. Although the uncertainties on photometric redshifts could result in the inclusion  in the sample of a few lower-redshift contaminants, we avoided applying color criteria to pre-select high-$z$ objects, in favor of completeness.
The {\it HST}/WFC3/F160W $H$-band image mosaic ($>5\sigma$ point source sensitivity for  $H_{160}< 27.7$, AB system) drizzled to a pixel scale of $0\farcs06$, was exploited to study galaxy morphology in the optical-rest frame with a very high resolution (FWHM$\sim0\farcs18 \simeq 1.5$~kpc). For more details on the observations and data reduction see  \citet{2011ApJS..197...35G} and  \citet{2011ApJS..197...36K}.

\subsection{MIPS fluxes}\label{sec:mipsdata}
The MIPS source extraction on the whole GOODS-S MIPS/24~$\mu$m image mosaic will be described in a future paper (Daddi et al., in prep.). Here we briefly summarize the main issues, useful for the purpose of this work.
The 24~$\mu$m/MIPS counterparts were identified, based on the IRAC (3.6~$\mu$m) prior positions, 
using a PSF fitting method, with GALFIT \citep[version 3, ][]{2010AJ....139.2097P}.
The measurement errors were calibrated on direct Monte Carlo simulations, inserting one artificial source at a time in the real image, and measuring it together with all other priors. 
The catalog was built in such a way that a MIPS flux, or a flux upper limit, in case of MIPS-undetection, is associated to each IRAC source. 
The use of IRAC priors helps with the deconfusion of objects blended in MIPS, but is still biased against sources which are also blended in IRAC.
In the D07 catalog, IRAC sources are associated to the $K$-band coordinates using an encircling radius of 2\farcs0. However, when the angular separation between the IRAC and $K$-band positions is $\Delta\theta_{K-IRAC}> 0\farcs5$, the galaxy is probably blended in IRAC (cf., D07), and generally disregarded. Since objects with $\Delta\theta_{K-IRAC}> 0\farcs5$ represent a not negligible fraction in our sample (i.e., 6/56), we used prior positions from {\it HST}/WFC3/F160W to deblend them, as described in Section~\ref{sec:subsamples} and Appendix~\ref{app:A}.

\begin{figure}
\includegraphics[width=\columnwidth]{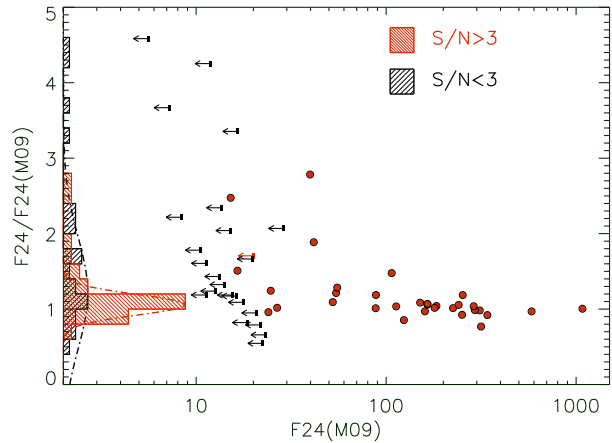}
\caption{The ratio between our 24~$\mu$m/MIPS fluxes (F24, Daddi et al. in prep.) and those from M09 (F24(M09)), as a function of F24(M09). For undetected sources (S/N$<3$) we plot 2 or 3$\sigma$ flux upper limits, as detailed in the text. Objects detected (undetected) in both catalogs are shown as red filled circles (black arrows). The object \#5530 detected in our catalog, but undetected in M09, is shown as a red arrow. The red and black shaded histograms (scaled down to arbitrary units), show the distributions of sources which are detected and undetected in our catalog, respectively. Gaussian fits to the histograms are shown as red and black dot-dashed lines.
}\label{fig:snr24um}

\end{figure}

We considered as MIPS-detection all sources with S/N$ \geq 3$ (i.e., detected above 3$\sigma$). For undetected sources, we derived upper limits from the flux errors at 2$\sigma$ or 3$\sigma$, for sources with S/N$< 1$ and $1\leq$S/N$< 3$, respectively.
In Figure~\ref{fig:snr24um} we compare our results with those from the catalog of \citet[][hereafter M09]{2009A&A...496...57M}. 
The figure shows the ratio between the MIPS fluxes from our catalog and those from the M09 catalog (F24/F24(M09)) as a function of F24(M09). For sources undetected in M09 we estimated 2$\sigma$ or 3$\sigma$ flux upper limits as we did for objects in our catalog. With one exception, all the sources detected in our catalog are also detected in M09 and are shown as red filled circles in Figure~\ref{fig:snr24um}. For sources undetected in both catalogs the black arrows show the corresponding upper limits in F24(M09)  and the vertical axis show the ratio of the upper limits in the two catalogs. The exception, object \#5530, is shown as a red arrow. The red (black) shaded histogram show the corresponding distributions of flux ratios (of the ratio of upper limits) for detected (undetected) sources. Gaussian fits to the histograms, shown as dot-dashed lines, give a mean ($\mu$) and standard deviation ($\sigma$) as $\mu\pm\sigma = 1.06\pm0.14$, and $1.21\pm0.74$, for detected, and undetected sources, respectively. The comparison shows that for sources detected above 3$\sigma$, the 24~$\mu$m/MIPS fluxes are in general agreement with those from M09. On average, also the upper limits are in fair agreement between the two catalogs, whereas for a small number of objects our upper limits are more conservative.

\section[MS at $\langle z=1.7\rangle $]{The Main Sequence (MS) of star-forming galaxies at $\mathbf{ \langle \lowercase{z}\rangle=1.7}$}\label{sec:MS}

It is now well established that a tight relation exists between the stellar mass ($M_{\star}$) and the star formation rate (SFR) for star-forming galaxies, both in the local Universe and at high redshift (\citealt[][hereafter D07]{2007ApJ...670..156D}, \citealt{2007A&A...468...33E}, \citealt{2007ApJ...660L..43N}, \citealt{2009ApJ...698L.116P}, \citealt{2011ApJ...730...61K, 2014MNRAS.443...19R,2015A&A...575A..74S}). At all redshifts the majority of star-forming galaxies follow this relation within a $\lesssim 0.3$ dex dispersion with a  small fraction of outliers \citep[cf.][]{2011ApJ...739L..40R}, thus it is commonly called the \emph{main sequence} (MS) of star-forming galaxies. 
 While the slope of the SFR-$M_*$ relation does not change much with redshift,  its zero-point, i.e., the specific SFR (sSFR=SFR/$M_*$) at fixed mass  gradually increases with redshift at least up to $z\sim 2.5$ \citep{2011ApJ...742...96W,2012ApJ...747L..31S,2012ApJ...754L..29W}. At higher redshifts, it is still debated whether the sSFR  remains constant or keeps (slightly) increasing \citep{2010ApJ...713..115G, 2013ApJ...763..129S,2014A&A...563A..81D,2013A&A...557A..66B,2014ApJ...781...34G}.  
So, massive galaxies at high redshifts formed stars and grew their mass at a much faster rate, fueled by a larger reservoir of fresh gas, with respect to local galaxies \citep[][]{2010ApJ...713..686D,2010Natur.463..781T}.  

To study our sample galaxies as a function of their position on the $M_{\star}$-SFR diagram, we define here the MS of star-forming galaxies at $\langle z\rangle\approx 1.7$, to be used as a reference in quantifying as sSFR/sSFR(MS) the deviation form it of any given galaxy.
 We adopted the following empirical relation from \citet{2012ApJ...757L..23B}, based on the \citet[][]{2012ApJ...747L..31S} parametrization of the sSFR evolution with redshift for MS galaxies, which in turn is derived by fitting a combination of mid-IR, UV, far-IR, and radio data: 

\begin{multline}
  {\rm sSFR}_{\rm MS}(z,M_*= {\rm sSFR}_{\rm MS,0})\times \left(\frac{M_{\star}}{10^{11}M_{\odot}}\right)^{\beta_{\rm MS}}\\
   \times(1+z)^{\gamma_{\rm MS}} ,\label{eq:sSFRev} 
\end{multline}

\noindent  
where sSFR$_{\rm MS,0}=0.063$ Gyr$^{-1}$ is the sSFR at $z=0$ for $M_{\star}=10^{11}~M_{\odot}$, $\beta_{MS}=-0.21$ is the slope of the sSFR-$M_{\star}$ relation derived by \citet{2011ApJ...739L..40R}, and $\gamma_{MS}=3$ describes the evolution of the normalization of the MS out to redshift $z=2.5$. All these parameters are given in \citet[][Table~1]{2012ApJ...757L..23B} assuming a \citet{1955ApJ...121..161S} IMF. 
From Eq.~\eqref{eq:sSFRev} it follows that the MS at $\langle z\rangle\approx 1.7$ 
is given by:
\begin{equation}
{\rm log\, SFR = -6.6 + 0.79\,log\, M_*}. \label{eq:MSz1e7}
\end{equation}
\noindent

\subsection[Stellar Mass]{Stellar Mass}\label{sec:mass} 
The stellar masses were computed based on the $K$-band magnitudes, according to the empirical relation derived in D04 and D07 (with a typical error of a factor of $\sim 2$), for homogeneity with the samples used to derive the reference main sequence \citep[i.e.][]{2011ApJ...739L..40R,2012ApJ...747L..31S, 2012ApJ...757L..23B}, in which stellar masses were derived using this method. 
In D04 the empirical relation was tested and calibrated on the results from best-fit Spectral Energy Distributions (SED) of \citet{2004A&A...424...23F}, using \citet{2003MNRAS.344.1000B} stellar population models (hereafter BC03), and Salpeter IMF. As shown in Section~\ref{sec:sedfitting}, we verified that the stellar masses derived with this method ($M_{\star \rm D04}$) agree well with those obtained from the SED fitting analysis (i.e., $M_{\star \rm  SED}$) also for our massive sample.

\subsection{Star Formation Rate}\label{sec:sfr}
By combining the deep PACS {\it Herschel} data with all the high-quality multi-band information at shorter wavelengths, we estimated  SFRs including both the contribution of the UV rest-frame (from the unobscured part of the light), and of the total IR emission, i.e., the light reprocessed by dust. The total SFR is then indicated as: 

\begin{equation}
    {\rm SFR(IR+UV) = SFR(IR) + SFR(UV).}
\end{equation}

\noindent
Following \citet{1998ARA&A..36..189K}, we inferred the UV contribution (SFR(UV)), from the 1500~\AA\ rest-frame luminosity ($L_{1500}$), which in turn was derived from the observed $B$-band magnitude, as done in D07. 

The IR contribution, SFR(IR), was derived from the total IR luminosity integrated over the full infrared spectrum, i.e., rest-frame L[8-1000~$\mu$m] (hereafter LIR) using the \citet{1998ARA&A..36..189K} conversion, for a Salpeter IMF (i.e., $SFR(IR)=1.7\times 10^{-10} \msun {\rm yr^{-1}} L_{\odot}^{-1}$). 
LIR was measured by fitting the {\it Spitzer}/MIPS and {\it Herschel}/PACS/SPIRE photometry (i.e., from 24 $\mu$m to 250 $\mu$m) to the SED templates for MS galaxies at $z =1.325-2.25$ from \citet{2012ApJ...760....6M}. The possible contribution of an obscured AGN was also taken into account in the fit, as in \citet{2011ApJ...739L..40R}.
The uncertainties on LIR were estimated from the $\chi^2$ variations following \citet{1976ApJ...210..642A} for the case of one single parameter.
This is a robust method to retrieve the LIR when a galaxy is detected in at least  two filters (i.e., MIPS/24~$\mu$m and one {\it Herschel} filter). Otherwise, it only provides an upper limit to LIR. 
Hence, the LIR of objects detected in MIPS, but undetected in {\it Herschel}, was estimated based on the 8 $\mu$m rest-frame luminosity ($L8=\nu \ L_{\nu}[8\mu\rm m]$), according to the empirical relation proposed by \citet{2011A&A...533A.119E} : 
\begin{equation}
{\rm LIR} = L8 \times 4.9^{+2.9}_{-2.2}.\label{eq3},
\end{equation}
where $1\sigma$ errors on the coefficient are indicated. For galaxies in the considered redshift range ($z=1.4-2$) the 8 $\mu$m rest-frame luminosity can be derived from the observed MIPS/24~$\mu$m flux ($F_{24}$), with a relatively small {\it k-correction} depending on the redshift ($0.8\lesssim k_{\rm corr}(z) \lesssim 1.9$), as :
\begin{equation}
L8=\nu_{24\mu \rm m}\ F_{24}4\pi DL(z)^2\ k_{\rm corr}(z), \label{eq4}  
\end{equation}
\noindent
where {\it DL(z)} is the Luminosity distance. The {\it k-correction} was derived using the \citet{2012ApJ...760....6M} templates for MS galaxies at $z =1.325-2.25$. For MIPS-detected/PACS-undetected objects, we verified that the MIPS-derived LIRs are consistent with the 2$\sigma$ upper limits derived from the IR SEDs. 

For objects undetected in MIPS, upper limits to the LIR, and correspondingly SFR(IR), were estimated from the MIPS/24~$\mu$m detection-limit (i.e., accurately measured on the MIPS images, as described in Section~\ref{sec:mipsdata}), using Equations~(\ref{eq3}) and~(\ref{eq4}). We ignore here a possible contribution to LIR coming from circumstellar dust in old stellar populations, which may be relevant for sub-MS galaxies. Hence, for such galaxies, the resulting SFR(IR+UV) should be regarded as an upper limit, while a better constraint on the SFR can be derived from optical/near-IR SED fitting (see Section~\ref{sec:sedfitting}).

\subsection[AGN]{Identification of AGN hosts}\label{sec:agn} 
We cross-correlated our sample with both the 4 Ms {\it Chandra} X-ray catalog \citep[][]{2011ApJS..195...10X}, and the VLA (1.4 GHz) radio catalog \citep{2008ApJS..179..114M}, using an encircling radius of 1\farcs0, and found 19 and 5 counterparts, respectively (cf., Table~\ref{tab:1}). 
All the radio-detected objects are also X-ray detected, at least in one of the three bands of the 4 Ms {\it Chandra} catalog (i.e., Soft: 0.5--2 kev, Hard: 2--8 kev, and Full: 0.5--8 kev). All the X-ray-detected objects in our sample had been classified as ``AGN candidates'' by \citet{2011ApJS..195...10X}. However, we considered as AGN hosts only those galaxies showing an excess in X-ray and/or radio luminosity with respect to the expected SFR, derived by summing the UV- and the IR-derived contributions, i.e., SFR(IR+UV). We used the \citet{2014MNRAS.437.1698M} calibration, relating the total, integrated X-ray luminosity LX[0.5---8 kev] to the galaxy SFR (LX/SFR=$(4.0\pm0.4)\times 10^{39}$ [erg\ s$^{-1}$/ $\msun$\ yr$^{-1}$]), and found that all the X-ray detected galaxies, but two (i.e., \#4705, and \#5503) show an X-ray excess (at least by a factor of 3).  
In a similar way, by adopting the conversion of \citet{2010A&A...518L..31I,2010MNRAS.402..245I}, we verified that all our radio-detected galaxies (\#686, \#3231, \#5503, \#5860, and \#5898) have a higher radio-emission (by a factor $> 2.5$) than what expected only from star-formation, thus revealing the AGN contribution. Since \#5503 shows radio-excess, but no X-ray excess, we consider this object just as an AGN candidate. We notice that this source also exhibits a peculiar, asymmetric, morphology, and a large offset ($\sim 1$~arcs) between the optical and the X/radio positions, indicating that it could be a merging system. In the following figures it will be flagged with a different symbol (large circle + four-pointed star) with respect to the other more secure AGNs, just marked as four-pointed stars. Based on the LX-SFR(IR+UV) comparison, object \#4705 is instead considered as a normal star-forming galaxy. We conclude that 18 X-ray detected AGNs (17 secure, and one candidate) are included in our massive sample (see Table~\ref{tab:1}). This roughly corresponds to $\sim$ 30\% of the galaxies with $M_*\geq 10^{11}~\msun$ at $1.4\leq z \leq2$ in GOODS-S. Interestingly, we will show in the following that AGN hosts are not only among star-forming galaxies, but they represent $\sim$ 22\% of the quenched sources.

\section{Source classification}\label{sec:subsamples}
\subsection{Position on the $\mathbf M_*$-SFR plane}\label{sec:class}
\begin{figure*}
\includegraphics[width=\textwidth]{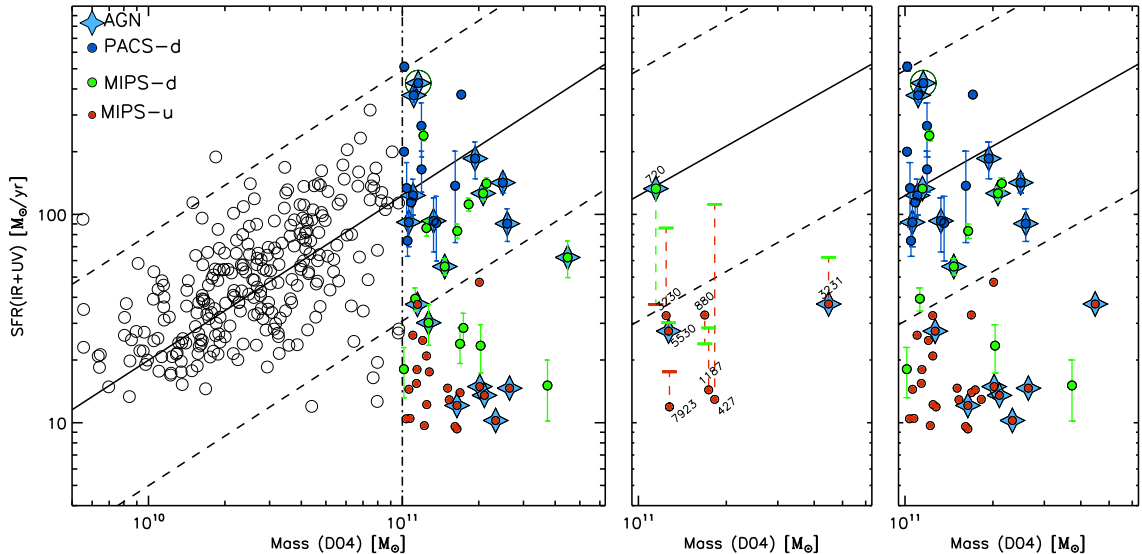}
\caption{In this figure we show the position of our massive sample galaxies on the MS at $\langle z\rangle=1.7$ (black solid line), before (left panel) and after (right panel) the MIPS-deblending procedure (see Appendix~\ref{app:A}). In the central panel we report the ``migration path'' of each deblended galaxy, from the position occupied before (horizontal bar, color coded as the object in the left panel), and after deblending (filled circle, color coded as the object in the right panel). The right panel shows the final positions with respect to the MS. 
The \pacsd, \mipsd, and \mipsu\ galaxies are displayed with different colors, as labeled in the figure, and explained in the text (Section~\ref{sec:subsamples}). Note that the red filled circles just represent 2, or 3$\sigma$ upper limits to the SFR for \mipsu\ galaxies (cf., Section~\ref{sec:mipsdata}, and ~\ref{sec:class}).
For comparison, lower-mass star-forming galaxies at $1.4\leq z \leq 2$, and detected at 24~$\mu$m/MIPS, from the parent sample, are shown the as black open circles.} 
\label{fig:pmulti}
\end{figure*}

The position occupied by our massive galaxies in the SFR-$M_*$ plot is shown in Figure~\ref{fig:pmulti}. The MS relation at $\langle z \rangle=1.7$, derived from Eq~\eqref{eq:MSz1e7}, is bounded by two dashed lines, representing offsets in SFR 4$\times$ above and 4$\times$ below the MS following \citet{2011ApJ...739L..40R}, and roughly corresponding to the $\sim 2.5\sigma$ dispersion. 
Here, and throughout all the paper, the sample is color coded based on the galaxy IR properties. Blue filled circles are objects detected at {\it Herschel} (PACS) wavelengths, and hereafter are indicated as the \pacsd\ sub-sample. 
Sources detected at 24~$\mu$m/MIPS, but not detected by {\it Herschel}, hereafter the \mipsd\ galaxies, are shown as green filled circles and have, on average, a lower specific SFR (sSFR) compared to the previous sub-sample. Then, objects which are not detected above 3$\sigma$ at 24~$\mu$m/MIPS (i.e., $F\nu_{24}\lesssim 20-25~\mu$Jy, on average) are  indicated as \mipsu\ galaxies, and the corresponding upper limits on their SFR are shown as red filled circles. Thus, this sub-sample should include quenched systems, or galaxies with a low residual of star-formation.  The 17 secure AGN hosts identified in Section~\ref{sec:agn} are highlighted by four-pointed stars, and the AGN candidate by a four-pointed star+circle. 
For comparison, we also show the parent sample of 24~$\mu$m/MIPS-detected star-forming galaxies at $1.4\leq z \leq 2$, for which the IR contribution to the SFR(IR+UV) was derived from Eq.~\eqref{eq3}, and~\eqref{eq4} (black open circles). 

As mentioned in Section~\ref{sec:mipsdata}, for completeness we also included in our sample the objects with $\Delta\theta_{K-IRAC}>0\farcs5$, which are blended in IRAC and, consequently, in MIPS. A close examination of {\it HST} and MIPS maps for all the 56 galaxies then revealed that 6 other  objects are blended and could have been misclassified (i.e., as \mipsd\ or \mipsu) by the automatic procedure associating the MIPS sources to $K$-selected counterparts.
Most of them reside in projected (or real) galaxy pairs, and in many cases both  objects in a pair are included in our massive sample (see Appendix~\ref{app:B}).
Hence, we attempted to deblend these objects to derive a more accurate estimate of their SFR, and re-classify them as MIPS-detected or undetected, if necessary, as briefly summarized below and further detailed in Appendix~\ref{app:A}). 

We used GALFIT to fit PSF models to the 24 $\mu$m maps with the F160W/WFC3 positions as priors, leaving the galaxy magnitude as the only free parameter to be recovered. This procedure worked quite smoothly in most cases but  some ``ambiguous'' objects are discussed in Section~\ref{sec:oddcases}.
In the right panel of Figure~\ref{fig:pmulti} we show the final SFR-$M_*$ diagram obtained after MIPS de-blending, while in the middle panel we show the `migration path' of each deblended galaxy from the position occupied before deblending (horizontal bar, color-coded as in the left panel) and after de-blending (filled circle, color coded as in the right panel). The `migration path' is also color-coded depending on the sub-sample to which the object belong after de-blending (e.g., migrations from the \mipsd\ to the \mipsu\ sub-sample are drawn in red, and vice-versa in green).
From Figure~\ref{fig:pmulti} it appears that several galaxies, previously classified as \mipsd, have been reclassified as \mipsu\ objects after deblending (\mipsd$\rightarrow$u in Table~\ref{tab:1}). 
This is the case for objects: \#427, \#880, \#1187, \#3230, \#3231, and \#5530. 
On the contrary, only galaxy \#720, previously indicated as \mipsu, has been reclassified as \mipsd\ (\mipsu$\rightarrow$d). Then, object \#7923 is confirmed to be a \mipsu\ galaxy, but with a lower limit on its 24~$\mu$m flux, hence on its  SFR. The de-blending procedure for each object is detailed in Appendix~\ref{app:A}.

All the \pacsd\ galaxies are within the MS band, while all the \mipsu\ galaxies are more than $4\times$ below it. 
The \mipsd\ sub-class instead includes objects  both on, and below the MS. Hence, some of the \mipsd\ galaxies could be quenching, or even quenched systems, judging from the relatively wide range that they cover in sSFR ($10^{-9}-10^{-10.4}~$yr$^{-1}$, i.e. from the typical MS rate down to $\sim 20\times$ lower). 
It is also possible that some \mipsd\ galaxy is instead undergoing a sort of rejuvenation process, caused by a merging/accretion event, although  we expect that  most such massive high-$z$ galaxies are evolving in the opposite direction. Finally, in some case the observed MIPS emission could be powered not (only) by star-formation, but (also) by obscured AGN, AGB stars, or old stellar population \citep{2009ApJ...700..161S}.

In the following sections (cf. Section~\ref{sec:sedfitting}, \ref{sec:colors}, and~\ref{sec:oddcases}), we combined the IR/X/radio information with the results from SED fitting, and two-color diagrams to better constrain the SFR of our galaxies and shed some light on the nature of some ambiguous objects.

\subsection[SED]{SED fitting}\label{sec:sedfitting} 
We built the multi-color SEDs by fitting the optical to IRAC/8~$\mu$m photometry using the {\it hyperzmass} software \citep{2000A&A...363..476B,2007A&A...474..443P}. For IRAC-blended objects we excluded the IRAC bands from the fit. However, we verified that the best-fit parameters do not change substantially using instead the IRAC fluxes from \citet[][]{2013ApJS..207...24G}, which are de-blended with the TFIT procedure. 
For galaxies at $1.4\leq z\leq1.7$, the 8~$\mu$m flux  was excluded from the SED fit, as at those redshifts it corresponds to a rest-frame part of the spectrum that may  include thermal dust and 3.3 $\mu$m PAH emission line, features which are not included in the used stellar population models. 
We adopted BC03 stellar population models, with a grid of star formation histories (SFH) including an SSP, a constant SFR, and exponentially-declining $\tau$-models ($\tau$=0.1, 0.3, 1, 2, 5, 10, 15, and 30 Gyr). To reduce the number of free parameters, we only used templates with solar metallicity. These adopted star formation histories are easy to handle but are certainly inappropriate for near-MS star-forming galaxies at these redshifts, where actually the SFRs must have {\it increased} with time (on average) as a result of the mere existence of the MS \citep{2009MNRAS.398L..58R,2010MNRAS.407..830M,2010ApJ...721..193P,2012ApJ...754...25R}. 
As shown by \citet{2010MNRAS.407..830M}, for star-forming galaxies, the {\it age} returned by the SED fit should be interpreted as the typical age of the stars contributing the bulk of the light, rather than the time since the beginning of star formation, as it formally represents in the fits. For this reason, to avoid estimating unrealistic low ages for such galaxies, in which the youngest stellar population outshines the older one, we imposed a lower age limit of 0.1 Gyr.

The {\it age} returned by the SED fits has yet a different physical meaning in the case of quenched galaxies, i.e., here assumed to be those for which only upper limits can be placed on SFR(UV+IR). As apparent from Table  \ref{tab:sedfit}, for many such galaxies the fits give $\tau=0$, which is to say that a single age, simple stellar population (SSP) provides the best fit. Correspondingly, the age can be interpreted as representative of the time elapsed since quenching. This is a fair interpretation also for the other cases, in which $\tau\ne 0$ but is $<<$ age, hence the current SFR is very low in the best fit exponentially declining model. In the following, we adopt this interpretation for the ages of quenched galaxies. 

\begin{figure*}
\includegraphics[width=0.35\textwidth]{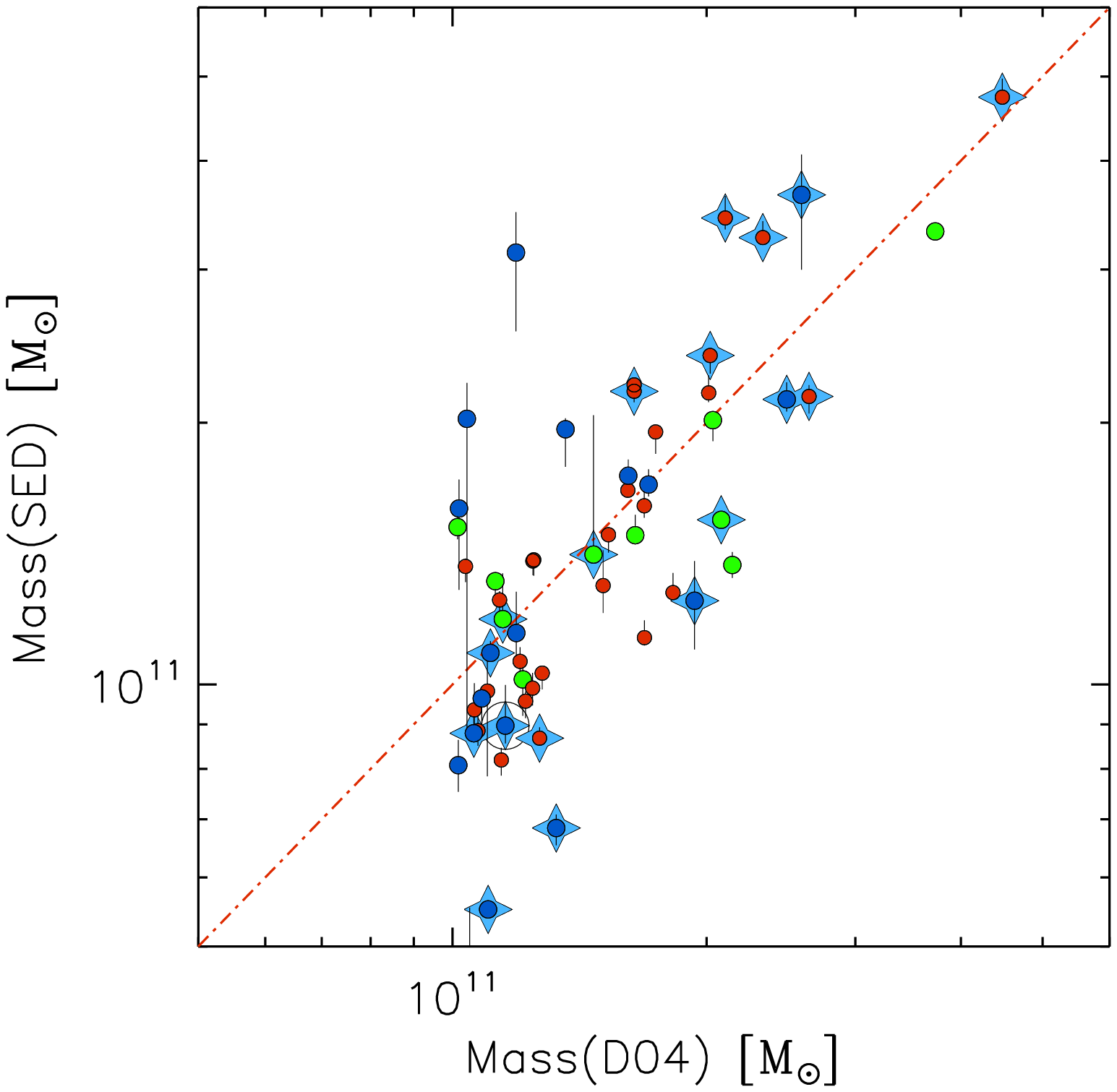}\includegraphics[width=0.35\textwidth]{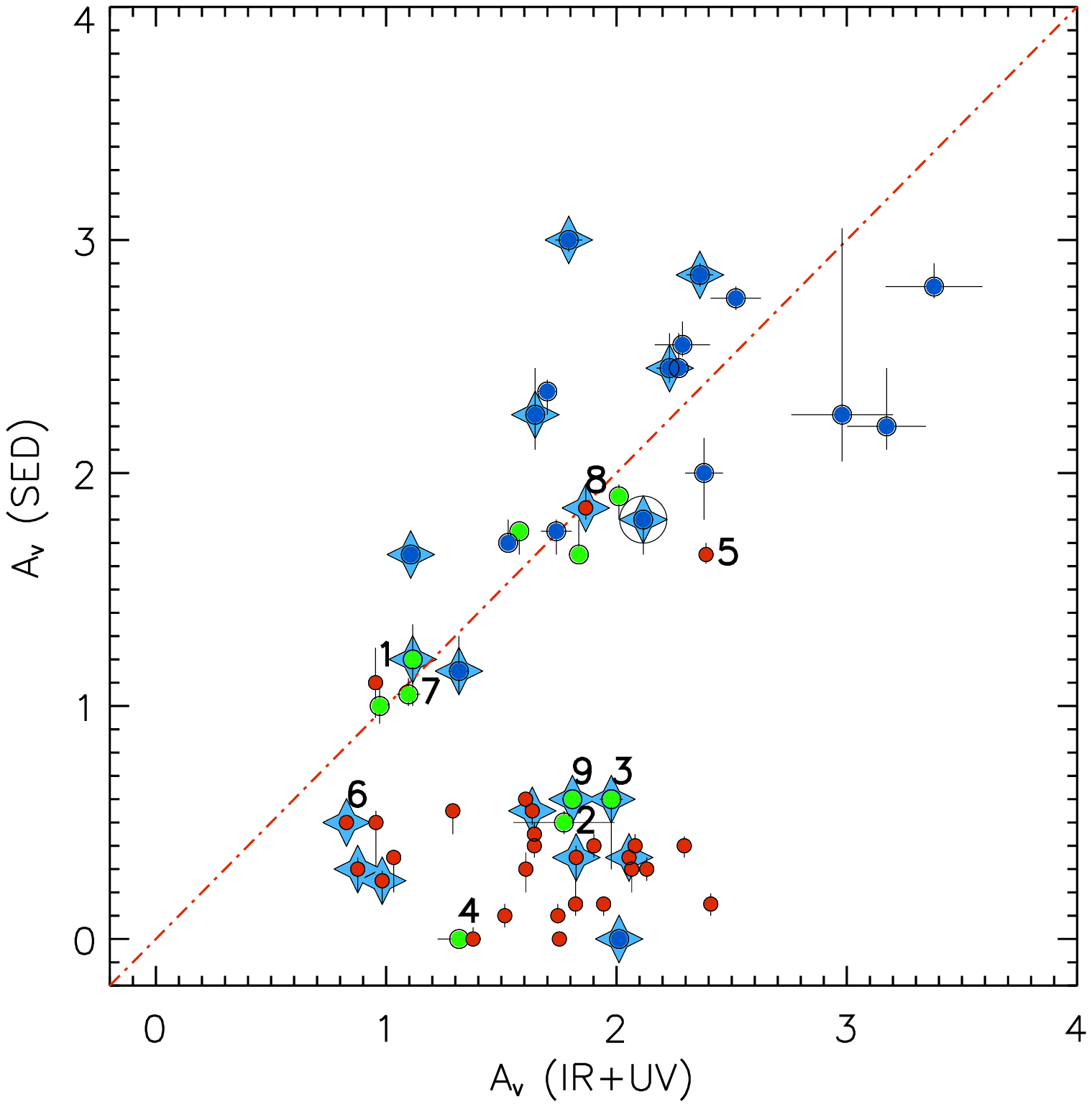}\includegraphics[width=0.35\textwidth]{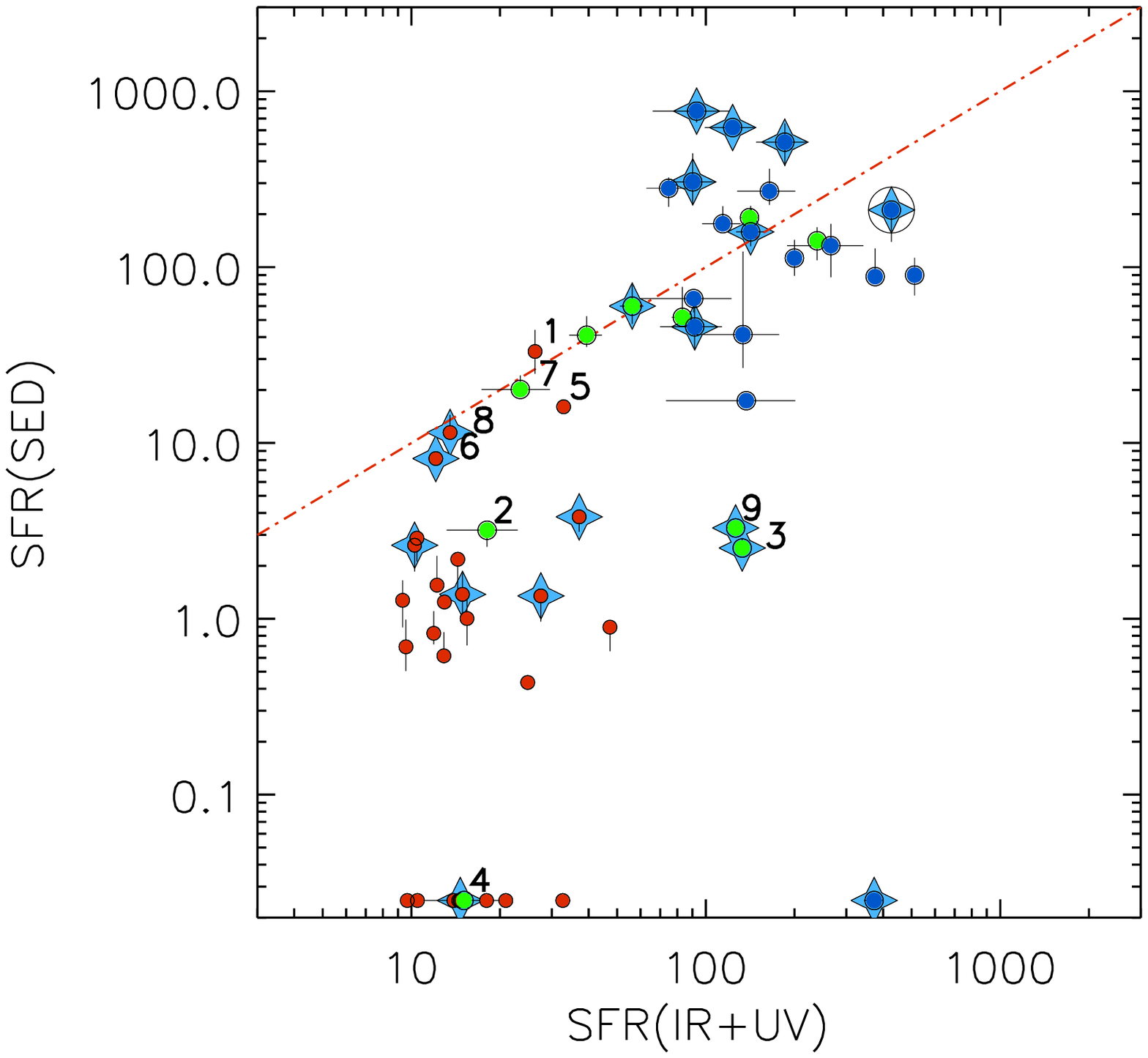} 
\caption{The SED fitting derived Mass, SFR, and reddening are compared with the same quantities, as derived with different methods. In all the panels symbols are like in Figure~\ref{fig:pmulti}. The numbers from 1 to 9, labeled in the middle and right panels, represent the odd cases discussed in Section~\ref{sec:oddcases}, and correspond to objects: \#428=1, \#557=2, \#720=3, \#856=4, \#880=5, \#2940=6, \#3853=7, \#5509=8, \#6193=9.  {\bf Left panel:}  Best-fit masses ($M_{*\rm SED}$) {\it vs} masses derived from $K$-band magnitude ($M_{*{D04}}$). {\bf Middle panel:} Best-fit reddening (A$_{\rm V,SED}$) {\it vs} $A_{\rm V}$(IR+UV), derived from the ratio between the total and UV uncorrected SFR. {\bf Right panel:} Best-fit SFR (SFR$_{SED}$){\it vs}  SFR(IR+UV), derived as explained in Section~\ref{sec:sfr}. }\label{fig:obsvssed}.
\end{figure*}

The SED fitting results are shown in Table \ref{tab:sedfit} and Figure~\ref{fig:obsvssed}, where symbols are like in Figure~\ref{fig:pmulti}.
In the left panel of Figure~\ref{fig:obsvssed} we compare the best-fit stellar masses derived using BC03 models, and Salpeter IMF ($M_{*\rm SED, BC03}$), with those derived following D04 and D07, as explained in Section~\ref{sec:mass} ($M_{\star \rm  D04}$). The two quantities agree well, with an average ratio close to unity ($\langle M_{\star \rm D04}/M_{\star \rm  SED, BC03} \rangle=1.05$) and a $1\sigma$ scatter of 0.16 dex. 
By inspecting the optical/nIR best-fit SEDs of the AGN hosts we found that most of them do not seems to suffer significant contamination by the AGN emission. On the contrary, two objects (i.e., \#1906, and \#6898) show a typical `IRAC power-law' SED, indicating that the AGN component is swamping the stellar emission \citep{2006ApJ...640..167A,2007ApJ...660..167D}. Hence, for these two objects, in particular, and also for all the AGN hosts with at least a MIPS detection (11/18), we verified that the adopted stellar masses ($M_{\star \rm  D04}$) are consistent, (within $\sim 0.25$ dex) with those derived by fitting the optical to far-IR SEDs with a modified version of the MAGPHYS software, which also includes the AGN warm dust component \citep{2008MNRAS.388.1595D,2013A&A...551A.100B}.

While the stellar mass inferred from SED fitting is relatively robust, both for quenched and star-forming galaxies \citep[cf.][]{2010MNRAS.407..830M}, the other SED-fitting derived parameters are affected by the well-known degeneracy between the galaxy {\it age} and reddening (A$_{\rm V, SED}$), then also affecting the measured SFR(SED).
However, if properly combined with multi-wavelength data (i.e., IR, radio, X), the SED fitting analysis can provide some insight on the nature of the studied galaxies. For instance, it can help to better constrain the reddening and SFR, for galaxies that are undetected in MIPS, and, in general, for galaxies below the MS. This is visible from the middle and right panels of Figure~\ref{fig:obsvssed}, where the best-fit reddening values and SFRs are compared with those derived from the IR and UV-uncorrected luminosities ($A_{\rm V, IR+UV}$ and SFR(IR+UV)). For almost all the \mipsu\ galaxies, the fit suggests lower values of $A_{\rm V, SED}$, and SFR(SED), with respect to the $A_{\rm V}$(IR+UV) and SFR(IR+UV) upper limits. This strengthen the interpretation of such galaxies as already quenched, or nearly so. When considering only the \pacsd\ and \mipsd\ galaxies, $A_{\rm V, SED}$ and $A_{\rm V}$(IR+UV) are on average in good agreement, with a relative difference of ($A_{\rm V, SED}-A_{\rm V}$(IR+UV))/A$_{\rm V}$(IR+UV))=-0.09 and $1\sigma$ scatter of $\sim$0.41. Some particular cases (flagged with numbers from 1 to 9) in the right and middle panels of Figure~\ref{fig:obsvssed} are discussed in Section~\ref{sec:oddcases}.

\subsection{Observed and rest-frame two-color plots}\label{sec:colors}
\begin{figure*}
\includegraphics[width=\textwidth]{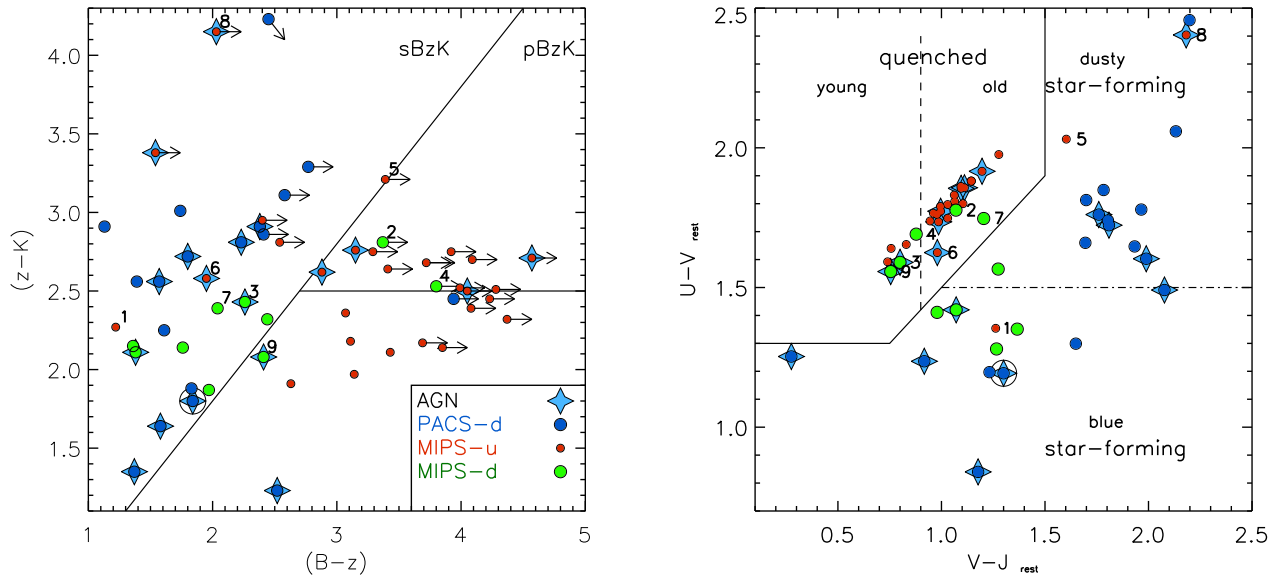}
\caption{BzK and UVJ diagrams for the three sub-samples, where \mipsu\ are the red filled circles, \mipsd\ are green filled circles and \pacsd\ are blue filled circles, as labeled in the right panel of the figure. X-ray sources and IRAC power-law are also highlighted by cyan stars. The odd cases discussed in Section~\ref{sec:oddcases} are flagged with numbers from 1 to 9, as in Figure~\ref{fig:obsvssed}.
}\label{UVJBzK}
\end{figure*}

In Figure~\ref{UVJBzK} we explore the observed and rest-frame colors of our sample in two color-color diagrams widely used in the literature to separate star-forming from passive galaxies at high redshifts, i.e., the $BzK$ plot \citep[][]{2004ApJ...617..746D} and the rest-frame $(U-V)$  vs. $(V-J)$ (hereafter, $UVJ$) plot \citep{2005ApJ...624L..81L, 2007ApJ...655...51W,2009ApJ...691.1879W}.
The different sub-samples are color-coded as labeled in the right panel, and detailed in Section~\ref{sec:class} and Figure~\ref{fig:pmulti}.  
The $BzK$ diagram is  a quite efficient criterion to separate galaxies at $1.4\leq z\leq 2.5$ from those at lower-redshift and among these  high-$z$ galaxies it separates the star-forming  ones (\sbzk, to be found in the left part of the diagram) from the passive ones  (\pbzk, located  within the wedge in the top-right side of the plot). Lower-$z$ systems are to be found in its bottom-right region of the plot. Rightwards arrows in the $BzK$ diagram indicate lower limits to the $B-z$ color, for objects undetected in $B$-band. These very red objects include both quenched galaxies (that would be classified as \pbzks\ with deeper $B$-band data) and highly-reddened dusty star-forming galaxies \citep{2014MNRAS.443...19R}.  
The $UVJ$ diagram (right panel) has the advantage of allowing to separate dusty star-forming from quenched galaxies at any redshift. In fact, since the dust-free quenched objects have bluer $V-J$ colors with respect to the extincted star-forming galaxies with similarly red $U-V$ colors, they can be isolated on the top-left box of the diagram. We note that galaxies in this region define a tight sequence, which is almost parallel to the diagonal border line of the quenched box. The extent of the ``quenched sequence'' mostly depends on the age range spanned by this population, which in our sample is $\sim 0.5-2$ Gyr. In fact, young and old quenched objects could be split according to their $V-J$ color along the sequence \citep[cf.][]{2010ApJ...719.1715W}. The tightness of the ``quenched sequence'' in the $UVJ$ plane is instead related to the fact that most of these systems are affected by a small amount of dust reddening. The $U-V$ color of star-forming galaxies depends on their dust extinction, and separates blue from red dusty objects \citep[cf.][]{2014ApJ...796...35F}.

In the left panel of Figure~\ref{UVJBzK} some objects do not match the $BzK$ criterion, i.e., occupy the lower part of the plot that should contain $z<1.4$ galaxies (namely, \#282, \#427, \#686, \#887, \#1187, \#1272, \#3230, \#4705, \#5556, \#6193, \#7491, \#7617, and \#7923). 
For some objects (i.e., those closer to the boundaries of the \pbzk\, or \sbzk\ regions) this could be due to even small photometric errors, and in one case to Type-1 AGN contamination (\#686).    
Then, 4 of the 7 \mipsu\ galaxies with the bluest $z-K$ colors (i.e., \#282, \#7491, \#7617, and \#7923), do not have spectroscopic identification, and could be at slightly lower redshift. 
However, as shown in Figure~8 of \citet{2004ApJ...600L.127D}, the colors of these objects are also consistent with those of relatively young ($\lesssim 2$~Gyr) passively evolving galaxies at $z\sim 1.4-1.5$, characterized by a rapidly declining SFR, which could be missed by a strict $BzK$ selection. It is worth noticing that most of these bluer \mipsu\ galaxies also lie in the young-quenched region of the $UVJ$ diagram, and have SEDs best-fitted with SSPs, or declining $\tau=0.1$ Gyr models with age $\sim 0.7- 1.3$ Gyr. Hence, we did not exclude any of these sources from the sample.     

In Figure~\ref{UVJBzK} the galaxy colors strongly correlate with their position relative to the MS: \mipsu\ galaxies are generally  in the ``quenched regions'' (\pbzk, and $pUVJ$), and \pacsd\ galaxies in the star-forming regions (\sbzk, and $sUVJ$), in both the panels. With two exceptions, the \mipsd\ objects lie in the star-forming region in the $BzK$ diagram, while half of them are in the quenched region in the $UVJ$ diagram. In the next section, we will discuss these objects, together with other peculiar cases. 

\subsubsection{A few odd cases}\label{sec:oddcases}
A few objects show $BzK$ or $UVJ$ colors and  best-fit SED which are apparently at odd with respect to their position relative to the MS. In some cases this is due to (even small) photometric errors, but in other cases this could indicate a different nature of these ``outliers'' (i.e., star-forming galaxies with a relatively old age, on the way to be quenched, or rejuvenated, AGN, etc.), as discussed below. To facilitate the identification of these objects in Figure~\ref{fig:obsvssed} and \ref{UVJBzK}, we flagged them with numbers from 1 to 9.

Object {\bf \#428} ({\bf `1'} in the figures) is a \mipsu\ galaxy, but shows colors similar to star-forming galaxies, i.e., lies in the \sbzk\ and $sUVJ$ regions. 
Judging from its appearance (see Figure~\ref{fig:lens}), this object may be  a lens Early-Type galaxy surrounded by a blue arc. So, the blue colors can be due  to the magnified background source when the photometry does not separate the arc from the  central galaxy. Indeed,  the rest-frame colors of the central galaxy,  having  masked out the blue arc as detailed in Section~\ref{sec:bsut},  are $(U-V)=1.93$ and $(V-J)=1.58$, consistent with those of a quenched object (see Section \ref{sec:lens}). 
 
There are two \mipsu\ objects in the dusty-star-forming box of the $UVJ$ plane.
They are {\bf \#880}, and {\bf \#5509} (flagged as {\bf `5'}, and {\bf `8'}), and in the $BzK$ diagram are shown as lower limit to $B-z$, being undetected in $B$-band. Their $BzK$ colors are compatible with those of old red galaxies, but the SED fitting analysis indicates a higher reddening and SFR(SED) with respect to other \mipsu\ objects (cf. Figure~\ref{fig:obsvssed} and Table~\ref{tab:sedfit}), yet still in agreement with the upper limits derived from the 24~$\mu$m/MIPS.  
In particular, since object \#880 is one of the MIPS-deblended sources (see Section~\ref{sec:mipsdata} and Appendix~\ref{app:A}), we do not exclude that it has a low residual star-formation (partly contributing to the MIPS blended source). However, it is also very close to the boundaries of the $pUVJ$ region, and, alternatively, its odd position in the $UVJ$ color diagram could just depend on small photometric errors.   
The AGN host \#5509 is undetected, or very faint, in all the optical bands. Its location in the dusty-$sUVJ$ could be due to residual obscured star-formation, which could also be responsible of the marginal 24~$\mu$m/MIPS detection of the object (2.5$\sigma$, see Section~\ref{sec:subsamples}, and Table~\ref{tab:1}). It is also possible that the AGN partly, or totally, contributes to the low MIPS emission of the source, and that its odd position on the $UVJ$ diagram is instead due to a higher redshift with respect to the photo-$z$ quoted here \citep[][for instance, \citealt{2010ApJS..187..560L} and \citealt{2014ApJ...796...60H} report $z_{\rm phot}\simeq 2.5$]{2006A&A...449..951G}. 

Object {\bf \#2940} (flagged as {\bf `6'}) is a \mipsu\ AGN host lying in the \sbzk\ region. It is in the quenched region of the $UVJ$ diagram, but its bluer $B-z$ color could indicate that a low residual star-formation is still present. The SFR inferred from the best-fit SED is $\sim 8$~\msun/yr, which is very close to the upper limit from MIPS (cf., the right panel of Figure~\ref{fig:obsvssed}).

Two of the \mipsd\ galaxies $>4\times$ below the MS, i.e., {\bf \#557} and {\bf \#856} (labeled as {\bf `2'}, and {\bf `4'}, respectively) lie in the quenched region, both in the $BzK$, and in the $UVJ$ diagram. Moreover, their best-fit SEDs indicate a very low SFR (i.e., SFR(SED)$<$3 \msun/yr), and modest extinction (cf. Table~\ref{tab:sedfit} and Figure~\ref{fig:obsvssed}). Hence, they are likely to be quenched galaxies, where the 24~$\mu$m/MIPS detection (slightly above $3\sigma$) could be due to old AGB stars, or alternatively to a highly obscured AGN, rather than to star-formation. In the following we consider them as quenched galaxies.

On the other hand, the third \mipsd\ galaxy below the MS, {\bf \#3853} (labeled as {\bf `7'}), seems to have a different nature with respect to the other two discussed above. It also lies in the quenched region of the $UVJ$ diagram, and has a red, $z-K$ color ($\simeq 2.4$), consistent with a relatively old age (1.14 Gyr, cf. Table~\ref{tab:sedfit}), but it is classified as a \sbzk, due to the bluer $B-z$ color.
This seems to indicate that the 24~$\mu$m/MIPS emission is plausibly powered by a low residual star-formation, as also confirmed by the SED fitting results in Figure~\ref{fig:obsvssed} ($A_V(SED)=0.99$, and SFR(SED)=20  \msun/yr, in agreement with that derived from MIPS, SFR(IR+UV)=$23 \pm 6$\msun/yr ). 
 
The \mipsd\ AGN objects {\bf \#720} and {\bf \#6193} (i.e., {\bf `3'} and {\bf `9'})  have observed colors typical of \sbzk\ but lie in the ``young-quenched'' region of the $UVJ$ diagram. In fact, they are redder with respect to the other \mipsd\ galaxies {\it on the} MS, and best-fit with declining SFR with $\tau= 0.1$ stellar population models, age=0.64 Gyr, low reddening, and SFR(SED)$\sim 2-3$~\msun/yr. 
These results do not agree with the high SFR(IR+UV)$> 100$ \msun/yr and large reddening ($\sim 1.8-2$ mag) derived from the 24~$\mu$m/MIPS emission of these galaxies. If the AGNs were responsible of most of the MIPS emission, both of these objects should be shifted below the MS in Figure~\ref{fig:pmulti}, entering the sub-sample of quenched galaxies. However,  it is not trivial to estimate  the fraction of the IR luminosity due to the AGN and to obscured star-formation, because  both objects are undetected in the far-IR. The bluer observed colors, and the younger best-fit {\it age} of these galaxies, with respect to other \mipsu\ quenched galaxies in the sample (including AGN hosts), seems to suggest that they are still star-forming, but maybe on the way to be quenched (possibly triggered by the AGN). 
On the contrary, the third AGN among the \mipsd\ galaxies, i.e., \#552, shows $BzK$/$UVJ$ colors and best-fit SED consistent with a moderate unobscured star-formation activity, in agreement with what expected based on the MIPS emission (SFR$_{SED}\approx$SFR(IR+UV)$\approx$56 \msun/yr).   

\subsubsection{Summary of the source classification}
As shown in the previous sections, for  galaxies lying more than $4\times$ below the MS the best-fit SEDs provide stricter upper-limits to their SFR, compared to the limits set by the 24~$\mu$m/MIPS data. Hence, hereafter we use these new upper-limits to the SFR for all the \mipsu\ galaxies, as well as  for the three \mipsd\ galaxies located below the MS (i.e., \#557, \#856, and \#3853, flagged as `2', `4', and `7' in Figures~\ref{fig:obsvssed}-\ref{UVJBzK}). 

\begin{figure*}
\includegraphics[width=0.6\textwidth]{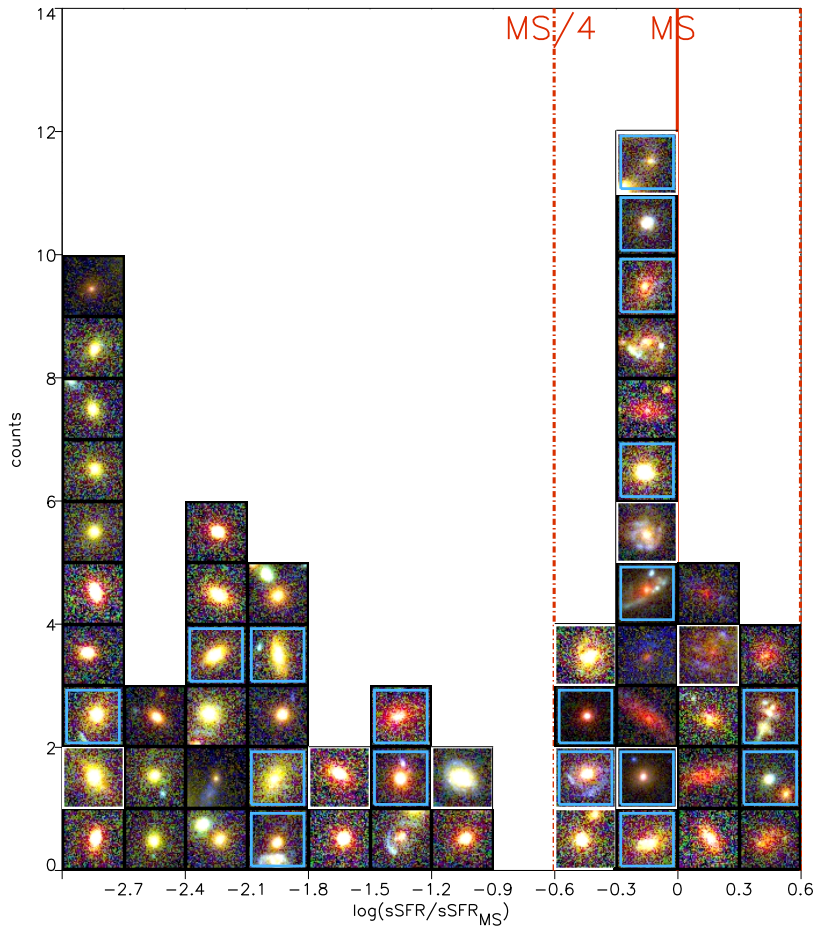}
\caption{The distribution of the sample  galaxies as a function of the `distance' from the main sequence, i.e., the log(sSFR/sSFR$_{\rm MS}$)). As explained in the text, hereafter, we used the SFR inferred from the best-fit SED (SFR(SED)) for all the galaxies below the MS strip. Within each log(sSFR/sSFR$_{\rm MS}$)) bin the distance from the MS also increases from top to bottom. On the contrary, in the first bin, which only includes galaxies best-fit with a SSP, and with SFR(SED)=0, objects are ordered randomly. The red solid and dot-dashed lines show the MS, and the loci 4$\times$ above and below it, respectively. The MIPS-detected (and PACS-undetected) galaxies are distinguished by the white frames, and AGN hosts by a cyan frame. The postage stamps are red-green-blue (R-G-B) composite images (R=F160W, G=F850LP, B=F435W) of 3 arcsec size.}\label{fig:histo_morph}
\end{figure*}

Figure~\ref{fig:histo_morph}, shows the distribution of the sources as a function of the ``distance'' from the main sequence, parametrized as the sSFR normalized to that predicted by Eq.~\eqref{eq:sSFRev} for a MS galaxy of the same mass, in logarithmic scale, i.e. log(sSFR/sSFR$_{\rm MS}$)). 
The red solid and dot-dashed lines in the figure show the average MS, and the offsets of $\pm 0.6$ dex, corresponding to $4\times$ above, and below it. The \mipsd\ galaxies (undetected in the FIR, hence which are not PACS-d galaxies) are highlighted with a white frame. Within the MS band all other galaxies are detected in the FIR, while outside this band none of the galaxies is detected. AGN hosts are identified by  a  cyan frame.
The first log(sSFR/sSFR$_{\rm MS}$) bin of the histogram  includes only galaxies which are best-fit by an SSP,  hence with SFR(SED)=0. For each of the remaining bins, the postage stamps are displayed in such a way that the distance from the MS also increases, i.e., log(sSFR/sSFR$_{\rm MS}$) decreases,  from top to bottom along the $y$ axis.

Out of the 56 most massive galaxies at $1.4\leq z \leq 2$ in GOODS-S, 25 objects ($\sim 45\%$) are on the MS, i.e., within  a factor of 4 ($\pm 0.6$ dex in Figure ~\ref{fig:histo_morph}) from the average MS relation at $z\simeq 1.7$, while 31 objects ($\sim 55\%$) are below it (log(sSFR/sSFR$_{\rm MS})<$-0.6).  
Several authors have found a {\it flattening} of the Main Sequence slope at the high-mass end \citep[e.g.,][]{2011MNRAS.417..289B,2011ApJ...742...96W,2012ApJ...754L..29W,2014MNRAS.443...19R}, interpreted as a further evidence that the mass-quenching process is more efficient on the most massive galaxies \citep[][]{2010ApJ...721..193P}. Also here the fraction of MS galaxies below the average relation (2/3) exceeds the fraction above it (1/3), although the sample is too small to establish the significance of this result. 
On the other hand, if the transition from star-forming to passive galaxies is linked to a morphological transformation, one cannot exclude that the appearance of passive bulges in the most evolved massive star-forming galaxies, contributing to the galaxy $M_*$ but not much to its SFR could flatten the MS slope at the highest masses \citep[][and Tacchella et al. 2014, submitted]{2014ApJ...788...11L,2014MNRAS.444.1660B}. 
Actually, a visual inspection of the image stamps in Figure~\ref{fig:histo_morph}, seems to confirm that many MS galaxies with $-0.6<$log(sSFR/sSFR$_{\rm MS})<0$ host a bulge. As shown in the following, this has been also confirmed through a 2D surface brightness fitting in the {\it HST}/WFC3 images (Section~\ref{sec:morphology}).

At the end of this exercise, i.e., after the de-blending procedure, and the use of SED fitting to better constrain the SFR for galaxies below the MS, we conclude that all the galaxies below the MS band have a sSFR more than 10$\times$ lower than that of MS galaxies in the same redshift range, and most of them $\sim 100$ to $\sim 1000$ times lower (Figure~\ref{fig:histo_morph}). Their global properties are also typical of quenched objects. This suggests that at $z\sim 1.4-2$ the mass-quenching may be a relatively fast process, much shorter than the time spent by these  galaxies in their massive phase  within the MS band (i.e., a transition time $<<\sim$1Gyr), as one may expect if quenching is caused by  the sudden expulsion of all the gas and dust from the galaxy. 

In summary, our sample of 56 galaxies splits in two main groups:
(i) the {\it MS galaxies}, included in the MS band ($-0.6<$log(sSFR/sSFR$_{\rm MS})<0$), and (ii) the {\it quenched galaxies}, with log(sSFR/sSFR$_{\rm MS})<-1$, which have a mass doubling time (the time needed by the galaxy to double its $M_*$, in case it continues to form stars at the present rate) $\geq 10$ Gyr. 
In our small sample a  well-defined, sizable class of objects {\it on the way to be quenched}  cannot be unambiguously recognized. Such transition  galaxies could be found among the {\it MS galaxies} with the lowest sSFR ($-0.6<$log(sSFR/sSFR$_{\rm MS})<0$), and/or among the {\it quenched galaxies} with the highest sSFR upper-limits ($-2\lesssim$ log(sSFR/sSFR$_{\rm MS})\lesssim -0.6$). Clearly a much bigger sample should be analyzed to adequately populate this class of objects which must exist and should allow to estimate the {\it quenching timescale}. However, the present investigation demonstrates that a careful, object-by-object study is necessary to identify bona-fide transition objects, given that blending may lead to gross errors in the estimated SFRs, especially for this class of galaxies (see Figure \ref{fig:pmulti}).

\begin{figure*}
\includegraphics[width=0.45\textwidth]{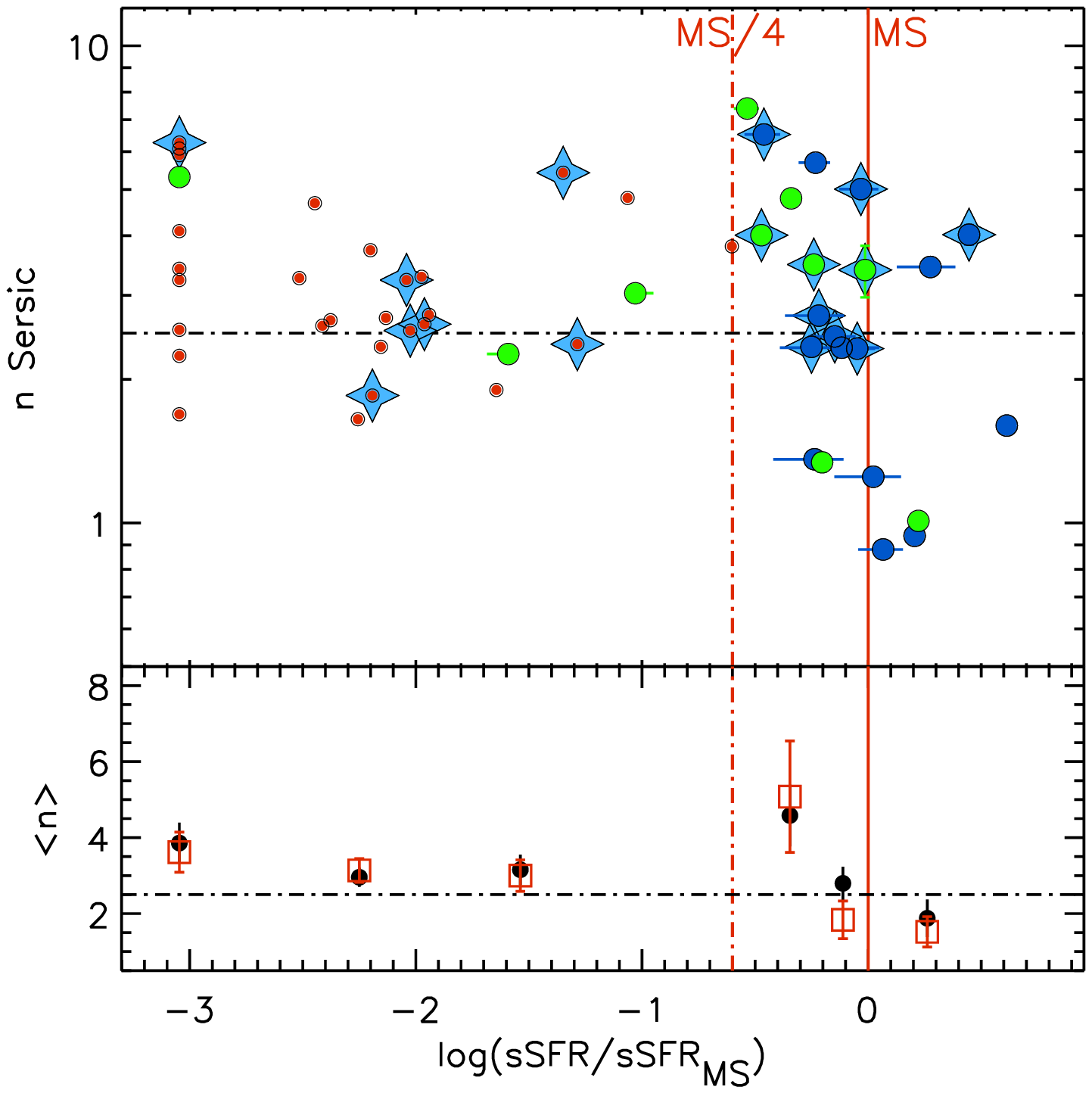}
\includegraphics[width=0.45\textwidth]{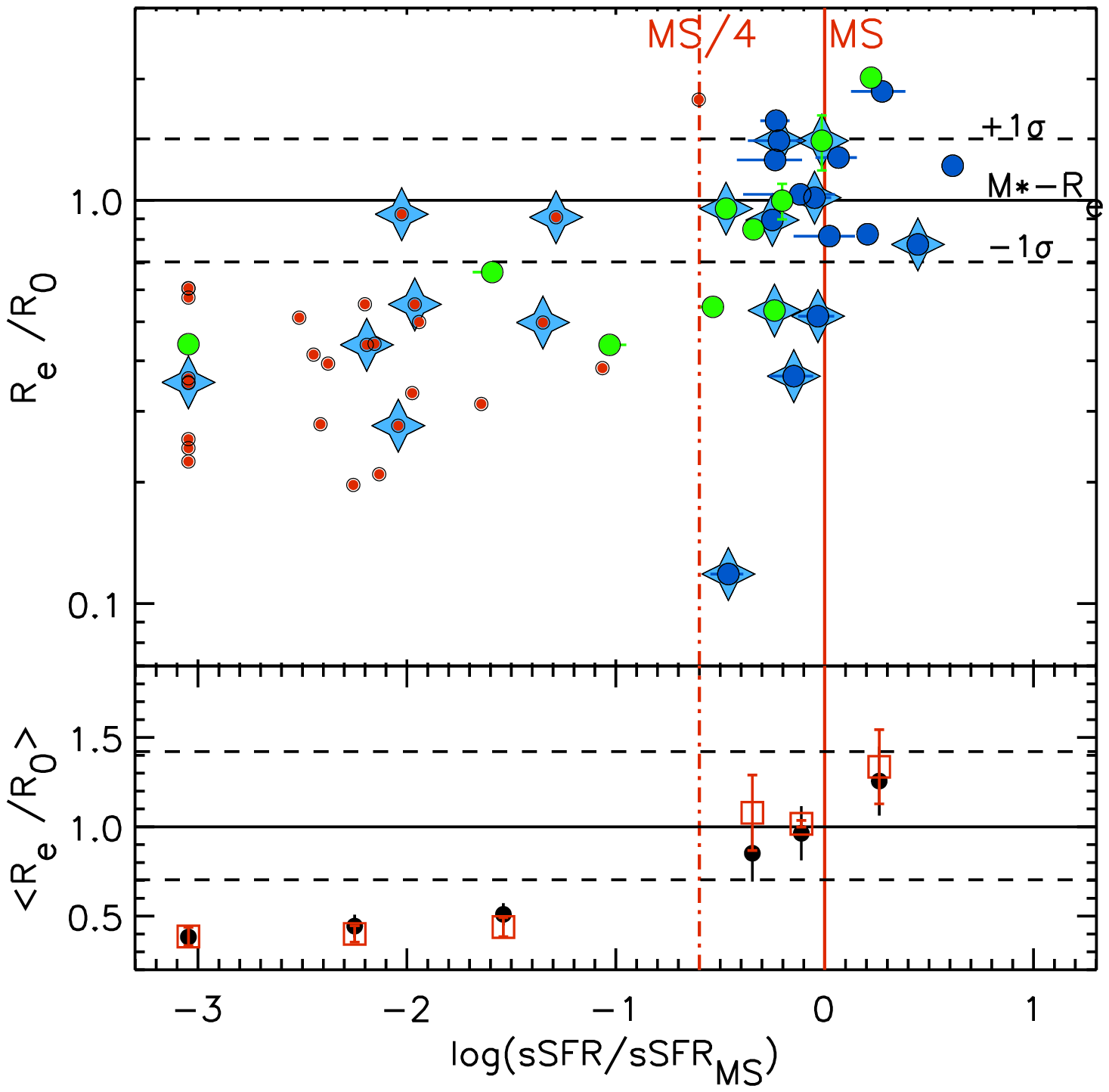}
\caption{{\bf Left:} Sersic index $vs$ distance from MS (i.e.,log(sSFR/sSFR$_{MS}$)). The vertical red dot-dashed line at log(sSFR/sSFR$_{MS})=-0.6$ marks the lower boundary of the MS band, and the horizontal $n=2.5$ line separates disk-dominated and bulge-dominated objects. {\bf Right:} The ``distance'' from the $M_*$-\re\ relation for local ETGs (horizontal solid line), expressed as $\re/R_0$, as a function of the distance from MS, log(sSFR/sSFR$_{MS}$). Symbols are like in the previous figures. {\bf Bottom panels:} Mean values of the same quantities, in log(sSFR/sSFR$_{\rm MS}$) bins chosen so to have $7-11$ objects per bin. Black filled circles are relative to all the objects within a bin, while red open squares are obtained by only considering galaxies without evidence of AGN.}\label{fig:n_RevsMSdist} 
\end{figure*}

\section{Morphological classification}\label{sec:morphology}
 
\subsection{Single-\sersic\ surface brightness fitting}\label{sec:sersic}

We used GALFIT to fit the surface brightness (SB) profile of our galaxies. 
Since the sample includes both star-forming and quenched  high-$z$ massive galaxies, for a first analysis we fit the SB profiles of each galaxy  in the F160W/WFC3 filter ($H$-band) with a single-\sersic\ function \citep{1968adga.book.....S}, leaving the \sersic\ index $(n)$ as a free parameter. 
In general, the surface brightness distribution of high-$z$ galaxies is far from following a regular \sersic\ profile, and especially so those of star-forming galaxies, that often are clumpy or merger-like systems \citep{2009ApJ...706.1364F,2009A&A...504..789E,2011ApJ...733..101G}. However, this method allows us  to classify objects as either Early-Type Galaxies (ETG, bulge-dominated, $n>2-2.5$) or Late-Type Galaxies (LTG, disk-dominated, $n<2-2.5$). Three out of the 56 galaxies have irregular morphology, and show strong residuals when fitted with a single \sersic\ profile, or with a bulge+disk model. They are the AGN `candidate' \#5503 (see Section \ref{sec:agn}), and the inactive galaxies \#3236, and \#5534, which have been excluded from the morphological analysis presented in the following.
To reliably measure the sky background, we run GALFIT on large images, of 30 arcsecond in size centered on each galaxy, and fit together all the brightest sources ($H<$23.5) included in the field (masking out the fainter ones). A prior run with SExtractor \citep{1996A&AS..117..393B} on the same images provided the morphological parameters to be set as starting point in the fit. 
The PSF model was built using the 10--20 closest unsaturated stars in the field. 
All the parameters, i.e., sky background, \sersic\ index ($n$), centroid, total magnitude ($\rm m_{160}$), half-light semi-major axis (\ae), position angle (PA), and axial ratio $(b/a)$, were left as free. 
The results are reported in Table~\ref{tab:morph} and Figure~\ref{fig:n_RevsMSdist}. 

The left panel of Figure~\ref{fig:n_RevsMSdist} shows the index $n$ as a function of the distance from the MS at $\langle z\rangle=1.7$, i.e., log(sSFR/sSFR$_{\rm MS}$). The \pacsd, \mipsd, and \mipsu\ sub-samples are shown in blue, green, and red, respectively, and the AGN highlighted with cyan four-point stars, such as in Figures~\ref{fig:pmulti}-\ref{UVJBzK}. The vertical solid, and dash-dotted lines represent the MS (Eq.~\eqref{eq:MSz1e7}), and its 4$\times$ lower boundary (labeled as ``MS/4''), respectively. 
The \sersic\ $n=2.5$ line shows the separation between ``Early-Type'' (or bulge-dominated) and ``Late-Type'' (or disk-dominated) galaxies.  
The right panel of the same figure, shows the distance from the MS {\it vs.} the distance from the $M_*$-\re\ relation of local ETGs \citep[i.e., the circularized effective radius, \re, normalized to the value at $z=0$, R$_{0}$, from][]{2003MNRAS.343..978S}. Here we use the $M_*$-\re\ relation of local ETGs as a reference for both the quenched and the star-forming galaxies in our sample, as $z\sim1.4-2$ star-forming galaxies with such high masses (and SFRs) will necessarily be quenched in a relatively short time (to avoid mass overgrowing), and then passively evolve down to $z=0$ without major further increase in mass. The bottom panels show the mean values of the same quantities, in log(sSFR/sSFR$_{\rm MS}$) bins chosen so to have $7-11$ objects per bin.  
Black filled circles are relative to all the objects within a bin, while red open squares are obtained by only considering galaxies without evidence of AGN.

From Figure ~\ref{fig:n_RevsMSdist}, it appears that almost all the {\it quenched galaxies} show an `early-type' morphology, with $n> 2-2.5$, and a circularized effective radius (\re) from 1.4 to 5 (2.5 on average) times smaller with respect to the local ETGs of the same mass, a well known property of high-$z$ quenched galaxies.
On the contrary, {\it MS galaxies} split in two groups, above and below the border-line at $n=2.5$. Almost all the MS galaxies with $n>2-2.5$ host an AGN, whereas all the AGN host have $n>2$ (but the AGN `candidate', \#5503, which has irregular morphology and is not shown in the figure). 
The high \sersic\ index of the AGN hosts could be due either to the nuclear emission of the AGN itself, detected as a central, unresolved point-source, or to a significant bulge component \citep[cf.,][]{2015A&A...573A..85R}. We investigate this issue in Section~\ref{sec:bsut}, by means of surface brightness profile decomposition, in bulge+disk, and bulge+disk+PSF components.

MS galaxies have in general size comparable with those of local ETGs, a part from 4 AGN hosts (i.e., \#1906, \#6898, \#6193, \#6352), and one normal galaxy (\#5415), which are comparable to the compact star-forming galaxies (CSFGs) identified by \citet[][hereafter, B13 and B14]{2013ApJ...765..104B, 2014ApJ...791...52B} at $1.4<z<3$, and considered the direct progenitors of compact quenched systems at similar redshifts. 
Since, as mentioned in Section~\ref{sec:sedfitting}, \#1906, and \#6898 are contaminated by the AGN in their photometric SEDs, it is not clear if their compact size are intrinsical, or just due to the AGN affecting the surface brightness profile (cf., also the discussion in Section~\ref{sec:bsut}). 
In any case, all these 5 objects have a relatively low sSFR, lying in the region $-0.6<$log(sSFR/sSFR$_{\rm MS})<0$ of Figure ~\ref{fig:n_RevsMSdist}, consistently with having started their quenching phase.  

Concerning the quenched population, our results are in agreement with the well-documented smaller size of the high-$z$ quenched galaxies with respect to the local ones \citep[e.g.][]{2005ApJ...626..680D,2006ApJ...650...18T, 2007MNRAS.382..109T,2007MNRAS.374..614L,2008ApJ...682..303M,2007ApJ...671..285T,2008A&A...482...21C, 2008ApJ...677L...5V,2008ApJ...687L..61B,2009ApJ...695..101D, 2009MNRAS.392..718S,2010ApJ...714L..79C,2013ApJ...765..104B,2013ApJ...764L...8B,2013ApJ...763...73S}. On the other hand, not all the quenched galaxies at high-$z$ have been found to be undersized, both in the field and in proto-clusters \citep{2010MNRAS.401..933M,2011MNRAS.412.2707S,2012ApJ...755...26O,2011MNRAS.414..445S,2013MNRAS.428.1715H}. 
It has been suggested that the average size growth of ETGs from $z=3$ to $z=0$ is not (or not only) due to the evolution of the individual galaxies (e.g., by mergers), but mostly (or partly) caused by the gradual addition of larger, newly-quenched galaxies at lower redshifts \citep{2010ApJ...712..226V,2013ApJ...775..106C,2013ApJ...777..125P,2013ApJ...773..112C}, although some authors have a different view (e.g., \citealt{2009ApJ...698.1232V}). 
In our sample, the scarcity of compact with respect to normal star-forming galaxies, could be interpreted as an indication that the majority of the incoming quenched galaxies at $z\lesssim 1.4$ are progressively larger in size with respect to the objects quenched at higher redshifts. 
However, the small size of the sample prevent us from providing new clues on this issue.

\subsection{ Bulge-Disk decomposition}\label{sec:bsut}
\begin{figure}
\begin{center}
\includegraphics[width=0.48\textwidth]{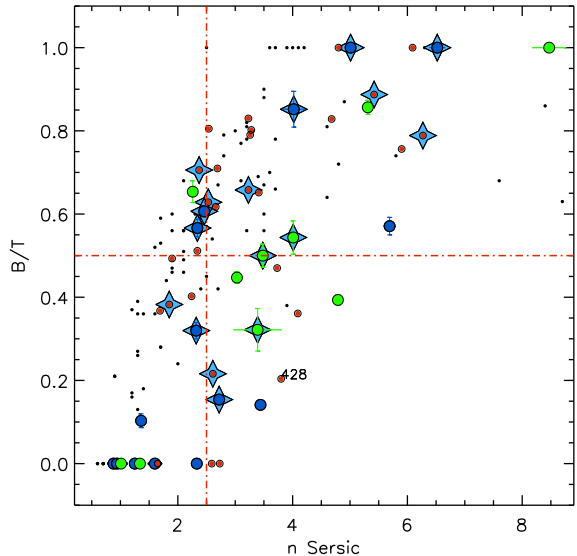}
\caption{{\bf Left:} Bulge to Total light ratio (B/T) $vs$ n Sersic. Symbols are like in the previous figures. The \citet{2012MNRAS.427.1666B} results for objects in the same mass and redshift range of our sample are also shown for comparison (small black dots).}\label{fig:B2Tvsn}
\end{center}
\end{figure}

We built two-component bulge+disk models, by fitting each galaxy with two \sersic\ components with $n=1$ (disk) and $n=4$ (bulge), both centered on the position derived in the previous single-\sersic\ fit. 
The bulge to total flux ratio (hereafter, B/T) gives an indication of the relevance of the bulge-component in the galaxy. 
We discarded those models in which unreliable parameter values were retrieved for some of the subcomponents, i.e., too large/too small effective radius (\re $> 4\farcs0$ or \re$< 0\farcs03$), or axis ratio $b/a <0.1$. In most cases, the component with unreliable parameters was also found to contribute less than 10\% to the total flux \citep[see][for a similar discussion]{2012MNRAS.427.1666B}. In summary, beyond the three objects with irregular morphology, the double-component model was discarded for 15 galaxies, of which 10 are classified as simple disks (B/T=0), and 5 as simple bulges (B/T=1), (see Table~\ref{tab:morph}).      
In Figure~\ref{fig:B2Tvsn} we show the relation between the $B/T$ and the single-\sersic\ $n$ parameter for our sample (color-coded as in the previous figures). Since \citet{2012MNRAS.427.1666B} also studied bulge+disk decomposition for similarly massive galaxies ($M_*> 1.7\times 10^{11} M_{\odot}$, Salpeter IMF) at $1<z<3$, for comparison, we also show in the figure objects included in the same redshift and mass range from that sample (black small points).  
As in the \citet{2012MNRAS.427.1666B} sample, also for our galaxies there is a good correspondence between the morphological classification based on the single-\sersic\ index, $n$, and on $B/T$, although with a considerable scatter. 
In  general, objects with $B/T\geq 0.5$ show $n \geq 2-2.5$, and those with $n<2-2.5$ show $B/T<0.5$. 
However, there are also some objects with a high $n$ and B/T$<0.5$, i.e., located in the bottom-right quadrant of Figure~\ref{fig:B2Tvsn}. Some of these objects 
have a relatively high $0.3\lesssim B/T\lesssim 0.5$, still consistent with the presence of a significant bulge, as suggested by their high \sersic\ index. 
For the remaining objects (two \pacsd\ galaxies, i.e.,\#3066, and \#6572, and three \mipsu\ galaxies, i.e., \#1084, \#1187, \#8121) the apparent incongruity between $n$ and $B/T$ may be due to the presence of an unresolved point-like source in the center, interpreted as a small bulge in the 2-component fit (\re$_{\rm Bulge}<1$~pixel), that could also be due to residual star formation or an AGN in the center. 
This alternative is supported by the fact that when the surface brightness  profiles of these galaxies  are fit with a single-\sersic\ model + central PSF, the resulting  \sersic\ index is $n\sim 1$ (a pure disk), and the point-like component is found to contribute  $\geq$ 10\% of the total galaxy flux (see Table~\ref{tab:morph}). 
Then, \#428 (flagged in the figure) is the already mentioned lens galaxy candidate (also highlighted in Figure~\ref{UVJBzK}, see also Section~\ref{sec:lens}). In this case the low B/T could be explained by the presence of the magnified background source, in the outer part of the galaxy, fit as a relatively bright disk.

\begin{figure}
\begin{center}
\includegraphics[width=0.5\textwidth]{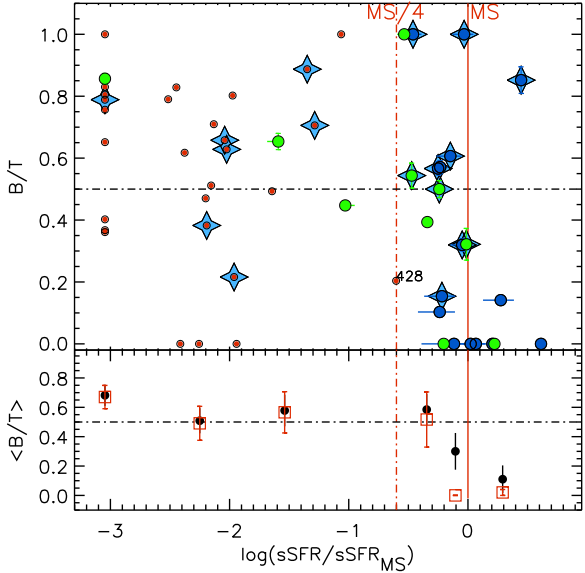}
\caption{Bulge to Total light ratio $vs$ distance from MS, log(sSFR/sSFR$_{\rm MS}$). 
Symbols are like in the previous figures. {\bf Bottom panel:} Mean B/T ratio in bins of log(sSFR/sSFR$_{\rm MS}$), as in Figure~\ref{fig:n_RevsMSdist}. Black filled circles are relative to all the objects within a bin, and red open squares are obtained by excluding the AGNs.}\label{fig:B2TvsMSdist}
\end{center}
\end{figure}

The top panel of Figure~\ref{fig:B2TvsMSdist} shows the bulge to total light ratio $B/T$ {\it vs} log(sSFR/sSFR$_{\rm MS}$) in logarithmic scale, and the bottom panel the average trend, $\langle B/T\rangle$, as in Figure~\ref{fig:n_RevsMSdist}. As mentioned for the $n$ index, most of the {\it quenched galaxies} have $B/T>0.5$, while the {\it MS galaxies} again split in two groups with $B/T$ $<0.5$ (64\%) or $>0.5$ (36\%) (when considering a boundary of $B/T=0.3$ they split almost in half). For galaxies within the MS band $\langle B/T\rangle$ seems to increase  with decreasing log(sSFR/sSFR$_{\rm MS}$) (from $\sim 0-0.2$ at log(sSFR/sSFR$_{\rm MS}>0$, to 0.5-0.7 at log(sSFR/sSFR$_{\rm MS}<0$), and then to remain almost constant below it, where almost all the galaxies are bulge-dominated. 
Figures~\ref{fig:B2Tvsn} and~\ref{fig:B2TvsMSdist} show that the high single-\sersic\ index measured for all the secure AGNs on the MS appears to be due to the presence of a significant bulge component ($B/T\geq 0.3-0.5$) in almost all of them (9/10). The only AGN with $B/T<0.3$ is \#3066, already discussed above. Since for most of the X-ray detected AGNs in our sample the nuclear emission does not dominate the photometric light in the optical-near-IR SED of the host galaxy, we expect a similarly low contamination also in the surface brightness profile. However, to rule out the possibility that the nuclear emission have biased the results towards higher \sersic\ indices and B/T ratios, we also tried to fit each object by adding a central point-source to the double component fit (i.e., bulge+disk+PSF). For most of the galaxies (50/56) the hypothetical point-source was found to contribute $<10\%$ to the total flux. Moreover, the B/T ratio derived from the three-components fit (bulge+disk+PSF) agrees very well with that derived from the two-components fit (bulge+disk), for all the objects (including the X-ray detected AGNs). This means that the nuclear emission, if present, does not appreciably affect the light profile of our galaxies. The only exception may be represented by the two `IRAC power-law' sources, i.e., \#1906, and \#6898, already discussed in Section~\ref{sec:sedfitting}, and~\ref{sec:sersic}. Their compact size, high \sersic\ $n$, and the fact that unreliable parameters were recovered for the disk component, both in the Bulge+Disk, and in the Bulge+Disk+PSF fits (B/T=1), seems to suggest that their light profile is outshined by the nuclear emission. Hence we cannot reliably estimate their morphological parameters.
However, for all the other AGNs we conclude that a real link should exist between the presence of an AGN and a relevant bulge component in the host galaxy. This fits well with the notion that galaxies and supermassive black holes (SMBH) grow together in order to establish the SMBH-bulge mass relation as observed in the local Universe \citep[e.g.,][]{1998AJ....115.2285M,2000ApJ...539L...9F,2007ApJ...670..173D}.
 The above analysis confirms  that {\it MS galaxies} with $M_*\geq 10^{11}~\msun$ show on average a more relevant bulge component with respect to their lower mass counterparts \citep{2014ApJ...788...11L,2014MNRAS.444.1660B,2014arXiv1411.7034T}. 
Moreover, since the most massive among star-forming galaxies are the first  to be quenched, this supports the idea that the {\it mass quenching} processes is strongly related to the bulge growth and to AGN activity, as also assumed  by some theoretical and semi-analytical models \citep{1998A&A...331L...1S,2005Natur.433..604D,2008MNRAS.391..481S}. This scenario is further supported  by the large fraction of AGNs that we found also among quenched galaxies, as discussed in Section \ref{sec:agn_q}.

\subsection{A double lens system?}\label{sec:lens}
\begin{figure*}

\includegraphics[width=0.2\textwidth]{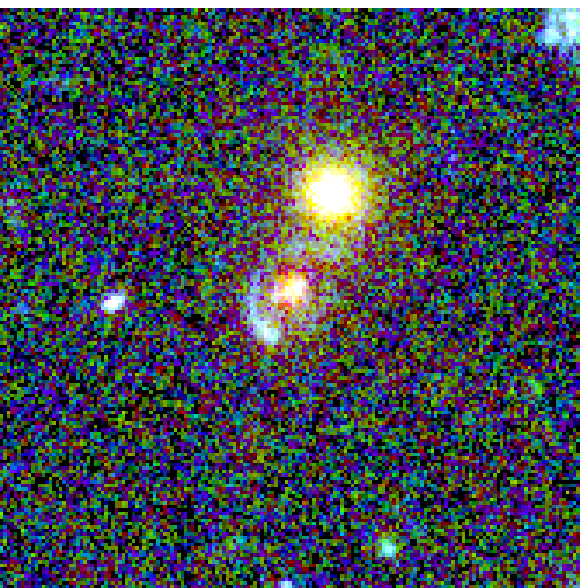}
\includegraphics[width=0.2\textwidth]{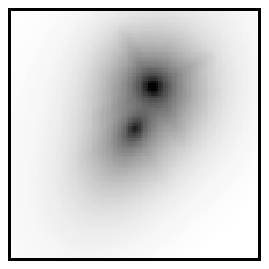}
\includegraphics[width=0.2\textwidth]{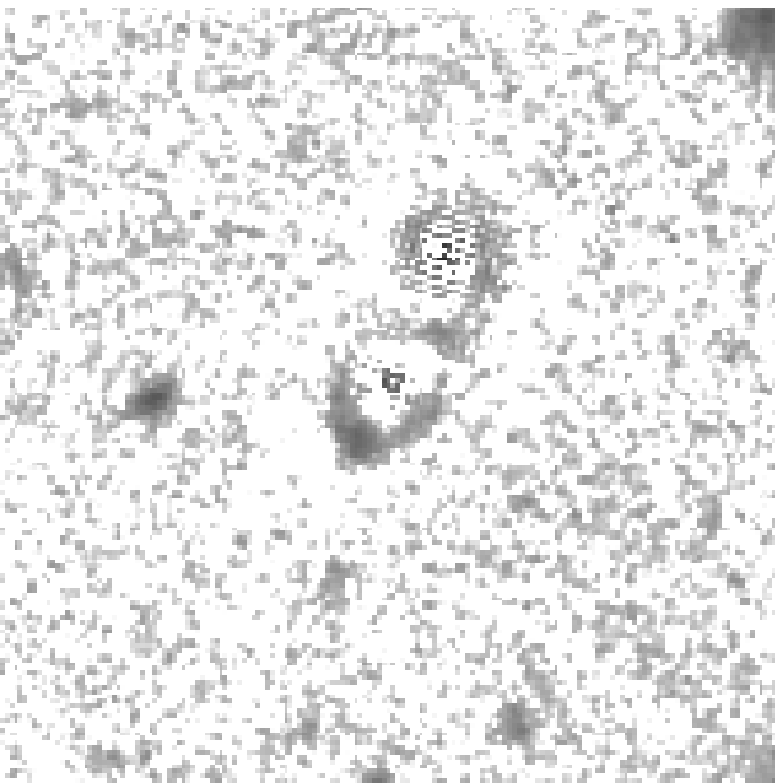}

\caption{From left to right we show : the RGB image (where R=$H$-band (F160W), G=$z$-band (F850LP), $B$-band (F435W) of HST), the GALFIT model in the F160W filter for single-\sersic\ fit, and the residual (image-model) in the same filter for objects \#427 and \#428.}\label{fig:lens}
\end{figure*}

The close pair of galaxies \#428 and \#427 represents a puzzling system as illustrated in Figure \ref{fig:lens} and we suspect that they  may form a double lens system. 
The red, compact galaxy \#428 is surrounded by a blue arc, which is highly reminiscent of a  section of an Einstein ring, as illustrated in the left panel of the figure. 
If this is indeed a lensing system, the  very massive ETG \#427, may contribute to further distort the image of a background source. 

Object \#427  at $z_{\rm spec}=1.427$ is spectroscopically classified as a passively evolving galaxy based on the CaH+K, MgII, and MgI absorption lines  \citep[][see Table~1]{2008A&A...478...83V}, while no spectroscopic confirmation exists for \#428 which has a $z_{phot}$=1.59. The small projected distance ($1\farcs88$, i.e., $\sim$16 kpc at this redshift) and the massive nature of the two galaxies makes it possible that they are a gravitationally bound pair at $z=1.427$. The central spheroid of \#428 is undetected in the $U$-band and close to the detection limit in the  {\it HST}/ACS $B$-band, while the outer arc is detected in all bands, hence it is not a $U$-band {\it dropout}. If indeed lensed, the  source should be at $2\lesssim z\lesssim 3$.  The radius of the blue arc is $R_{\rm arc}=0\farcs8$, somewhat larger than expected for the Einstein radius of an isolated $M_{\rm tot}\sim 40\times 10^{11}~M_{\odot}$ lens at $z=1.427$ and a $z\sim 2-3$ background source, for which the Einstein radius would be $R_E=0\farcs33-0\farcs6$. This is assuming  a dark matter contribution $\sim 40$ times the stellar mass of the galaxy. Hence, the unusual configuration of the system, with only half visible ring around object \#428 (and such large $R_{E}$), could be due to the presence of the massive \#427 neighbor perturbing  the gravitational field and producing a warped and atypical source image.

The further exploration of the various components of this system has been pursued by fitting the surface brightness distribution of \#428  and \#427 with both  single-\sersic\ function and with bulge+disk models (Section~\ref{sec:morphology}). From the single-\sersic\ fit we obtained a high \sersic-index ($n=3.8$) and a relatively large \re$_{,circ}$ (5.8 kpc) for \#428, and $n=2.73$ and \re$_{,circ} = 2.4$ kpc for \#427. The corresponding model is shown in the middle panel of Figure \ref{fig:lens}. Once the single-\sersic\ models have  been subtracted, strong ring-like residuals are visible in the outskirts of  both galaxies as shown in the right panel of Figure~\ref{fig:lens}. From this figure it also appears that the bottom-right part of the blue ring of \#428 emerges in the residual image, with an evident brighter ``blob'' SE of the galaxy. A smaller ring also appears in the residual image of \#427,  along with a structure apparently bridging these two rings. The bulge+disk decomposition resulted in a very faint  and small bulge ($B/T\sim 0.2, \re_{,circ,bulge}\sim 0.5$ kpc) for \#428, and a bright and large disk \re$_{,circ,disk}\sim$5.2 kpc), but produced similarly strong residuals. This shows that an exponential disk is not suitable to fit the outer structure in \#428. 

Thus, from these evidences we cannot conclude whether we are dealing with a double-lens system, or whether this complex structures are the result of tidal interactions. What is more certain is that the blue ring is a distinct, star-forming galaxy, either being lensed or torn apart by \#428. 
We then  derived more reliable morphological parameters and total photometry for the supposed lens galaxy by masking-out on the {\it HST} $H$-band image the outer arc with an annulus of internal and external radii of $r_{\rm in}=0\farcs6$ and $r_{\rm out}=1\farcs2$, respectively.  By fitting again the galaxy with a single-\sersic\ function we then derived nearly the same \sersic\ index, ($n\sim 3.8$), but a smaller effective radius, \re$_{,circ}$=3.9 kpc. Using  the same masking we have also determined the total magnitudes in all the other available {\it HST} optical (i.e., $B$,$V$,$i$,$z$) and near-IR ($Y$,$J$) filters  and all the other GALFIT parameters to those derived in the $H$-band. 
As shown in the Appendix (Figure~\ref{appenfig:427_428}), a blended detection at MIPS/24~$\mu$m is also associated to the system with the MIPS peak being clearly shifted toward S-E with respect of  both  the IRAC  and  the  {\it HST}/WFC3 centroids.  Hence, the 24 $\mu$m flux most likely  comes from the blue arc, rather then from either of the two galaxies \#427 and \#428,  confirming  their quenched nature.

We fit the new HST optical/near-IR photometry with an SED including relatively low reddening ($A_V<1$), as inferred from the MIPS upper limit (assuming MIPS non detection, cf. Section~\ref{sec:sedfitting}, and Figure~\ref{fig:obsvssed}). The derived sSFR ($log(sSFR/sSFR_{MS})=-1.4$) is consistent with that of a purely quenched galaxy. 
Whatever the nature of the \#427/\#428 pair, either a lensing or a tidally interacting system, it clearly deserves further study.

\section{AGN and quenching}\label{sec:agn_q}
Out of the 56 massive galaxies in this study, 18 (or 17) are hosting an AGN, as revealed by their X-ray and radio detection (cf. Section 3.3). This fairly high $\sim 30\%$  fraction is  not uncommon among massive galaxies at these high redshifts \citep[e.g.,][]{2007ApJ...670..173D,2009ApJ...693..447F}. 
AGN-driven, high velocity nuclear outflows are found to be ubiquitous among massive ($\gtrsim 10^{11}\,\msun$) star-forming, near-MS galaxies (F\"orster Schreiber et al. 2014; Genzel et al. 2014). Given their extension (few kpc) such outflows must have fairly long lifetimes ($\gtrsim 10^8$ yrs) and therefore be detectable even if the powering AGN is caught in an off phase, due to its short term variability. This may provide further evidence for SMBHs being  responsible for regulating the growth of the galaxies and vice versa.
It also provides circumstantial support to the notion of AGN feedback as a plausible  physical process leading to  mass quenching. However, evidence of AGN feedback, e.g., in the form of fast nuclear outflow, is not necessarily evidence for AGN quenching, as pointed out in the introduction.

Particularly interesting is the fact that the AGN fraction is quite high not only among MS galaxies (10-11/25, $\sim 40\pm 10\%$) but also among sub-MS galaxies, which are either quenching or already quenched (7/31 $\sim 22\pm 7\%$). One would have expected that fading of star formation should be accompanied by fading of nuclear activity as well. Why this is not the case? One possibility is the following. Galaxies may be quenched because at some point they are evacuated of virtually all their gas content, or gas supply from the circumgalactic medium is somehow suppressed. Yet, this does not mean they will remain gas free forever: even in absence of gas infall from the environment. The mere gas return from dying stars is sufficient to replenish the gas reservoir of galaxies over relatively short time scales, with gas accumulating in the potential well. Even if supernovae of Type Ia may provide enough energy to drive galactic winds and outflows in less massive galaxies, in the most massive quenched galaxies the gas accumulation leads inevitably to a central cooling catastrophe and the establishment of an inflow towards the center \citep{1991ApJ...376..380C}. What happens as a result of such a runaway cooling has been explored with hydrodynamical simulations by \citet{1997ApJ...487L.105C,2001ApJ...551..131C} and more recently by \citet[][and references therein]{2012MNRAS.427.2734N}.
The cooling gas is found to feed the central SMBH and its radiative and mechanical feedback is able to temporarily revert the inflow into an outflow, possibly even restoring a global galactic wind. As the galaxy is evacuated of its gas, the AGN subsides and gas starts to accumulate again until the next cooling catastrophe takes place and another cycle begins. The duty cycle of this intermittent behavior is shorter the higher the rate of mass return from dying stars, which decreases with increasing age of a stellar population. Thus, the galaxies with the youngest stellar populations among quenched galaxies, have the highest rate of mass return, and therefore the shortest AGN duty cycle. At least qualitatively, this scenario accounts for the high frequency ($\sim 22\%$) of AGN among the quenched/quenching galaxies in our sample. Actually, their low level of star formation, rather than to quenching, could also be due to minor star-formation activity that may accompany the central cooling catastrophe. This intermitting AGN activity is actually necessary to keep quenched a galaxy once it has been quenched in the first place. 

We point out that a large fraction of X-ray detected AGNs in quenched galaxies at $1.5\leq z\leq 2.5$ has been also recently uncovered by other authors \citep[][]{2013PASJ...65...17T,2013ApJ...764....4O,2013ApJ...770L..39W}. In particular \citet{2013ApJ...764....4O} found that the fraction of luminous AGN ($LX_{0.5-8keV}>3 \times 10^{42} erg s^{-1}$) among quenched and star-forming galaxies with $M_*>5\times 10^{10}~\msun$ are comparable (i.e., 19$\pm$ 9\%, and 23$\pm$5\%, respectively) in the 464.5 arcmin$^2$ CDF-S survey, which currently  provides the deepest  0.5-8 keV data. They also found an X-ray excess by stacking the X-ray undetected quenched galaxies, and claim that 70\%-100\% of them should host a high- or low-luminosity AGN ($LX_{0.5-8keV}\sim 10^{42} erg s^{-1}$). 
The larger fraction of luminous X-ray AGNs that we found among star-forming galaxies, with respect to that among quenched galaxies, mostly depends on the different selection criteria that we used to separate the two classes, not only based on $UVJ$ rest-frame colors, but also on the sSFR (cf. Section~\ref{sec:subsamples}). By adopting a pure $UVJ$ color-criterion one finds that the luminous AGN fraction among quenched and star-forming galaxies, are $\sim$  26$\pm$8\%, and  28$\pm$9\%, respectively,  nearly as in \citet{2013ApJ...764....4O}. 

In our massive sample at $1.4\leq z\leq 2$, AGNs quenched galaxies only account for $\sim 5\pm 5\%$ of the total X-ray luminosity density. This finding agreees well with recent results from \citet[][]{2015ApJ...800L..10R} at similar redshift, and \citet{2014MNRAS.440..339G} at $z<1.2$, showing that the bulk of the accretion density of the Universe seems to be associated to normal star-forming (i.e., MS) galaxies.
For MS galaxies we estimated an average ${\rm log( L_X /SFR)=41.62\pm0.08}$, which is from 0.6 to 0.8 dex higher than the average values found by \citet{2012ApJ...753L..30M}, and \citet{2015ApJ...800L..10R} for similarly massive MS galaxies at $z\sim 2$. This discrepancy comes from the average X-ray luminosity being derived by combining both detected and stacked X-ray signals, while here we only consider X-ray detected objects. A better agreement is found with the average $L_{\rm X}/SFR$ ratio derived by \citet{2015ApJ...800L..10R} for X-ray detected objects in their two higher mass bins (i.e., $log(L_{\rm X} /SFR)=41.35\pm0.05$, $41.38\pm0.05$, reported to a Salpeter IMF, G. Rodighiero private communication). The residual discrepancy ($\sim0.3$ dex) should be attributed to the different depths of the X-ray surveys in the GOODS and COSMOS fields.

\section{A lesser impact of Environment on quenching at $\mathbf z=1.4-2$}\label{sec:environment_q}

\subsection{The $z\sim 1.61$ overdensity}

According to the \cite{2010ApJ...721..193P} phenomenological model, environment quenching is not expected to play a major role at the redshifts and stellar masses explored in this paper. The existence of a diffuse, and well documented sheet-like overdensity at $z\sim 1.61$ on the entire GOODS-South field \citep{2007ApJ...671.1497C,2009A&A...504..331K,2011ApJ...743...95G} offers the opportunity to test this expectation by comparing the abundance of quenched galaxies, bulge-dominated galaxies, and X-ray detected AGNs, within and outside the overdensity. 
We note that \citet{2007ApJ...671.1497C} isolated a symmetric structure corresponding to the highest density peak, centered approximately at R.A.$ = 03^h$:32$^m$:29.28$^s$, DEC.$ = -27^{\circ}$:42$'$:35.99$''$, and with an extension of  $\sim 3\times3$ Mpc comoving, embedded in such a wide wall-like or filamentary overdensity. 
They considered as possible members of such cluster-like structure only the objects within a square of side $2R_A$ (where $R_A=2.14$~Mpc is the Abell radius). 

Here we consider as part of the overdensity all the galaxies with spectroscopic redshift in the range $1.6\leq z_{spec}\leq 1.62$ (19 sources), or photometric redshift in the range $1.5\leq z_{phot}\leq 1.7$ (8 sources). Then, we also tested the results by only considering objects in the highest density peak as in \citet{2007ApJ...671.1497C} (16 in total, 11 of which spectroscopically confirmed).
The results are summarized in the following, where the fractions relative to the highest density peak are shown in parenthesis.

\begin{itemize}
\item The fraction of quenched galaxies (i.e. galaxies with log(sSFR/sSFR$_{\rm MS})<-1$) within the overdensity is $59\pm 9\%$ ($68\pm 12\%$),  
which is comparable to the fraction at all other redshifts, outside the overdensity, which is $52\pm 9\%$ ($50\pm8\% $).

\item The fractions of bulge-dominated objects ($B/T\geq 0.5$) within, and outside the overdensity are also comparable within the errors, being $59\pm 9\%$ ($62\pm 12\%$), and $41\pm 9\%$ ($45\pm 8\%$), respectively. They are also comparable to the fraction of bulge-dominated galaxies found by Bruce et al. (2012) in the same mass and redshift range, in the UDS field ($55\pm 9\%$ at $1.4\leq z\leq2$, or $65\pm 9\%$ at $1.5\leq z\leq1.7$).

\item Out of our 18 X-ray detected AGN hosts, 12 (6) are in the overdensity, and 6 (12) in the field. The relative fractions of X-ray detected AGN are $\sim 44 \pm 9 \%$ ($37 \pm 12 \%$) within the overdensity and $\sim 20\pm 7 \%$($30\pm 8 \%$) in the field.  

\end{itemize}
In the overdensity there appears to be a slightly higher fraction (at the $\sim 1\sigma$ level) of quenched, bulge-dominated and X-ray detected galaxies, but given that these properties correlate with each other we cannot claim to have unambiguously detected an environmental effect. Overall, this indicates a  lesser role for  {\it environment quenching} with respect to the {\it mass quenching} within  the studied redshift and mass ranges, where no morphology-density relation is yet observed. 
Concerning the X-ray detected AGN, we found different results when considering the more diffuse sheet-like structure, and the highest density peak, since 6 AGNs are at a distance of $\sim 8'-10'$ ($\sim$ 4-5 Mpc, southward) from the cluster center identified by \citet{2007ApJ...671.1497C}. This discrepancy is probably due to statistical uncertainties  due to the small number of objects in  our sample. So, our data do not allow us to establish whether or not the overdensity has an effect  on the frequency of AGN activity, of the kind proposed by some authors \citep{2005ApJ...623L..81R,2012AdAst2012E..32F}.

\begin{figure} 

\includegraphics[width=\columnwidth]{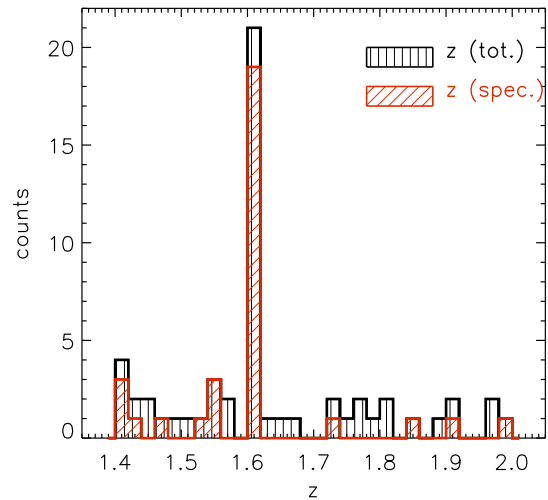}
\caption{Redshift distribution of the sample. The diagonal-shaded red histogram shows the distribution for spectroscopic redshift only, whereas the vertical-shaded distribution refers  to the addition of galaxies with only photometric redshift}\label{fig:z_hist}
\end{figure}

\section{Discussion and Conclusions}\label{sec:conclusions}
This paper is meant to be a step towards  shedding new  light on the physical process(es) responsible for the suppression of the star-formation in the most massive galaxies at high-redshift, leading to the build up of the high-mass end of the quenched population observed in the local universe (i.e., the so-called {\it mass quenching}).
To this purpose, we took advantage of the high quality WFC3/{\it HST} near-IR imaging (CANDELS), as well as of the multi-wavelength data (from X-ray to radio) available in the GOODS-S field, to study a mass-complete sample of galaxies at $1.4\leq z\leq2$ with $M_*\geq 10^{11}~\msun$. 
At  this cosmic epoch the SFR density and the AGN activity reached their peak and, at the same time, the quenching of star-formation started to be very efficient in galaxies with mass $\gtrsim 10^{11}~\msun$.
Hence, we chose this mass and redshift ranges so to deal only with galaxies which are either already quenched, or which have initiated the quenching process or soon will do so. The purpose is indeed to study the interconnection between {\it mass quenching}, and AGN, star-formation activity, and morphology, especially with the aim at identifying transition objects that have  left the main sequence since a short time. Such transition objects are certainly rare compared to fully star-forming, main sequence galaxies or fully quenched ones, hence simple automatic criteria for identifying them are likely to be contaminated by these more common neighbors.

Although many studies have been performed in similar mass and redshift ranges, most of them were based on very large samples, which while  providing statistical significance could lead to miss relevant  information if automated procedure are used for the analysis. Instead in  this work we study galaxies one by one, so to reliably derive their specific SFR by addressing blending problems in mid-/far-IR data, and comparing results from different diagnostics. 
The studied  sample consists of 56 galaxies, culled from the $K$-selected D07 optical-to-IRAC catalog, so to include all the galaxies with $M_*\geq 10^{11}$ M\sun\ (Salpeter IMF) and spectroscopic or photometric redshifts at  $1.4\leq z\leq 2$. The optical/IRAC data were complemented with {\it Spitzer}/MIPS and {\it Herschel}/PACS/SPIRE data. 
The main results can be summarized as  follows:  

\begin{itemize}
\item By comparing MIPS and high-resolution {\it HST} images, we identified sources which are blended in {\it Spitzer} data, and used F160W/WFC3/{\it HST} prior positions to estimate the amount of flux associated with each blended galaxy. We found that the automated MIPS counterpart association based on IRAC priors had failed for 7/56 ($\sim 12\%$) sources, which had been erroneously considered as MIPS-detected, or, just in one case, MIPS-undetected. Based on these results, we derived the relative fractions of objects which are PACS-detected (\pacsd, 18/56), MIPS-detected/PACS-undetected (\mipsd, 10/56), and MIPS/PACS-undetected (\mipsu, 28/56).  While $\sim 12\%$ may appear a small contamination, it is worth emphasizing that these reclassified galaxies represent almost  the totality of
potential transition/quenching objects in the sample.

\item We examined the position of our galaxies on the $M_*-$SFR plane, relative to the {\it main sequence} (MS) of star-forming galaxies at $\langle z\rangle = 1.7$ (the mean redshift of the sample), formally including all the galaxies with a sSFR within a factor of 4 above and below the average relation (the ``MS band''). To accurately establish the ``degree of quenching'' of each object, we first derived the SFR as the sum of the UV and IR contributions (SFR(IR+UV)), and then cross-checked the results using optical-to-IRAC SEDs, and two-colors $BzK$ and $UVJ$ diagrams. 
Consistently with the idea that the {\it mass quenching} process is more efficient in the most massive galaxies, we observe a turn down of the average SFR at such high-masses. In fact, 31/56  galaxies ($\sim 55\%$) are located below the lower boundary of the ``MS band'', and 2/3 of the MS galaxies lie below the average $M_*-$SFR relation. 
All the galaxies below the ``MS band'' turned out to have a specific SFR more than 10$\times$ lower with respect to the average relation (log(sSFR/sSFR$_{\rm MS})<-1$), as well as properties typical of already {\it quenched} objects, such as SED, colors and spectra, when available). 
This suggests that the {\it mass quenching} may be a very fast process in massive galaxies, being able to shut down the star-formation in relatively short timescales ($<<1$ Gyr), though a much bigger sample should be analyzed in a similar fashion to achieve firmer conclusions in this respect.    

\item We fitted the  surface brightness of galaxies  with both a single-\sersic\ function and with a double \sersic, in the latter case to model bulge+disk profiles. We found that the bulge to total light ratio ($B/T$), used as a proxy for the importance of the bulge component in a galaxy, provides results in general agreement with those obtained from the single-\sersic\ index ($n$). 
In the full sample $\sim$68\% of the objects have a relevant bulge component ($B/T>0.3$), and $\sim$ 50\% are bulge-dominated ($B/T>0.5$). These findings are in agreement with recent results from Bruce et al. (2012, 2014), showing that while at $z>2$ galaxies with $M_*> 10^{11}~\msun$ are mostly disk-dominated, at $1<z<2$ they are primarily bulge+disk systems. 
In particular, $\sim$84\% (26/31) of the quenched galaxies have a relevant bulge component, and 61\% (19/31) are bulge-dominated. 
On the other hand, about half of the MS galaxies have $B/T>0.3$, and $\sim$36\% (9/25) are bulge-dominated. The average $\langle B/T\rangle$ seems to increase within  the MS, with decreasing sSFR, and then remain almost constant below it. Although we do not have enough statistics to establish the solidity of this finding, this fits well with recent observational results showing that the bulge growth is directly related to the quenching of star-formation \citep[e.g.][]{2014ApJ...785...75G, 2014ApJ...788...11L,2014arXiv1411.7034T}. 

\item We identified 17 secure and one candidate AGN hosts in our sample (i.e.$\sim$ 30\%), both among MS (11/25, $\sim 44\%$) and among quenched (7/31, $\sim 22\%$) galaxies, by comparing the SFR derived from {\it Chandra}/X-ray and VLA/radio data with those inferred from IR+UV, and SED fitting. Although the fraction of AGNs among quenched galaxies is quite high, their contribution to the X-ray luminosity density is small, representing only $5\pm 5\%$ in our sample of $M_*\geq 10^{11}~\msun$ at $z=1.4-2$, in agreement with recent results from \citet{2015ApJ...800L..10R}.
In general, we found that the presence of the AGN correlates with a steep central profile, with \sersic\ $n\geq 2-2.5$, and $B/T\geq 0.3-0.5$, not only in quenched galaxies, but also in almost all the MS galaxies. This is consistent with the idea that the galaxy bulge and the central SMBH grow together since  early  epochs, as expected based on the bulge mass-SMBH relation observed in the local Universe.
Moreover, it seems that AGNs are much more frequent in galaxies with $-0.6<$log(sSFR/sSFR$_{\rm MS})<0$ than in those with higher specific SFR ($0<$log(sSFR/sSFR$_{\rm MS})<0.6$), suggesting  that the AGN is somehow related to the decline of the SFR.
This findings  provide some support to the notion  that AGN feedback may  be crucial in shutting down the star-formation in the most massive galaxies, as assumed in some theoretical and semi-analytical models.  

\item With respect to the local ETGs, the majority of the quenched galaxies (28/31, $\sim$90\%) are undersized (by a factor $\sim$ 2.5, on average), while the majority of the MS galaxies (20/25, $\sim$ 80\%) are comparable to them in size, in agreement with what found in previous works. 
All the 5/25 compact MS galaxies are similar to the ``compact star-forming galaxies" (CSFG)  studied by Barro et al. (2013, 2014), showing high central concentration ($n>2.5$, and $B/T\geq 0.5$) and a sSFR slightly lower with respect to the average MS relation. Moreover, 4 out of 5 are AGN hosts, although in two objects the AGN dominates the photometric light, possibly affecting the derived galaxy size. All these characteristics suggest that these objects will be quenched soon, originating compact quenched remnants, as suggested by Barro et al.  

\item  In our sample 27 out of 56 galaxies  are included in the well-known  overdensity at $z=1.61$ in the GOODS-S field, hence we compared the relative abundances of quenched galaxies, bulge-dominated ($B/T\geq0.5$) galaxies, and X-ray AGNs, within and outside the overdensity.  
This test showed that no appreciable difference exists in the relative fractions of these three sub-populations in the different environments. The result does not change significantly when the test is repeated by considering as part of the overdensity only the objects included in the highest-density region ($\sim 3\times 3$ Mpc) identified by Castellano et al. (2007).
This suggests that the environment should have a minor (if not negligible) role in triggering the quenching of the star formation in the most massive galaxies at $z=1.4-2$, as predicted by the Peng et al. (2010, 2012) phenomenological model.
\end{itemize}   

Perhaps the most relevant lesson learned from this  {\it pilot} investigation is that careful object-by-object analysis is necessary to properly identify candidate galaxies on their way to be quenched, whose number and structure can be important for the identification of the physical mechanism(s) responsible for the quenching of star formation in massive high-redshift galaxies.

\clearpage

\begin{deluxetable}{lllllllllllll}
\tabletypesize{\footnotesize}
\rotate
\tablecaption{Massive Sample: multi-wavelength and spectroscopic information from GOODS/CDFS/CANDELS Surveys.\label{tab:1}}
\tablewidth{0pt}
\tablehead{
\colhead{ID} &
 \colhead{RA} &
 \colhead{DEC} &
 \colhead{z} &
 \colhead{zq\tablenotemark{(1)}} &
 \colhead{Ref\tablenotemark{(2)}} &
 \colhead{$F_{\rm 24}$} &
 \colhead{$F_{\rm 24,new}$} &
 \colhead{$F_{\rm 1.4GHz}$\tablenotemark{(3)}} &
 \colhead{$F_{\rm Hard-X}$\tablenotemark{(4)}} &
 \colhead{{\bf $L_X$\tablenotemark{(4)}}} &
 \colhead{class(X)/(o)\tablenotemark{(5)}}&
 \colhead{sub-sample}\\
\colhead{} &
 \colhead{[deg]} &
 \colhead{[deg]} &
 \colhead{} &
 \colhead{} &
 \colhead{} &
 \colhead{[$\mu$Jy]} &
 \colhead{[$\mu$Jy]} &
 \colhead{[$\mu$Jy]} &
 \colhead{[$10^{-16} \frac{\rm erg}{\rm cm^2\,s}$]} &
 \colhead{[$10^{42} \frac{\rm erg}{\rm s}$]} &
 \colhead{} &
 \colhead{} 
}
\startdata
  282 & 53.061039   & -27.69352 & 1.44 & 0 & b, CaHK       & $<17.22$           &                  &  &  &  &          & \mipsu \\
  427 & 53.173473   & -27.697678 & 1.427 & 1 & b, CaHK-MgII-MgI & 155.19 $\pm$ 10.64 & $<17.66$  &  &  &  &          &\mipsd$\rightarrow$u   \\
  428 & 53.1736946 & -27.6981106 & 1.59  & 0 &                  & $<31.92$ & $<21.01$\tablenotemark{(8)}&  &  &  &         &\mipsu  \\
  552 & 53.1411934 & -27.7011375 & 1.604 & 1 & a                & 66.1 $\pm$ 6.02    &                  &      &$<4.2$   &  11.0$\pm$3.2  &   AGN        &\mipsd \\
  557 & 53.1763992 & -27.7011547 & 1.981 & 1 & a                & 24.89 $\pm$ 6.73   &                  &  &  &  &         & \mipsd \\
  686 & 53.104889   & -27.705238 & 1.617 & 1 & d                & 564.94 $\pm$ 19.92 & 423.76 $\pm$ 11.71 & 88.5 $\pm$ 14.6 & 27.0$\pm$1.7 &  77.0$\pm$2.1 & QSO-1/BLAGN&\pacsd \\
  687 & 53.1046181 & -27.7054272 & 1.609 & 1 & b         & $<59.76$           &  &   &  &  &       &\mipsu \\ 
  720 & 53.0620193 & -27.706501  & 1.89 & 0 &                   & $<28.23$ & 204.38 $\pm$ 9.41 &   & 3.8$\pm$1.6  & 8.9$\pm$3.3  &  AGN       &\mipsu$\rightarrow$d \\
  848 & 53.1085548 & -27.7101555 & 1.609 & 1 & a                & $<18.75$\tablenotemark{(6)}           &                   &   &  &  &     &\mipsu  \\
  856 & 53.1830406 & -27.7090015 & 1.76 & 0 &                   & 23.02 $\pm$ 7.37   &                   &   &  &  &     &\mipsd  \\
  880 & 53.1496506 & -27.7113838 & 1.612 & 1 & a                & 30.7 $\pm$ 6.0     & $<41.55$ &   &  &  &    & \mipsd$\rightarrow$u \\
  887 & 53.179775   & -27.711674 & 1.416 & 1 & a                & $<19.34 $          &          &  &  &  &     &\mipsu  \\
  903 & 53.089447 & -27.7115822  & 1.72 & 0 &                   & 242.9 $\pm$ 14.82  &          &   &  &  &    &\pacsd  \\
  947 & 53.1164207 & -27.7127037 & 1.61 & 1 & a                 & $<19.78 $\tablenotemark{(6)}&   &   &  &  &     &\mipsu  \\
  986 & 53.1512337 & -27.713728  & 1.609 & 1 & a                & $<18.75$\tablenotemark{(7)}&   &   &$<3.4$  & 5.7$\pm$1.5  & AGN     &\mipsu \\
  1084 & 53.1527634 & -27.7162361 & 1.614 & 1 & a               & $<12.33 $\tablenotemark{(6)}&   &   &  &  &    &\mipsu  \\
  1187 & 53.1279221 & -27.7188625 & 1.66 & 0 &                  & 37.61 $\pm$ 6.55          &$<19.18$\tablenotemark{(6)} &   &  &  &     &\mipsd$\rightarrow$u  \\
  1272 & 53.130501 & -27.72118 & 1.415 & 1 & a                  & $<25.76$\tablenotemark{(6)}&    &  &  &  &    & \mipsu \\
  1510 & 53.188858 & -27.725603 & 1.618 & 0 &                   & 305.44 $\pm$ 16.63 &          &  &  &  &     & \mipsd  \\
  1906 & 53.0454712 & -27.7375183 & 1.613 & 1 & c               & 289.09 $\pm$ 16.27 &          & 3312.0 $\pm$ 13.3 & 95.0$\pm$2.9 &  240.0$\pm$4.0 & QSO-1/BLAGN& \pacsd\\
  2450 & 53.1746445 & -27.7533722 & 1.848 & 1 & a               & $<15.35$    &                 &   &  &  &     &\mipsu  \\
  2470 & 53.0274162 & -27.7538891 & 1.63 & 0 &                  & 71.05 $\pm$ 18.29  &          &   &  &  &   &\pacsd  \\
  2940 & 53.1410179 & -27.7667179 & 1.903 & 1 & b, MgII         & $<17.61$\tablenotemark{(6)} &   &   &$<1.2$  &  1.8$\pm$0.7 &  AGN     &\mipsu \\
  3066 & 53.1655693 & -27.7698498 & 1.55 & 2 & b, OII-(MgII)    & 227.82 $\pm$ 14.21  &         &   & 28.0$\pm$1.7 &  60.0$\pm$2.2 & AGN    &\pacsd  \\
  3230 & 53.043789 & -27.774668 & 1.615 & 2 & c                   & 110.77$\pm$ 9.46 & $<41.66$ &  &  &  &    & \mipsd$\rightarrow$u \\
  3231 & 53.0449028 & -27.7743626 & 1.61 & 1 & c                & 78.47 $\pm$ 15.55 & $<46.43$  &   &$<2.2$  & 4.4$\pm$0.9  & AGN     &\mipsd$\rightarrow$u \\
  3236 & 53.0781097 & -27.7742271 & 1.729 & 1 & c               & 254.29 $\pm$ 15.24 &          &   &  &  &     & \pacsd \\
  3258 & 53.0521927 & -27.7747669 & 1.605 & 1 & c               & $<19.06 $\tablenotemark{(7)}&  &   & 2.0$\pm$0.8  & 4.9$\pm$1.3  &  AGN    &\mipsu \\
  3853 & 53.125061  & -27.7907791 & 1.553 & 1 & c, MgII-AlII{\tiny(red)} & 27.15 $\pm$ 7.42  &         &   &  &  &      &\mipsd  \\
  4310 & 53.0446396 & -27.8054123 & 1.61  & 0 &                 & $<13.43$\tablenotemark{(6)}&    &   &  &  &      & \mipsu  \\
  4406 & 53.101093 & -27.808559 & 1.97 & 0 &                    & 116.99 $\pm$ 8.68  &          &   &  &  &     &\pacsd  \\
  4705 & 53.066715 & -27.816507 & 1.413 & 1 & b, [OII]-D4000    & 174.75 $\pm$ 11.67 &          &   &$<1.2$  & $<$1.0&    & \pacsd \\
  5149 & 53.1717644 & -27.825676  & 1.74  & 0 &                 & $<19.18$           &          &   &  &  &     & \mipsu \\
  5415 & 53.0677795 & -27.8320599 & 1.79  & 0 &                 & 57.23 $\pm$ 7.51  &       &   &  &  &     &\mipsd \\
  5503 & 53.1572495 & -27.8335114 & 1.619 & 1 & b, OII{\tiny(noisy)} & 232.03 $\pm$ 14.39&          & 169.8 $\pm$ 12.5 &$<1.2$  & 1.1$\pm$0.5  & AGN?   & \pacsd \\
  5509 & 53.0578232 & -27.8334866 & 1.65 & 0 &                  & $<18.38$\tablenotemark{(7)}&  &   & 5.7$\pm$0.8 & 39.0$\pm$3.6 & AGN    & \mipsu \\
  5530 & 53.0476723 & -27.835022  & 1.51 & 0 &                & 34.12 $\pm$ 8.19     & $<34.68$\tablenotemark{(6)} &    & 2.5$\pm$0.6&  3.3$\pm$0.4 & AGN    & \mipsd$\rightarrow$u \\
  5534 & 53.07267   & -27.8341789 & 1.616 & 1 & b, [OII]{\tiny(faint)}  & 313.95$\pm$ 16.8     &       &   &  &  &     & \pacsd \\
  5556 & 53.150166 & -27.834509 & 1.612 & 1 & b, [OII]-MgII     & $<31.10$             &       &   &  &  &    &  \mipsu \\
  5564 & 53.1271667 & -27.8345394 & 1.81 & 0 &                 & $<15.28$\tablenotemark{(6)}&   &   &  &  &    &\mipsu  \\
  5860 & 53.1314774 & -27.8412781 & 1.613 & 1 & b, OII-MgII     & 191.47 $\pm$ 12.52   &       & 58.8 $\pm$ 12.8 & 1.2$\pm$0.5  &  2.1$\pm$0.7 & AGN     &\pacsd \\
  5974 & 53.07164   & -27.8436356 & 1.9 & 0 &        	        & 176.35 $\pm$ 11.75   &       &    &  &  &    &\pacsd  \\
  6071 & 53.1988525 & -27.8438873 & 1.57 & 0 &        	        & 299.89 $\pm$ 16.52   &       &    & 12.0$\pm$1.6  &  38.0$\pm$4.6  &  AGN    & \pacsd \\
  6193 & 53.076344 & -27.848677 & 1.536 & 1 & d                 & 158.05 $\pm$ 10.79   &       &  & 62.0$\pm$2.8 &   170.0$\pm$7.2& AGN-2/HEX& \mipsd \\
  6224 & 53.1370697 & -27.8500233 & 1.45 & 0 &                  & 105.85 $\pm$ 8.17    &       &    &  &  &    & \pacsd \\
  6352 & 53.06007   & -27.8529701 & 1.544 & 1 & d               & 89.08 $\pm$ 7.43     &       &    &  6.6$\pm$1.1 &  15.0$\pm$2.2  & AGN-2/HEX& \pacsd \\
  6572 & 53.077282  & -27.8595829 & 1.96 & 0 &                  & 299.89 $\pm$ 16.52   &       &    &  &  &   &\pacsd  \\
  6647 & 53.168438  & -27.8618145 & 1.77 & 0 &                  & $<32.90$\tablenotemark{(6)}&   &    &  &  &    & \mipsu  \\
  6898 & 53.1573715 & -27.8700867 & 1.603 & 1 & e               & 1085.15 $\pm$ 25.54       &  & 112.7 $\pm$ 12.7 & 98.0$\pm$2.8 &  250.0$\pm$3.7 & QSO-1/HEX&\pacsd  \\
  7077 & 53.2080154 & -27.8743706 & 1.609 & 1 & b               & 104.89 $\pm$ 8.13         &  &    &  &  &   &\mipsd   \\
  7116 & 53.1824493 & -27.8753071 & 1.48 & 0 &                  & 163.92 $\pm$ 11.1         &  &    &  &  &     &\pacsd  \\
  7491 & 53.148407 & -27.885511 & 1.604 & 0 &                  & $<12.27$\tablenotemark{(6)} &  &    &  &  &     &\mipsu  \\
  7617 & 53.099564 & -27.890564 & 1.57 & 0 &                    & $<18.21$\tablenotemark{(6)} &  &    &  &  &     & \mipsu \\
  7923 & 53.079632 & -27.901369 & 1.4 & 0 &        	        & $<50.40$                  & $<17.12$\tablenotemark{(6)}&  &  &  &   &\mipsu  \\
  8121 & 53.2112389 & -27.9082909 & 1.81 & 0 &                  & $<15.71$\tablenotemark{(6)} &  &    &$<4.8$  & 4.8$\pm$1.6  &  AGN    &\mipsu   \\
  8569 & 53.121048  & -27.9285126 & 1.469 & 1 &                 & 182.92 $\pm$ 12.09        &  &    &  &  &    & \mipsd \\
\enddata
\tablenotetext{(1)}{Redshift quality flag: 0=photometric, 1=spectroscopic, secure, 2=spectroscopic, likely (in agreement within 68\% confidence limits with the photometric redshift).}
\tablenotetext{(2)}{References for spectroscopic redshifts: (a)GMASS (KurK et al. 2012), (b)FORS2 (v3.0, Vanzella et al. 2008), (c)K20 (Mignoli et al. 2005), (d)X-ray CDFS (Szokoly et al. 2004), (e)VIMOS (Popesso et al. 2009; Balestra et al. 2009).} 

\tablenotetext{(3)}{VLA Radio integrated flux Data from \citet{2008ApJS..179...71K}, ATCA survey, \citet{2006AJ....132.2409N}, and \citet{2006AJ....131.1216A}}
\tablenotetext{(4)}{Data from 4 Ms {\it Chandra} X-ray catalog \citep{2011ApJS..195...10X}: Flux Hard-X band(2--8 kev), and total integrated Full-band (0.5--8 kev) luminosity.}
\tablenotetext{(5)}{Objects identified as AGN, based on X-ray/Radio data as explained in Section~\ref{sec:agn}. X-ray/Optical classification from Szokoly et al. (2004) is also reported, when available: class(X)=X-ray class (AGN-1,2, or QSO-1,2); class(o)=classification only based on optical spectra (HEX= object with unresolved emission lines and exhibiting high ionization lines, or emission line ratio indicating AGN activity. These objects are dominantly optical type-2 AGNs or QSOs; BLAGN= objects with emission lines broader than 2000 km/s. This classification implies an optical type-1 AGN or QSO).}
\tablenotetext{(6)}{2$\sigma$ upper limits in MIPS for objects with a S/N $<1$ (3$\sigma$ upper limits are used for the remaining objects).}
\tablenotetext{(7)}{Marginal MIPS detection, i.e., 2.5$<$S/N$<$3, treated as 3$\sigma$ upper limit.}
\tablenotetext{(8)}{This galaxy has been classified as \mipsu, being very likely a lens system, whose MIPS flux (Flux(24 $\mu$m)$_{new}$=152.12 $\pm$ 7.0), derives from a background source likely at $z\sim 2-2.5$ (see Appendix~\ref{app:A}).}
\end{deluxetable}                                                                                                                                
   
\clearpage
\begin{deluxetable}{ccccccccc}
\tabletypesize{\scriptsize}
\tablecaption{SFR, Stellar Mass, and SED fitting parameters (BC03, models, Salpeter IMF).\label{tab:sedfit}}
\tablewidth{0pt}
\tablehead{
\colhead{ID} & 										           
  \colhead{SFR(IR+UV)} &
  \colhead{$M_*$(D04)} &
  \colhead{$M_*$(SED)} &
  \colhead{Age} &
  \colhead{$A_{\rm V}$} &
  \colhead{SFR(SED)} &
  \colhead{$\tau$} &
 \colhead{$\tau_{\rm min}-\tau_{\rm max}$}  \\
\colhead{} & 																  
  \colhead{[$M_{\odot}/yr$]} &
  \colhead{[$M_{\odot}$]} &
  \colhead{[$M_{\odot}$]} &
  \colhead{[Gyr]} &
  \colhead{[mag]} &
  \colhead{[$M_{\odot}/yr$]} &
  \colhead{[Gyr]} &
  \colhead{[Gyr]}
}
\startdata
282 &$<$12.89          &     11.18&     11.17$^{+0.01}_{-0.02}$& 2.10$^{+0.00}_{-0.10}$& 0.10$^{+0.05}_{-0.05}$&  0.62 $_{ -0.03 }^{+ 0.22 }$& 0.3 &0.3-0.3\\
427 &$<$12.95          &     11.26&     11.11$^{+0.02}_{-0.01}$& 0.72$^{+0.47}_{-0.00}$& 0.50$^{+0.05}_{-0.19}$&  1.25 $_{ -0.02 }^{+ 0.07 }$& 0.1 &0.1-0.1\\
428 &$<$26.31          &     11.04&     10.99$^{+0.03}_{-0.01}$& 3.75$^{+0.25}_{-1.55}$& 1.10$^{+0.15}_{-0.15}$&  33.11 $_{ -8.33}^{+ 10.82}$& 30.0 & 3.0-1000\\
552 &56.23$\pm$5.09    &     11.17&     11.15$^{+0.16}_{-0.01}$& 0.72$^{+1.18}_{-0.08}$& 1.20$^{+0.15}_{-0.20}$& 59.99 $_{-12.36 }^{+20.61 }$& 0.3 &0.3-1\\  
557 &18.08$\pm$4.89    &     11.01&     11.18$^{+0.02}_{-0.01}$& 1.61$^{+0.07}_{-0.11}$& 0.50$^{+0.05}_{-0.05}$&  3.19 $_{ -0.62 }^{+ 0.13 }$& 0.3 &0.3-0.3\\
686 &332.14$\pm$9.39   &     11.04&     11.04$^{+0.01}_{-0.00}$& 0.45$^{+0.00}_{-0.01}$& 0.00$^{+0.03}_{-0.00}$&  0.00 $_{ -0.00 }^{+ 0.00 }$& 0.0 &0.0-0\\  
687 &$<47.23$          &     11.30&     11.34$^{+0.02}_{-0.01}$& 2.10$^{+0.10}_{-0.08}$& 0.30$^{+0.05}_{-0.05}$&  0.90 $_{ -0.24 }^{+ 0.04 }$& 0.3 &0.3-0.3\\
720 &132.89$\pm$6.46   &     11.06&     11.08$^{+0.05}_{-0.01}$& 0.64$^{+0.57}_{-0.00}$& 0.60$^{+0.01}_{-0.30}$&  2.53 $_{ -0.03 }^{+ 0.36 }$& 0.1 &0.1-0.2\\
848 &$<$14.65          &     11.18&     11.11$^{+0.04}_{-0.03}$& 1.28$^{+0.16}_{-0.00}$& 0.40$^{+0.05}_{-0.05}$&  0.00 $_{ -0.00 }^{+ 0.00 }$& 0.0 &0.0-0.1\\
856 &15.09$\pm$4.90    &     11.57&     11.52$^{+0.01}_{-0.00}$& 1.28$^{+0.00}_{-0.01}$& 0.00$^{+0.02}_{-0.00}$&  0.00 $_{ -0.00 }^{+ 0.00 }$& 0.0 &0.0-0\\  
880 &$<$32.88          &     11.23&     11.21$^{+0.02}_{-0.01}$& 1.14$^{+0.00}_{-0.00}$& 1.65$^{+0.05}_{-0.04}$& 16.06 $_{ -0.50 }^{+ 0.64 }$& 0.3 &0.3-0.3\\
887 &$<$13.94          &     11.23&     11.05$^{+0.02}_{-0.01}$& 1.28$^{+0.00}_{-0.00}$& 0.10$^{+0.05}_{-0.02}$&  0.00 $_{ -0.00 }^{+ 0.00 }$& 0.0 &0.0-0\\  
903 &164.43$\pm$36.70  &     11.08&     11.06$^{+0.05}_{-0.03}$& 0.18$^{+0.11}_{-0.02}$& 2.75$^{+0.05}_{-0.05}$&269.90 $_{-44.11 }^{+92.28 }$& 0.1 &0.1-0.3\\
947 &$<$15.44          &     11.06&     11.10$^{+0.01}_{-0.02}$& 1.90$^{+0.10}_{-0.10}$& 0.30$^{+0.05}_{-0.10}$&  1.00 $_{ -0.30 }^{+ 0.35 }$& 0.3 &0.3-0.3\\
986 &$<$14.65          &     11.42&     11.33$^{+0.01}_{-0.02}$& 1.28$^{+0.00}_{-0.00}$& 0.35$^{+0.03}_{-0.05}$&  0.00 $_{ -0.00 }^{+ 0.00 }$& 0.0 &0.0-0\\  
1084 &$<$9.58          &     11.21&     11.22$^{+0.02}_{-0.01}$& 2.10$^{+0.10}_{-0.10}$& 0.15$^{+0.10}_{-0.05}$&  0.69 $_{ -0.19 }^{+ 0.30 }$& 0.3 &0.3-0.3\\
1187 &$<$14.38         &     11.24&     11.26$^{+0.01}_{-0.01}$& 1.61$^{+0.04}_{-0.00}$& 0.45$^{+0.02}_{-0.06}$&  3.80 $_{ -0.54 }^{+ 0.09 }$& 0.3 &0.3-0.3\\
1272 &$<$18.00         &     11.06&     10.91$^{+0.01}_{-0.02}$& 1.28$^{+0.00}_{-0.00}$& 0.40$^{+0.04}_{-0.05}$&  0.00 $_{ -0.00 }^{+ 0.00 }$& 0.0 &0.0-0\\  
1510 &238.82$\pm$13.20 &     11.08&     11.01$^{+0.02}_{-0.04}$& 0.23$^{+0.03}_{-0.02}$& 1.90$^{+0.05}_{-0.10}$&141.22 $_{-31.95 }^{+27.38 }$& 0.1 &0.1-0.1\\
1906 &123.32$\pm$24.14 &     11.04&     10.74$^{+0.01}_{-0.00}$& 0.10$^{+0.00}_{-0.00}$& 1.65$^{+0.02}_{-0.00}$&621.53 $_{-15.98 }^{+19.18 }$& 1000.0 & 1.0-1000\\
2450 &$<$10.44         &     11.02&     11.14$^{+0.01}_{-0.02}$& 1.61$^{+0.09}_{-0.00}$& 0.55$^{+0.02}_{-0.10}$&  2.87 $_{ -0.75 }^{+ 0.06 }$& 0.3 &0.3-0.3\\
2470 &133.70$\pm$43.62 &     11.02&     11.31$^{+0.04}_{-0.36}$& 3.50$^{+0.25}_{-3.27}$& 2.25$^{+0.80}_{-0.20}$& 41.31 $_{-14.64 }^{+81.14 }$& 3.0 &0.1-1000\\
2940 & $<$12.10        &     11.22&     11.30$^{+0.04}_{-0.02}$& 1.28$^{+0.03}_{-0.14}$& 0.60$^{+0.25}_{-0.04}$& 12.67 $_{ -0.43 }^{+ 9.25 }$& 0.3 &0.3-0.3\\
3066 &93.05$\pm$27.05  &     11.12&     10.84$^{+0.02}_{-0.02}$& 0.10$^{+0.01}_{-0.00}$& 3.00$^{+0.00}_{-0.05}$&770.96 $_{105.34 }^{+14.06 }$& 1000.0 &0.3-1000\\
3230 &$<$32.67         &     11.09&     11.00$^{+0.02}_{-0.02}$& 1.28$^{+0.00}_{-0.00}$& 0.15$^{+0.05}_{-0.05}$& 0.00  $_{ -0.00 }^{+ 0.00 }$& 0.0 &0.0-0\\  
3231 &$<$37.14         &     11.65&     11.68$^{+0.02}_{-0.01}$& 1.90$^{+0.06}_{-0.00}$& 0.55$^{+0.05}_{-0.06}$&  3.80 $_{ -0.67 }^{+ 0.19 }$& 0.3 &0.3-0.3\\
3236 &114.20$\pm$16.96 &     11.03&     10.98$^{+0.01}_{-0.04}$& 0.20$^{+0.01}_{-0.02}$& 1.70$^{+0.10}_{-0.02}$&176.47 $_{10.98  }^{+45.19 }$& 0.1 &0.1-0.1\\
3258 &$<$14.91         &     11.31&     11.38$^{+0.01}_{-0.02}$& 2.00$^{+0.10}_{-0.10}$& 0.35$^{+0.05}_{-0.10}$&  1.38 $_{ -0.41 }^{+ 0.49 }$& 0.3 &0.0-0.3\\
3853 &23.44$\pm$6.11   &     11.31&     11.30$^{+0.01}_{-0.02}$& 1.14$^{+0.00}_{-0.04}$& 1.05$^{+0.07}_{-0.05}$& 20.14 $_{ -1.09 }^{+ 4.05 }$& 0.3 &0.3-0.3\\
4310 &$<$10.49         &     11.03&     10.95$^{+0.02}_{-0.02}$& 1.28$^{+0.00}_{-0.00}$& 0.40$^{+0.05}_{-0.05}$&  0.00 $_{ -0.00 }^{+ 0.00 }$& 0.0 &0.0-0\\  
4406 &137.21$\pm$63.97 &     11.21&     11.24$^{+0.02}_{-0.01}$& 1.14$^{+1.31}_{-0.00}$& 2.55$^{+0.10}_{-0.05}$& 17.38 $_{ -0.39 }^{+ 0.77 }$& 0.3 &0.3-0.3\\
4705 &199.93$\pm$5.83  &     11.01&     10.91$^{+0.03}_{-0.03}$& 0.23$^{+0.03}_{-0.02}$& 2.80$^{+0.10}_{-0.05}$&112.53 $_{23.24  }^{+30.33 }$& 0.1 &0.1-0.1\\
5149 &$<$12.22         &     11.10&     11.14$^{+0.00}_{-0.02}$& 1.80$^{+0.07}_{-0.12}$& 0.40$^{+0.10}_{-0.05}$&  1.55 $_{ -0.06 }^{+ 0.72 }$& 0.3 &0.3-0.3\\
5415 &39.38$\pm$5.00   &     11.05&     11.12$^{+0.00}_{-0.03}$& 0.81$^{+0.05}_{-0.09}$& 1.00$^{+0.10}_{-0.08}$& 40.98 $_{ -5.79 }^{+11.56 }$& 0.3 &0.3-0.3\\
5503 & 426.87$\pm$9.27 &     11.06&     10.95$^{+0.05}_{-0.02}$& 0.18$^{+0.05}_{-0.01}$& 1.80$^{+0.03}_{-0.15}$&211.19 $_{72.04  }^{+23.34 }$& 0.1 &0.1-0.1\\
5509 &$<$13.54         &     11.32&     11.54$^{+0.03}_{-0.01}$& 3.75$^{+0.00}_{-0.25}$& 1.85$^{+0.10}_{-0.05}$& 11.46 $_{ -0.34 }^{+ 2.99 }$& 1.0 & 1.0-1\\ 
5530 &$<$27.52         &     11.10&     10.94$^{+0.01}_{-0.01}$& 1.70$^{+0.10}_{-0.02}$& 0.30$^{+0.05}_{-0.10}$&  1.35 $_{ -0.38 }^{+ 0.13 }$& 0.3 &0.3-0.3\\
5534 &375.88$\pm$10.78 &     11.23&     11.23$^{+0.02}_{-0.01}$& 0.32$^{+0.00}_{-0.03}$& 2.45$^{+0.15}_{-0.03}$& 88.29 $_{ -2.15 }^{+39.43 }$& 0.1 &0.1-0.1\\
5556 &$<$24.83         &     11.08&     11.03$^{+0.02}_{-0.01}$& 0.81$^{+0.00}_{-0.00}$& 0.45$^{+0.04}_{-0.03}$&  0.43 $_{ -0.01 }^{+ 0.02 }$& 0.1 &0.1-0.1\\
5564 &$<$9.34          &     11.22&     11.34$^{+0.01}_{-0.02}$& 2.00$^{+0.10}_{-0.07}$& 0.30$^{+0.07}_{-0.10}$&  1.27 $_{ -0.38 }^{+ 0.38 }$& 0.3 &0.3-0.3\\
5860 &141.95$\pm$15.23 &     11.40&     11.33$^{+0.02}_{-0.01}$& 0.29$^{+0.00}_{-0.03}$& 2.45$^{+0.15}_{-0.06}$& 158.60$_{-27.17 }^{+64.16 }$& 0.1 &0.1-0.1\\
5974 & 511.86$\pm$24.04&     11.01&     11.20$^{+0.03}_{-0.09}$& 0.64$^{+0.08}_{-0.35}$& 2.20$^{+0.25}_{-0.10}$& 90.06 $_{-21.08 }^{+22.89 }$& 0.3 &0.1-0.3\\
6071 &185.39$\pm$37.17 &     11.29&     11.10$^{+0.05}_{-0.06}$& 0.20$^{+0.12}_{-0.07}$& 2.85$^{+0.05}_{-0.05}$&513.46 $_{-138.94}^{+141.29}$& 0.3 &0.1-1000\\
6193 & 126.22$\pm$8.65 &     11.32&     11.19$^{+0.00}_{-0.00}$& 0.64$^{+0.00}_{-0.01}$& 0.60$^{+0.00}_{-0.01}$&  3.28 $_{ -0.02 }^{+ 0.00 }$& 0.1 &0.1-0.1\\
6224 & 74.75$\pm$11.88 &     11.02&     10.68$^{+0.06}_{-0.07}$& 0.20$^{+0.08}_{-0.06}$& 2.35$^{+0.05}_{-0.10}$&280.64 $_{-60.14 }^{+42.58 }$& 1000.0 &0.3-1000\\
6352 &91.69$\pm$21.73  &     11.03&     10.94$^{+0.03}_{-0.00}$& 0.32$^{+0.04}_{-0.03}$& 1.15$^{+0.15}_{-0.09}$& 45.71 $_{-12.08 }^{+20.56 }$& 0.1 &0.1-0.1\\
6572 &265.86$\pm$77.08 &     11.08&     11.50$^{+0.05}_{-0.09}$& 3.00$^{+0.00}_{-1.20}$& 2.00$^{+0.15}_{-0.20}$&132.45 $_{-44.91 }^{+43.37 }$& 30.0 & 3.0-1000\\
6647 &$<$20.92         &     11.10&     11.14$^{+0.01}_{-0.02}$& 1.28$^{+0.04}_{-0.00}$& 0.60$^{+0.00}_{-0.05}$&  0.00 $_{ -0.00 }^{+ 0.00 }$& 0.0 &0.0-0.0\\
6898 &90.28$\pm$16.12  &     11.41&     11.56$^{+0.05}_{-0.09}$& 1.43$^{+0.47}_{-0.79}$& 2.25$^{+0.20}_{-0.15}$&305.12 $_{-78.62 }^{+138.28}$& 10.0 & 1.0-1000\\
7077 &83.23$\pm$6.59   &     11.22&     11.17$^{+0.02}_{-0.00}$& 0.36$^{+0.00}_{-0.04}$& 1.65$^{+0.15}_{-0.01}$& 51.81 $_{ -0.54 }^{+25.23 }$& 0.1 &0.1-0.1\\
7116 &90.94$\pm$31.19  &     11.13&     11.29$^{+0.01}_{-0.04}$& 4.00$^{+0.00}_{-0.75}$& 1.75$^{+0.05}_{-0.10}$& 66.21 $_{-10.96 }^{+ 6.91 }$& 1000.0 & 10.0-1000\\
7491 &$<$9.68          &     11.09&     10.98$^{+0.01}_{-0.02}$& 1.28$^{+0.00}_{-0.00}$& 0.15$^{+0.03}_{-0.05}$&  0.00 $_{ -0.00 }^{+ 0.00 }$& 0.0 &0.0-0.0\\  
7617 &$<$ 14.47        &     11.03&     10.97$^{+0.03}_{-0.03}$& 1.28$^{+0.23}_{-0.00}$& 0.00$^{+0.00}_{-0.00}$&  0.00 $_{ -0.00 }^{+ 0.00 }$& 0.0 &0.0-0.0\\
7923 &$<$11.91         &     11.11&     11.01$^{+0.00}_{-0.02}$& 1.90$^{+0.01}_{-0.10}$& 0.00$^{+0.05}_{-0.00}$&  0.83 $_{ -0.11 }^{+ 0.28 }$& 0.3 &0.3-0.3\\
8121 &$<$10.25         &     11.37&     11.51$^{+0.02}_{-0.01}$& 1.90$^{+0.10}_{-0.03}$& 0.25$^{+0.04}_{-0.10}$&  2.62 $_{ -0.76 }^{+ 0.12 }$& 0.3 &0.3-0.3\\
8569 &140.69$\pm$9.35  &     11.33&     11.14$^{+0.01}_{-0.01}$& 0.23$^{+0.03}_{-0.01}$& 1.75$^{+0.02}_{-0.10}$&191.17 $_{-44.35 }^{+14.01 }$& 0.1 &0.1-0.1\\

\enddata
\tablecomments{The second column shows the total SFR(IR+UV), where the IR contribution is derived from the IR SEDs, 24~$\mu$m/MIPS flux, or 24~$\mu$m/MIPS upper limit, for \pacsd, \mipsd, and \mipsu\ galaxies, respectively, as detailed in Section~\ref{sec:sfr}. $M_*$(D04), and $M_*$(SED) are stellar masses derived as in D04, and from the best-fit SED, respectively. Other best-fit SED quantities follow; in the order: mean stellar-age, reddening ($A_{\rm V}$), SFR, declining $\tau$, and $\tau$ 90\% confidence interval. All the uncertainties on the SED-derived quantities are within the 90\% confidence level, and inferred from the $\chi^2$ test. Typical uncertainties on $M_*$(D04) are around $40\%$ (cf., D04).}
\end{deluxetable}

\clearpage
\begin{deluxetable}{cccccccc}
\tabletypesize{\scriptsize}
\tablecaption{Morphological parameters from the GALFIT 2D SB fitting on the WFC3/f160W ($H$-band) images. \label{tab:morph}}
\tablewidth{0pt}
\tablehead{
\colhead{ID} &
\colhead{mag} &
\colhead{n} &
\colhead{\recirc} &
\colhead{B/T} &
\colhead{PSF/T (sS+PSF)} &
\colhead{$log \frac{sSFR}{sSFR(MS)}$} &
\colhead{$\frac{R_{\rm e,circ}}{R_0}$}\\
\colhead{} &
\colhead{[AB]} &
\colhead{} &
\colhead{[kpc]} &
\colhead{} &
\colhead{} &
\colhead{} &
\colhead{}}
\startdata

282 & 21.31 $\pm$ 0.01 & 4.68 $\pm$ 0.08 & 1.63 $\pm$ 0.02 & 0.83 $\pm$ 0.01 & $<0.1$             & -2.45       $^{+0.16}_{   -0.02}$  & 0.41\\ 
427 & 20.91 $\pm$ 0.01 & 3.73 $\pm$ 0.04 & 2.4  $\pm$ 0.02 & 0.47 $\pm$ 0.02 & 0.15 $\pm$ 0.006   & -2.20       $^{+0.02}_{   -0.01}$  & 0.55\\ 
428 & 21.65 $\pm$ 0.02 & 3.8$\pm$ 0.08 & 5.8$\pm$ 0.17 & 0.2$\pm$0.01  & 0.13 $\pm$ 0.01       & -0.6       $^{+0.08}_{   -0.06}$ & 1.78\\ 
552 & 21.18 $\pm$ 0.01 & 4.01 $\pm$ 0.09 & 3.66 $\pm$ 0.08 & 0.54 $\pm$ 0.04 & $<0.1$             & -0.47       $^{+0.04}_{   -0.04}$  & 0.95\\ 
557 & 22.28 $\pm$ 0.01 & 2.26 $\pm$ 0.04 & 2.07 $\pm$ 0.02 & 0.65 $\pm$ 0.03 & $<0.1$             & -1.59       $^{+0.02}_{   -0.08}$  & 0.66\\ 
686 & 21.73 $\pm$ 0.02 & 4.02 $\pm$ 0.19 & 2.55 $\pm$ 0.05 & 0.85 $\pm$ 0.04 & 0.22 $\pm$ 0.01    & 0.45        $^{+0.03}_{   -0.03}$  & 0.78\\ 
687 & 21.37 $\pm$ 0.01 & 2.66 $\pm$ 0.02 & 1.8 $\pm$ 0.0 & 0.62 $\pm$ 0.02 & $<0.1$               & -2.38       $^{+0.02}_{   -0.12}$  & 0.39\\ 
720 & 21.72 $\pm$ 0.09 & 3.39 $\pm$ 0.42 & 4.71 $\pm$ 0.73 & 0.32 $\pm$ 0.05 & $<0.1$             & -0.01       $^{+0.02}_{   -0.02}$  & 1.41\\ 
848 & 21.58 $\pm$ 0.01 & 3.23 $\pm$ 0.05 & 0.88 $\pm$ 0.01 & 0.83 $\pm$ 0.02 & $<0.1$             & $<$-3.0                       &  0.23\\ 
856 & 20.45 $\pm$ 0.01 & 5.31 $\pm$ 0.04 & 2.85 $\pm$ 0.02 & 0.86 $\pm$ 0.02 & 0.10 $\pm$ 0.01    & $<$-3.0                       & 0.44\\ 
880 & 21.91 $\pm$ 0.01 & 4.8 $\pm$ 0.16 & 1.59 $\pm$ 0.03 & 1.0 $\pm$ 0.0 & $<0.1$                & -1.06       $^{+0.02}_{   -0.01}$  & 0.38\\ 
887 & 21.04 $\pm$ 0.01 & 5.9 $\pm$ 0.08 & 2.38 $\pm$ 0.03 & 0.76 $\pm$ 0.02 & $<0.1$              & $<$-3.0                       & 0.57\\ 
903 & 22.18 $\pm$ 0.01 & 0.88 $\pm$ 0.01 & 4.36 $\pm$ 0.03 & 0.0 $\pm$ 0.0 & $<0.1$               & 0.07        $^{+0.10}_{   -0.10}$  & 1.28\\ 
947 & 21.83 $\pm$ 0.01 & 2.69 $\pm$ 0.05 & 0.7 $\pm$ 0.0 & 0.71 $\pm$ 0.03 & $<0.1$               & -2.13       $^{+0.15}_{   -0.13}$  & 0.21\\ 
986 & 20.96 $\pm$ 0.01 & 6.27 $\pm$ 0.09 & 1.89 $\pm$ 0.02 & 0.79 $\pm$ 0.02 & $<0.1$             & $<$-3.0                       & 0.35\\ 
1084 & 21.63 $\pm$ 0.01 & 2.59 $\pm$ 0.04 & 1.13 $\pm$ 0.01 & 0.0 $\pm$ 0.0 & 0.24 $\pm$ 0.01     & -2.41       $^{+0.19}_{   -0.12}$  & 0.28\\ 
1187 & 21.5 $\pm$ 0.01 & 2.73 $\pm$ 0.02 & 2.11 $\pm$ 0.01 & 0.0 $\pm$ 0.0 & 0.18 $\pm$ 0.01      & -1.94       $^{+0.01}_{   -0.14}$  & 0.50\\ 
1272 & 21.74 $\pm$ 0.01 & 2.54 $\pm$ 0.03 & 2.02 $\pm$ 0.02 & 0.81 $\pm$ 0.02 & $<0.1$            & $<$-3.0                       & 0.61\\ 
1510 & 21.61 $\pm$ 0.01 & 1.01 $\pm$ 0.01 & 6.94 $\pm$ 0.08 & 0.0 $\pm$ 0.0 & 0.10 $\pm$ 0.05     & 0.22        $^{+0.02}_{   -0.02}$  & 2.02\\ 
1906 & 21.39 $\pm$ 0.01 & 5.01 $\pm$ 0.05 & 1.68 $\pm$ 0.01 & 1.0 $\pm$ 0.0 & $<0.1$              & -0.03       $^{+0.09}_{   -0.09}$  & 0.52\\ 
2450 & 22.12 $\pm$ 0.01 & 1.9 $\pm$ 0.03 & 0.99 $\pm$ 0.0 & 0.49 $\pm$ 0.02 & $<0.1$              & -1.64       $^{+0.01}_{   -0.11}$  & 0.31\\ 
2470 & 22.63 $\pm$ 0.01 & 1.25 $\pm$ 0.02 & 2.58 $\pm$ 0.02 & 0.0 $\pm$ 0.0 & 0.10 $\pm$ 0.01     & 0.02        $^{+0.14}_{   -0.14}$  & 0.82\\ 
2940 & 21.26 $\pm$ 0.01 & 5.42 $\pm$ 0.09 & 2.03 $\pm$ 0.03 & 0.89 $\pm$ 0.02 & 0.17 $\pm$ 0.01   & -1.35       $^{+0.01}_{   -0.01}$  & 0.50\\ 
3066 & 21.6 $\pm$ 0.01 & 2.72 $\pm$ 0.04 & 5.11 $\pm$ 0.08 & 0.15 $\pm$ 0.01 & 0.10 $\pm$ 0.01    & -0.22       $^{+0.13}_{   -0.13}$  & 1.41\\ 
3230 & 21.69 $\pm$ 0.01 & 3.41 $\pm$ 0.03 & 1.23 $\pm$ 0.01 & 0.65 $\pm$ 0.04 & 0.14 $\pm$ 0.05   & $<$-3.0                       & 0.35\\ 
3231 & 20.42 $\pm$ 0.01 & 2.53 $\pm$ 0.01 & 6.62 $\pm$ 0.03 & 0.63 $\pm$ 0.01 & $<0.1$            & -2.02       $^{+0.02}_{   -0.08}$  & 0.92\\ 
3236 & ...             & ...            & ...           & ...            & ...              & -0.06       $^{+0.06}_{   -0.06}$  & ...\\ 
3258 & 21.17 $\pm$ 0.01 & 1.85 $\pm$ 0.01 & 2.01 $\pm$ 0.01 & 0.38 $\pm$ 0.02 & $<0.1$            & -2.19       $^{+0.15}_{   -0.13}$  & 0.44\\ 
3853 & 21.06 $\pm$ 0.01 & 3.03 $\pm$ 0.02 & 2.02 $\pm$ 0.01 & 0.45 $\pm$ 0.01 & 0.14 $\pm$ 0.01   & -1.03       $^{+0.09}_{   -0.02}$  & 0.44\\
4310 & 21.98 $\pm$ 0.01 & 2.24 $\pm$ 0.06 & 0.78 $\pm$ 0.01 & 0.4 $\pm$ 0.07 & 0.21 $\pm$ 0.02    & $<$-3.0                       & 0.24\\ 
4406 & 23.2 $\pm$ 0.02 & 2.33 $\pm$ 0.08 & 4.2 $\pm$ 0.14 & 0.0 $\pm$ 0.0 & 0.13 $\pm$ 0.01       & -0.12       $^{+0.20}_{   -0.20}$  & 1.04\\ 
4705 & 22.08 $\pm$ 0.01 & 0.94 $\pm$ 0.01 & 2.58 $\pm$ 0.01 & 0.0 $\pm$ 0.0 & $<0.1$              & 0.21        $^{+0.01}_{   -0.01}$  & 0.82\\ 
5149 & 21.84 $\pm$ 0.01 & 3.28 $\pm$ 0.05 & 1.17 $\pm$ 0.0 & 0.8 $\pm$ 0.04 & $<0.1$              & -1.97       $^{+0.20}_{   -0.02}$  & 0.33\\ 
5415 & 21.45 $\pm$ 0.02 & 7.39 $\pm$ 0.12 & 1.8 $\pm$ 0.03 & 1.0 $\pm$ 0.0 & $<0.1$               & -0.53       $^{+0.06}_{   -0.06}$  & 0.54\\ 
5503 & ...             & ...            & ...            & ...          & ...                & 0.49        $^{+0.01}_{   -0.01}$  & ...\\ 
5509 & 22.1 $\pm$ 0.01 & 2.37 $\pm$ 0.04 & 4.26 $\pm$ 0.07 & 0.71 $\pm$ 0.03 & $<0.1$             & -1.29       $^{+0.11}_{   -0.01}$  & 0.91\\ 
5530 & 21.91 $\pm$ 0.01 & 3.23 $\pm$ 0.06 & 0.98 $\pm$ 0.0 & 0.66 $\pm$ 0.02 & $<0.1$             & -2.04       $^{+0.04}_{   -0.12}$  & 0.28\\ 
5534 & ...         & ...      & ...      & ...     & ...             & 0.30        $^{+0.01}_{   -0.01}$  &... \\
5556 & 21.35 $\pm$ 0.01 & 3.26 $\pm$ 0.03 & 1.75 $\pm$ 0.01 & 0.79 $\pm$ 0.03 & $<0.1$            & -2.52       $^{+0.02}_{   -0.01}$  & 0.51\\ 
5564 & 21.53 $\pm$ 0.01 & 2.34 $\pm$ 0.02 & 1.8 $\pm$ 0.01 & 0.51 $\pm$ 0.03 & $<0.1$             & -2.15       $^{+0.13}_{   -0.13}$  & 0.44\\ 
5860 & 21.32 $\pm$ 0.01 & 2.34 $\pm$ 0.02 & 4.61 $\pm$ 0.03 & 0.57 $\pm$ 0.02 & $<0.1$            & -0.25       $^{+0.05}_{   -0.05}$  & 0.89\\ 
5974 & 22.44 $\pm$ 0.01 & 1.6 $\pm$ 0.02 & 3.81 $\pm$ 0.05 & 0.0 $\pm$ 0.0 & 0.12 $\pm$ 0.01      & 0.61        $^{+0.02}_{   -0.02}$  & 1.22\\ 
6071 & 21.45 $\pm$ 0.01 & 2.32 $\pm$ 0.02 & 4.56 $\pm$ 0.05 & 0.32 $\pm$ 0.02 & $<0.1$            & -0.05       $^{+0.09}_{   -0.09}$  & 1.02\\ 
6193 & 20.69 $\pm$ 0.01 & 3.48 $\pm$ 0.05 & 2.49 $\pm$ 0.03 & 0.5 $\pm$ 0.03 & 0.10 $\pm$ 0.01    & -0.24       $^{+0.03}_{   -0.03}$  & 0.53\\ 
6224 & 21.63 $\pm$ 0.01 & 5.69 $\pm$ 0.11 & 5.01 $\pm$ 0.15 & 0.57 $\pm$ 0.02 & $<0.1$            & -0.23       $^{+0.07}_{   -0.07}$  & 1.58\\ 
6352 & 21.19 $\pm$ 0.01 & 2.46 $\pm$ 0.02 & 1.17 $\pm$ 0.0 & 0.61 $\pm$ 0.01 & $<0.1$             & -0.15       $^{+0.10}_{   -0.10}$  & 0.37\\ 
6572 & 22.01$\pm$  0.01 & 3.44 $\pm$ 0.07 & 6.36 $\pm$ 0.17 & 0.14 $\pm$ 0.01 & 0.12 $\pm$ 0.01   & 0.28        $^{+0.13}_{   -0.13}$  & 1.86\\ 
6647 & 22.05 $\pm$ 0.01 & 1.69 $\pm$ 0.03 & 1.27 $\pm$ 0.01 & 0.37 $\pm$ 0.02 & $<0.1$            & $<$-3.0                       &0.36\\ 
6898 & 21.28 $\pm$ 0.01 & 6.52 $\pm$ 0.14 & 0.62 $\pm$ 0.01 & 1.0 $\pm$ 0.0 & 0.5 $\pm$ 0.01      & -0.46       $^{+0.08}_{   -0.08}$  & 0.12\\ 
7077 & 21.29 $\pm$ 0.01 & 4.79 $\pm$ 0.1 & 3.47 $\pm$ 0.08 & 0.39 $\pm$ 0.01 & 0.18 $\pm$ 0.01    & -0.34       $^{+0.03}_{   -0.03}$  & 0.85\\ 
7116 & 21.8 $\pm$ 0.01 & 1.36 $\pm$ 0.02 & 4.63 $\pm$ 0.06 & 0.1 $\pm$ 0.02 & $<0.1$              & -0.24       $^{+0.15}_{   -0.15}$  & 1.26\\ 
7491 & 21.53 $\pm$ 0.01 & 6.09 $\pm$ 0.15 & 0.88 $\pm$ 0.01 & 1.0 $\pm$ 0.0 & $<0.1$              & $<$-3.0                       & 0.26\\ 
7617 & 21.6 $\pm$ 0.01 & 4.09 $\pm$ 0.1 & 1.93 $\pm$ 0.03 & 0.36 $\pm$ 0.02 & 0.22 $\pm$ 0.01     & $<$-3.0                       & 0.60\\
7923 & 22.07 $\pm$ 0.01 & 1.65 $\pm$ 0.08 & 0.7 $\pm$ 0.01 & 0.0 $\pm$ 0.0 & 0.27 $\pm$ 0.05      & -2.26       $^{+0.14}_{   -0.06}$  & 0.20\\ 
8121 & 21.13 $\pm$ 0.01 & 2.61 $\pm$ 0.04 & 2.74 $\pm$ 0.04 & 0.15 $\pm$ 0.01 & 0.15 $\pm$ 0.01  & -1.96       $^{+0.02}_{   -0.13}$  & 0.55\\ 
8569 & 20.73 $\pm$ 0.0 & 1.34 $\pm$ 0.06 & 4.74 $\pm$ 0.48 & 0.0 $\pm$ 0.0 & $<0.1$               & -0.20       $^{+0.03}_{   -0.03}$  & 1.00\\
\enddata
\tablecomments{The total magnitude, Sersic index, and circularized effective radius are from the single-\sersic\ fit (see Section~\ref{sec:sersic}). For each object we also report the Bulge to Total flux ratio (B/T) from the composite Bulge+Disk model, discussed in Section~\ref{sec:bsut}, and the fractional contribution to the total flux of a hypothetical central point-source added to the single-\sersic\ fit, PSF/T(sS+PSF), when it is $>10\%$. The last two quantities are the distance from the MS in log-scale, as shown in Figures~\ref{fig:histo_morph}-\ref{fig:B2TvsMSdist}, and and the distance from the $M_*-$size relation of local ETGs, shown in Figures~\ref{fig:n_RevsMSdist}, and \ref{fig:B2TvsMSdist}.} 

\end{deluxetable}

\begin{twocolumn}
\section*{Acknowledgments}
This work is based on observations taken by the CANDELS Multi-Cycle Treasury Program with the NASA/ESA HST, which is operated by the Association of Universities for Research in Astronomy, Inc., under NASA contract NAS5-26555.
We thank Mattia Negrello and Mara Salvato for useful discussions. 

\bibliography{new_ref}

\bsp

\end{twocolumn}
\label{lastpage}

\appendix

\section[]{MIPS Deblending}\label{app:A}
In this appendix we detail the procedure used to derive the MIPS flux (and SFR) estimates for blended sources in our sample, whose results are shown in Section~\ref{sec:subsamples} and Figure~\ref{fig:pmulti}. 
Since the 24~$\mu$m/MIPS FWHM ($=5\farcs9$) is substantially larger than the size of our galaxies, each blended system includes all the objects closer than $\sim3''$ (HWHM) to the centroid of the MIPS emission. We used GALFIT to decompose the blended sources by fitting PSF models on the fixed F160W/WFC3 positions, leaving the galaxy magnitude as the only free parameter to be recovered. 
To avoid to have residual flux in the neighborhood, the sources at a distance $3-4''<r< 20''$ from the galaxy(ies) of interest were also fit together, but fixing all the parameters (i.e., position and magnitudes) to the known values from the available MIPS catalog.    
In the following we discuss the results derived for each object, or pair (when both the galaxies in a blended pair are included in our sample). We considered as MIPS-detected (to be included in \mipsd\ subsample) objects with SNR$>3$, and treated the others as upper limits (at 2$\sigma$ or 3$\sigma$, for SNR$<1$ and $\geq 1$, respectively).   
\begin{itemize}

\item {\bf Objects \#427 and \#428.} In the MIPS catalog based on IRAC positions the MIPS counterpart had been wrongly assigned to \#427, which is not compatible with such high MIPS flux, being spectroscopically classified as Early-Type galaxy as proved by the CaHK,MgII,MgI absorption lines (GOODS) and lack of the O[II] emission line (expected for a star-forming galaxy in the (VLT/FORS2) observed wavelength-range \citep{2008A&A...478...83V}. Moreover, the results of PSF fitting on the image confirm that object \#427 has a low 24~$\mu$m flux (F24$< 26~\mu$Jy), while most of the MIPS flux should come from object \#428. However, as detailed in Section~\ref{sec:lens}, this is one of the most peculiar case of blending in our sample, being consistent with a double lens system. In this perspective the MIPS flux should be associated to a lensed background star-forming galaxy, probably at $z\sim 2-3$. 

\item {\bf Objects \#686 and \#687.} These galaxies constitute a gravitationally bound system, as proved by their spectroscopic redshifts and mutual distance (Appendix~\ref{app:B}). The PSF fitting on the {\it HST}/WFC3 positions suggested that most of the MIPS emission comes from the AGN host \#686, although a small percentage of the flux may be associated to the companion (\#687). However, since both the red colors of \#687, and the SED fitting results, suggest old age ($\sim 2$~Gyr), and very low reddening ($A_V=0.3$ mag), and SFR ($\sim 1$~\msun/yr), we considered this object as \mipsu (cf. Table~\ref{tab:sedfit}). We notice that, also if considered as MIPS-detection, \#687 would remain a ``border-line'' galaxy more than 4$\times$ below the MS, since the MIPS flux derived from PSF fitting ($F_{24}=65.2 \pm 13.8~\mu$Jy) is very close to the quoted $3\sigma$ upper limit ($F_{24}<59.76~\mu$Jy). 

\item
{\bf Object \#720.} This galaxy has a near-IR/optical bright and massive ($M_*\sim5\times10^{11}~M_{\odot}$) neighbor (i.e., object \#721), not included in our sample  since its photometric redshift is slightly lower than the cut used in our sample-selection, i.e., $z_{phot}=1.36$. In the GOODS catalog \#721 had been identified as the only counterpart of the bright MIPS source. On the contrary, our deblending procedure suggested that most of the MIPS flux should be associated to object \#720 ($F_{24}=204.4\pm 9.4~\mu$Jy), with a smaller fraction of the flux ($F_{24}=83.5\pm 10 ~\mu$Jy) coming from \#721. 
The presence of peculiar tidal tail seems to suggest that this could be also a merging system, but this hypothesis is not supported by spectroscopic information.      

\item 
{\bf  Object \#880.} The detection from the GOODS catalog (SNR$>3$) was replaced with an upper limit, since the MIPS source seems to be the result of two blended marginal detection, as also suggested by the elongated shape of the source in the MIPS image.  

\item 
{\bf  Object \#1187.} As for object \#880, this MIPS detection appeared to be the sum of two or three blended partial detection's. Hence, the MIPS flux was replaced with an upper limit, consistently with the ``quenched nature'' suggested by the optical/near-IR colors of the galaxy.

\item 
{\bf  Objects \#3230 and \#3231.} These objects are blended with a third lower-mass ($\sim 2\times10^{10}~M_{\odot}$) neighbor (\#3232, $z_{phot}\sim 2$), not included in the MIPS GOODS catalog, which instead seems to be responsible of the majority of the MIPS flux ($F_{24}=109.6\pm 29.3 ~\mu$Jy). As detailed in Appendix~\ref{app:B}, while the object in the middle could be just a projected neighbor, \#3230 and \#3231 are spectroscopically confirmed at the same redshift. As also proved by their spectra, \#3230 and \#3231 seems to be quenched galaxies, with MIPS fluxes $< 46~\mu$Jy and $ <62~\mu$Jy (sSFR$\lesssim 10^{-10}yr^{-1}$), respectively.

\item 
{\bf Object \#5530.} This MIPS source, previously considered as detected above 3$\sigma$, was decomposed in three different sources, based on the WFC3 positions. A marginal 2$\sigma$ detection is found for the SE component, while the MIPS flux of \#5530 is below the 2$\sigma$ upper limit.

\item 
{\bf  Object \#5415.} This galaxy was considered as MIPS-detected in the automated MIPS-source extraction procedure, based on IRAC priors. However, it is blended with a close object (i.e., \#5414, at $z_{phot}=1.96$, and not included in our sample because of its mass two times below the mass limit). The PSF fitting method reveals a secure MIPS detection (SNR$\sim4$) for the neighbor, and a marginal detection (SNR$\sim2.5$) for \#5415. In this case, we classified \#5415 as \mipsd, due to its blue colors (typical of $sBzK$ and blue star-forming $UVJ$), and to the SED fitting results, showing reddening $A_V\sim 1$, and $SFR(SED)$ comparable with SFR(IR+UV), derived from MIPS (cf. Table~\ref{tab:sedfit}). 

\item 
{\bf  Object \#7923.} This galaxy is part of a complex blended system, and does not have MIPS counterparts within 2 arcsec in the MIPS catalog based on IRAC positions (the closest one is 2.4 arcsec away, SE, to this object). The MIPS flux decomposition based on WFC3 positions confirmed that this galaxies is not detected in MIPS, with a SNR$<1$, thus in Figure~\ref{fig:pmulti} we replaced the $3\sigma$ upper limit with a $2\sigma$ upper limit .

\end{itemize}

\begin{figure}
\begin{center}
\includegraphics[width=0.21\textwidth]{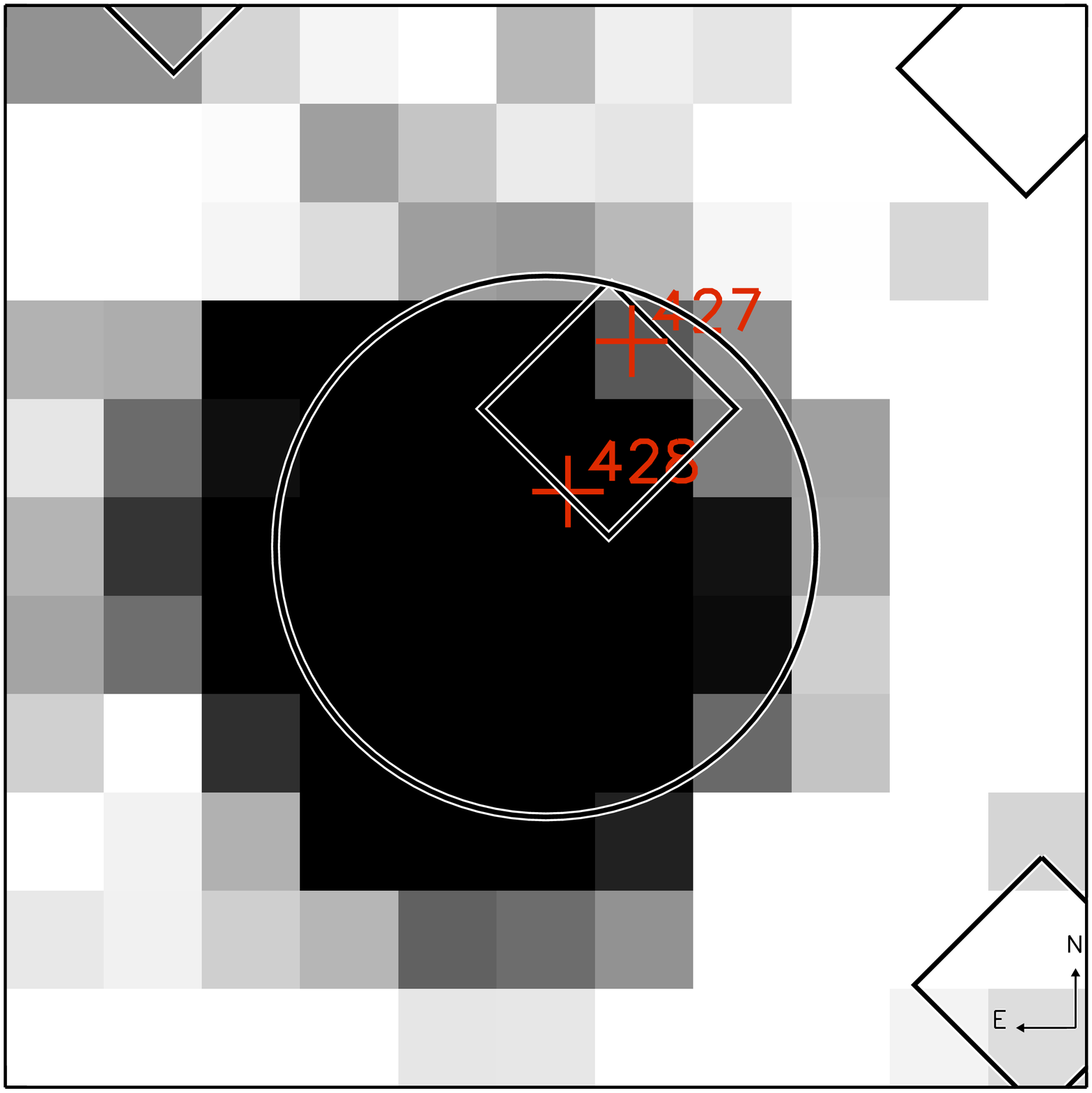}\includegraphics[width=0.21\textwidth]{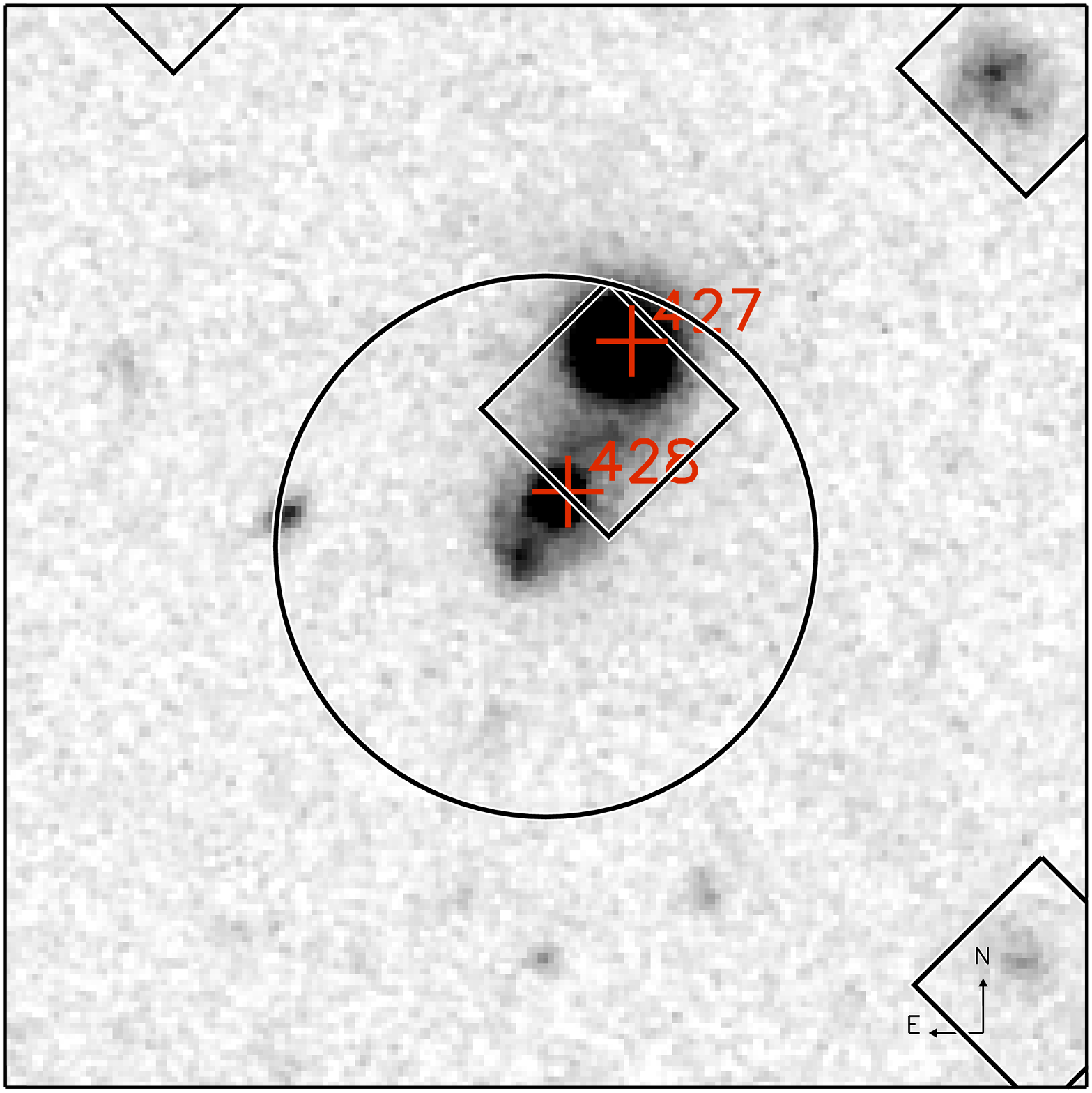}
\caption{ Comparison between MIPS/24$~\mu$m and HST/WFC3 $H$-band (F160W) images for objects \#427 and \#428. Both cutouts are 12 arcs in size. The red crosses correspond to WFC3/HST positions, while the large open circle shows the MIPS FWHM (5\farcs9), centered on the peak of the 24~$\mu$m/MIPS emission.  
The open diamonds mark the positions of IRAC sources in the field.}\label{appenfig:427_428}
\end{center}
\end{figure}

\begin{figure}
\begin{center}
\includegraphics[width=0.21\textwidth]{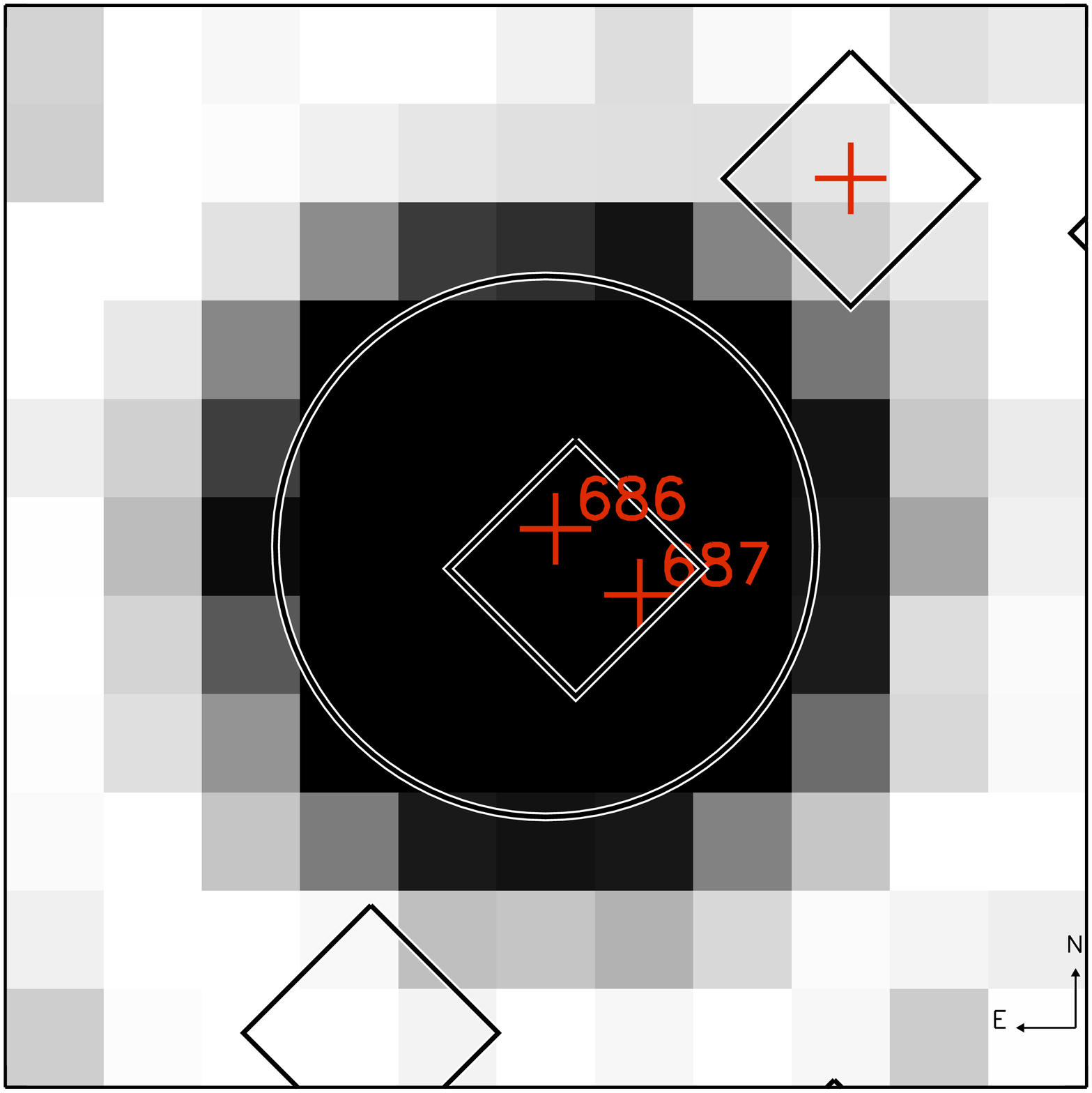}\includegraphics[width=0.21\textwidth]{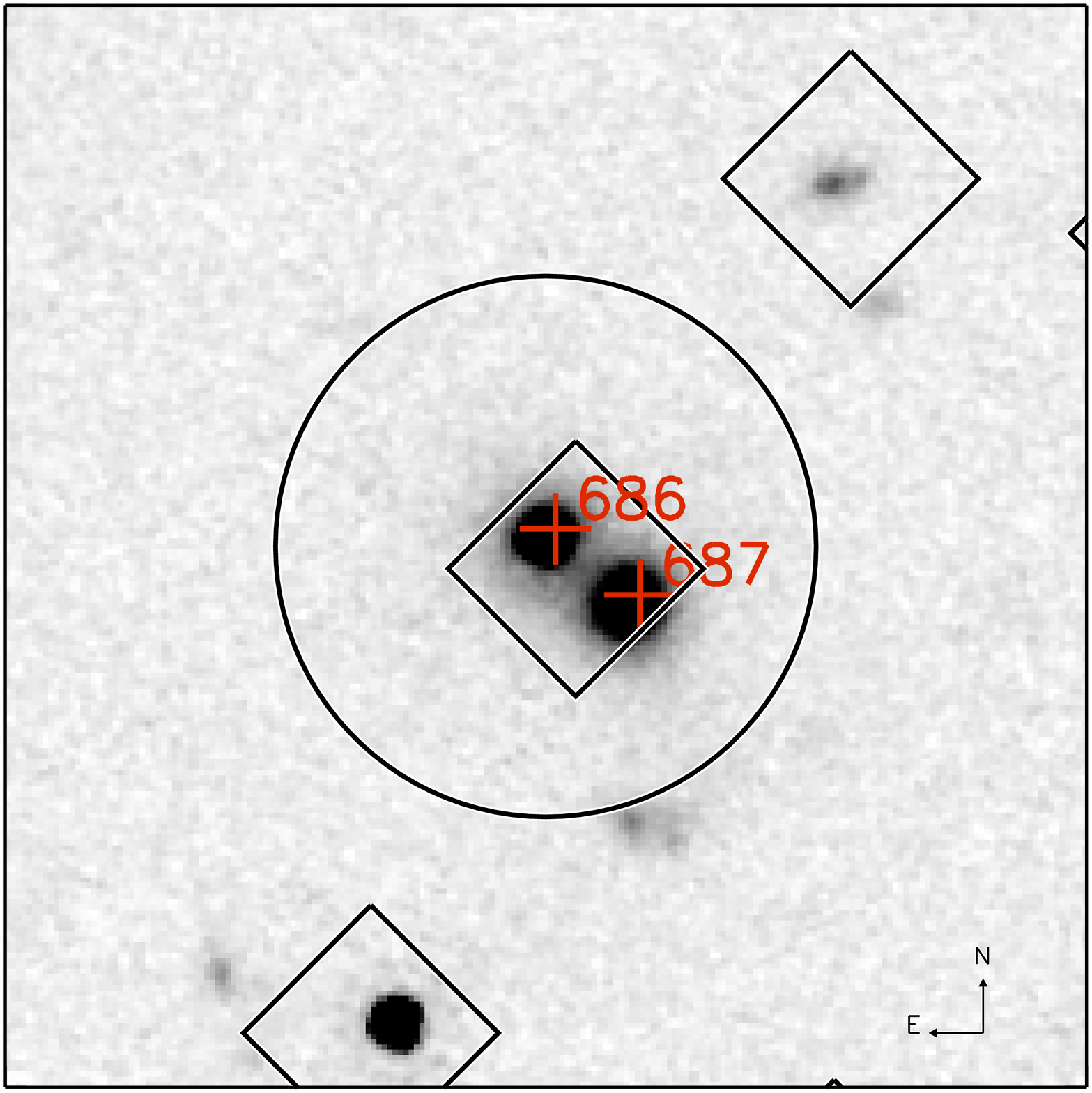}
\caption{Same as figure~\ref{appenfig:427_428}, for objects \#686 and \#687.}\label{appenfig:686_687}
\end{center}
\end{figure}

\begin{figure}
\begin{center}
\includegraphics[width=0.21\textwidth]{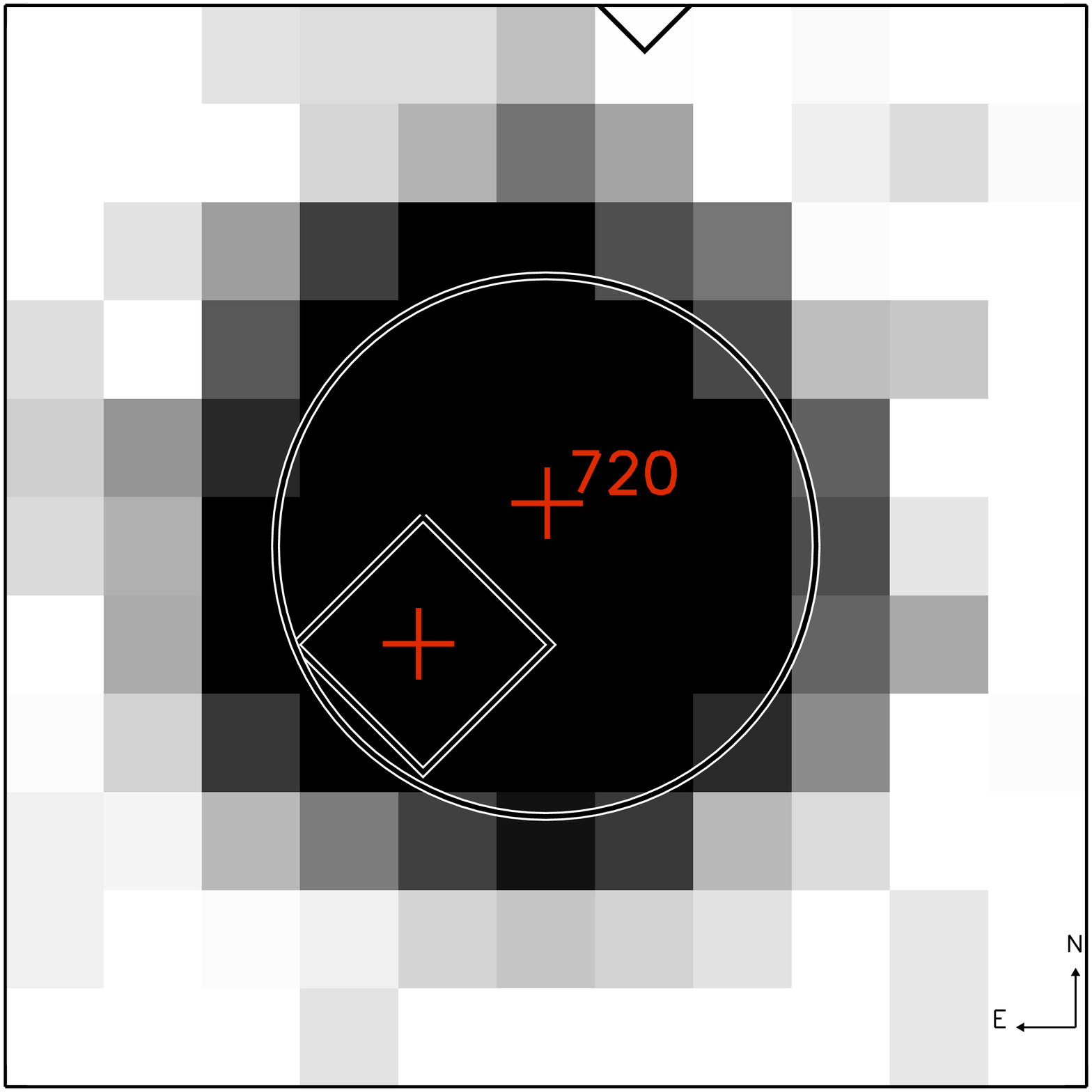} \includegraphics[width=0.21\textwidth]{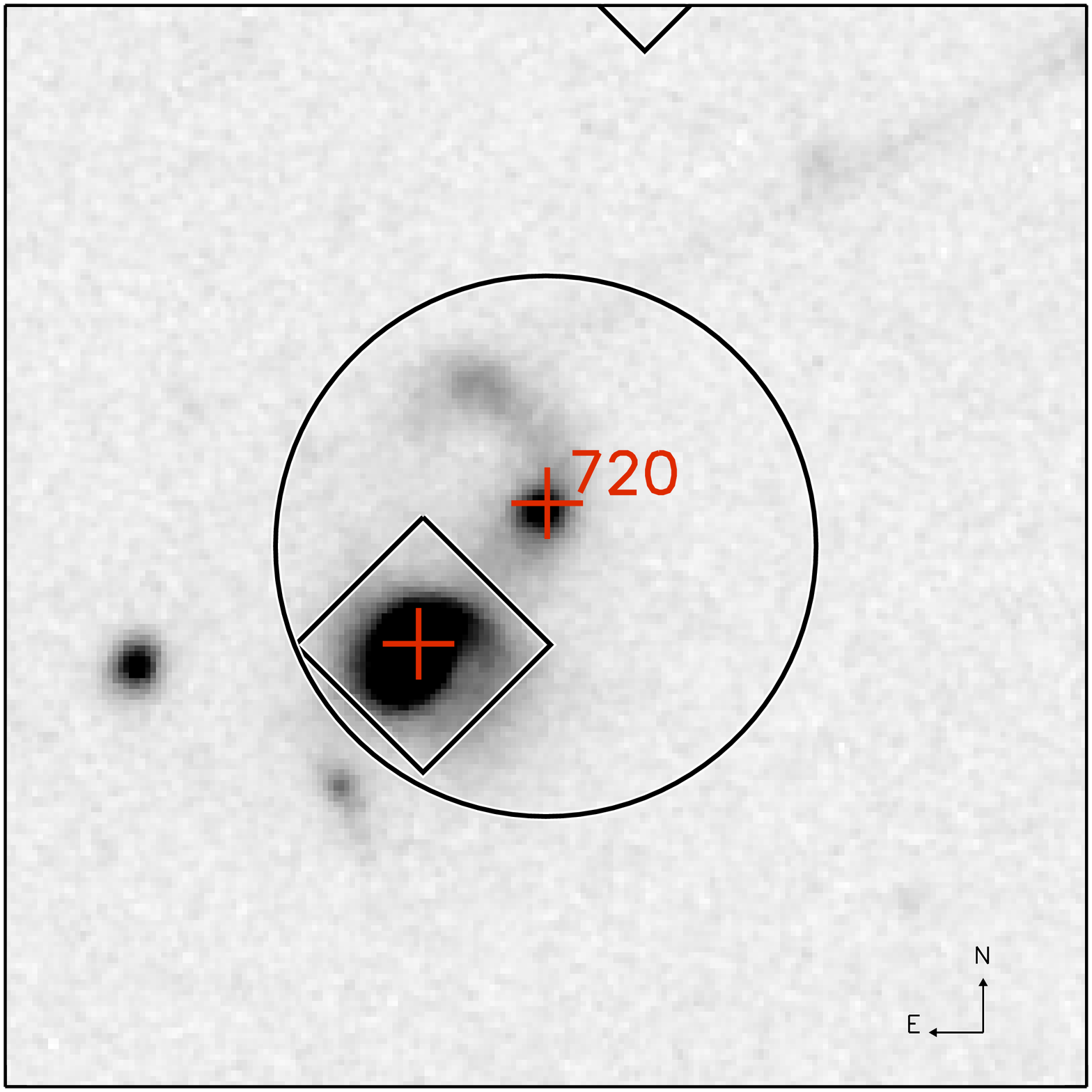}
\caption{Same as figure~\ref{appenfig:427_428}, for object \#720.}\label{appenfig:720}
\end{center}
\end{figure}

\begin{figure}
\begin{center}
\includegraphics[width=0.21\textwidth]{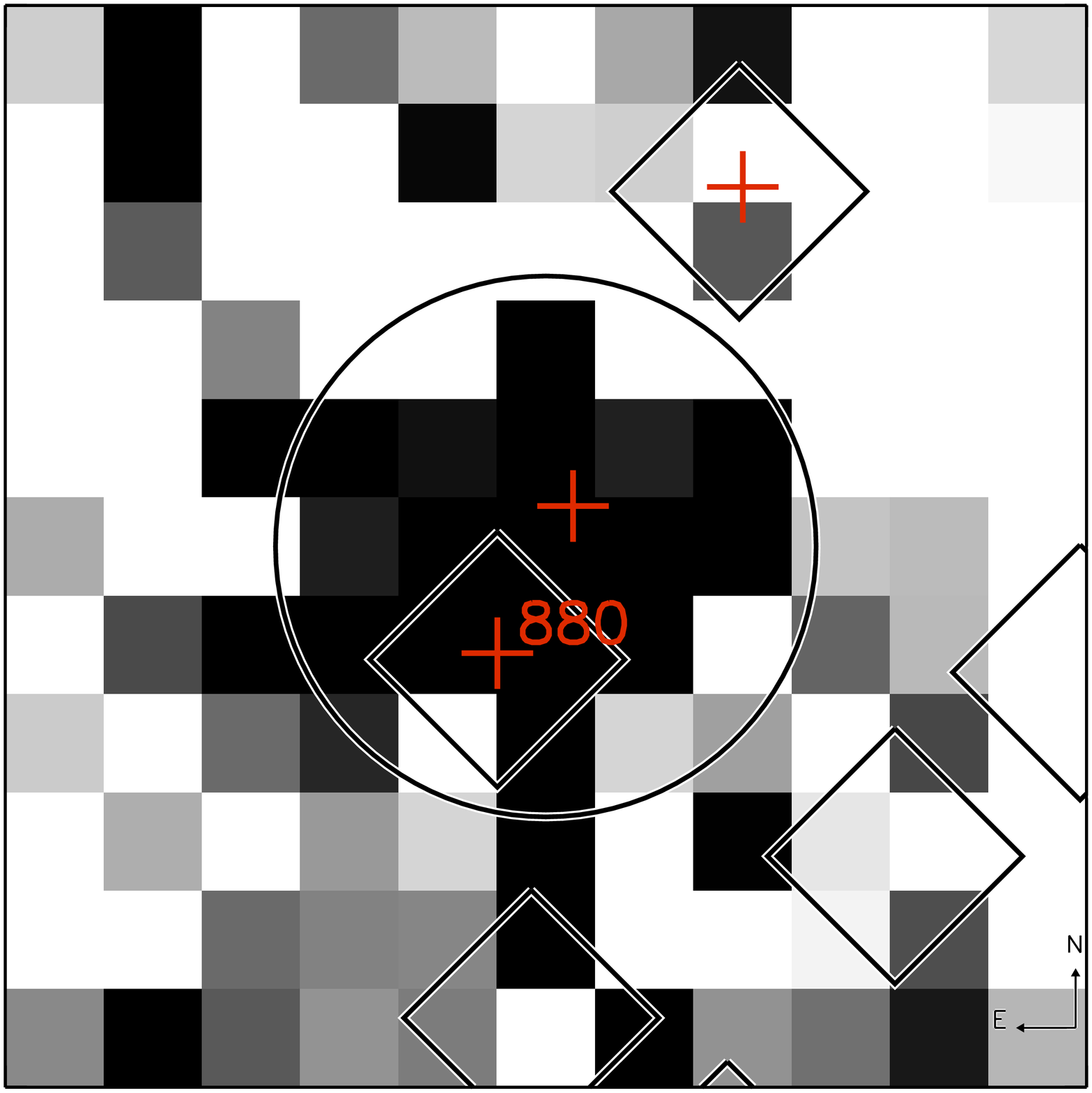} \includegraphics[width=0.21\textwidth]{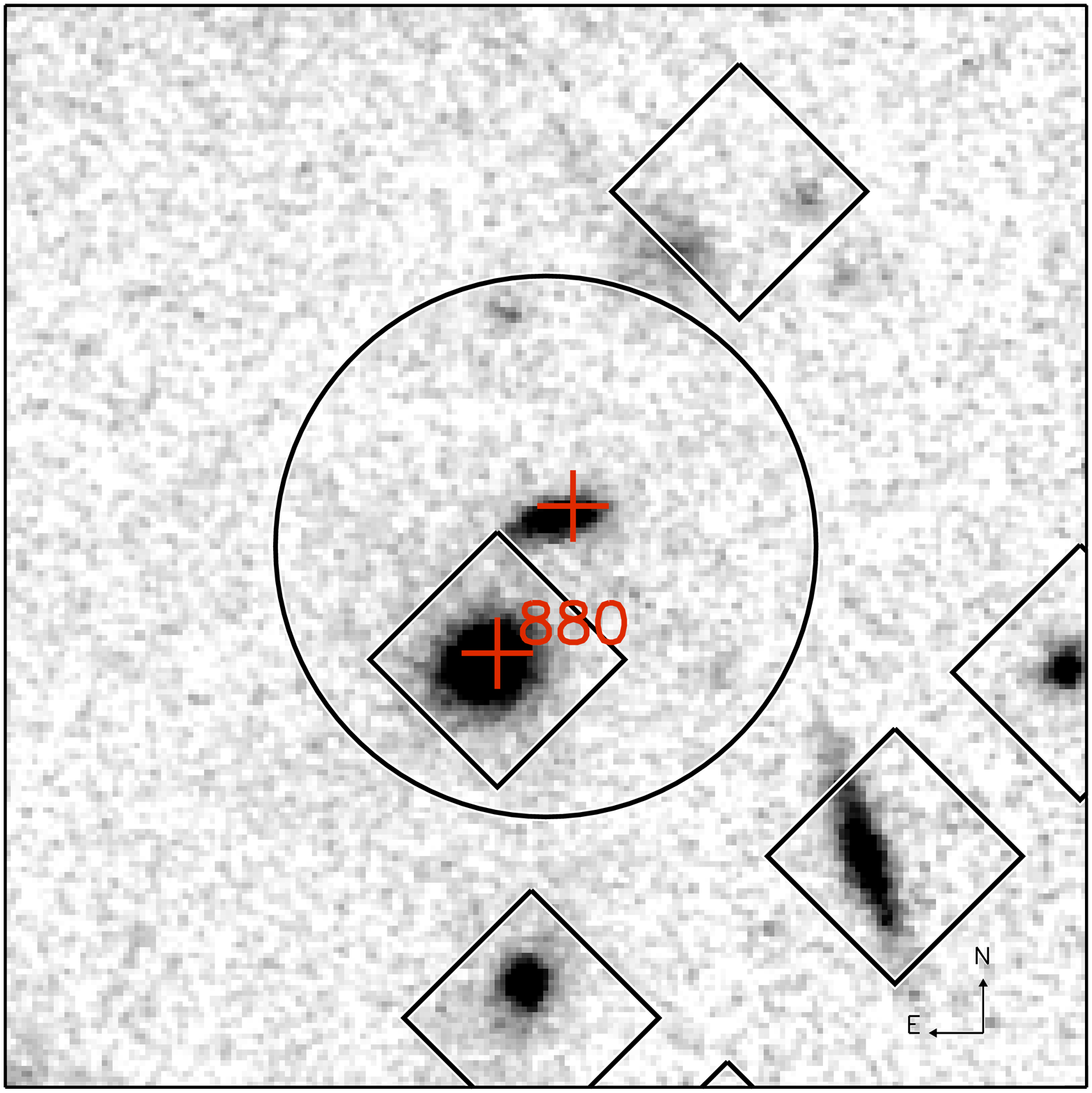}
\caption{Same as figure~\ref{appenfig:427_428}, for object \#880.}\label{appenfig:880}
\end{center}
\end{figure}

\begin{figure}
\begin{center}
\includegraphics[width=0.21\textwidth]{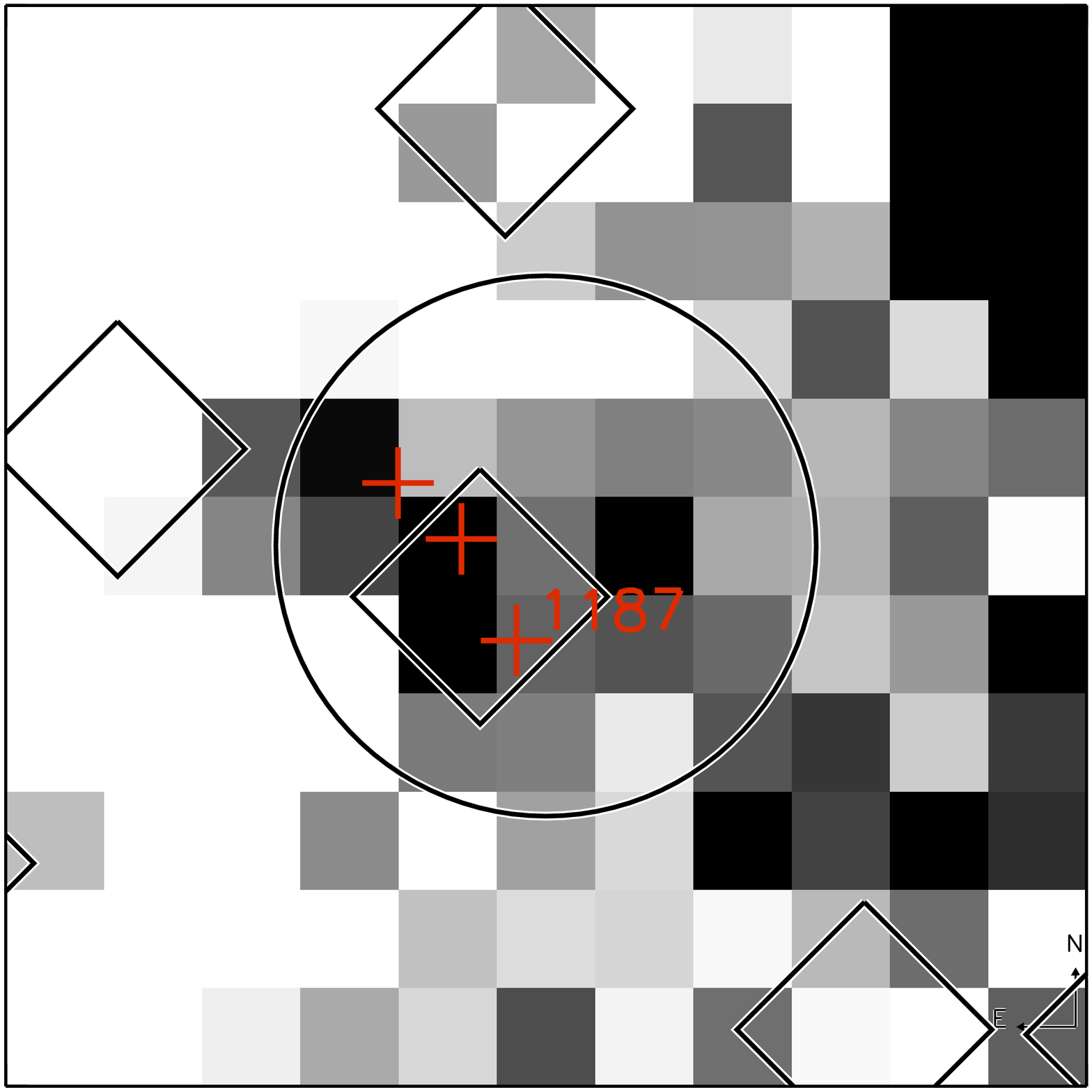}\includegraphics[width=0.21\textwidth]{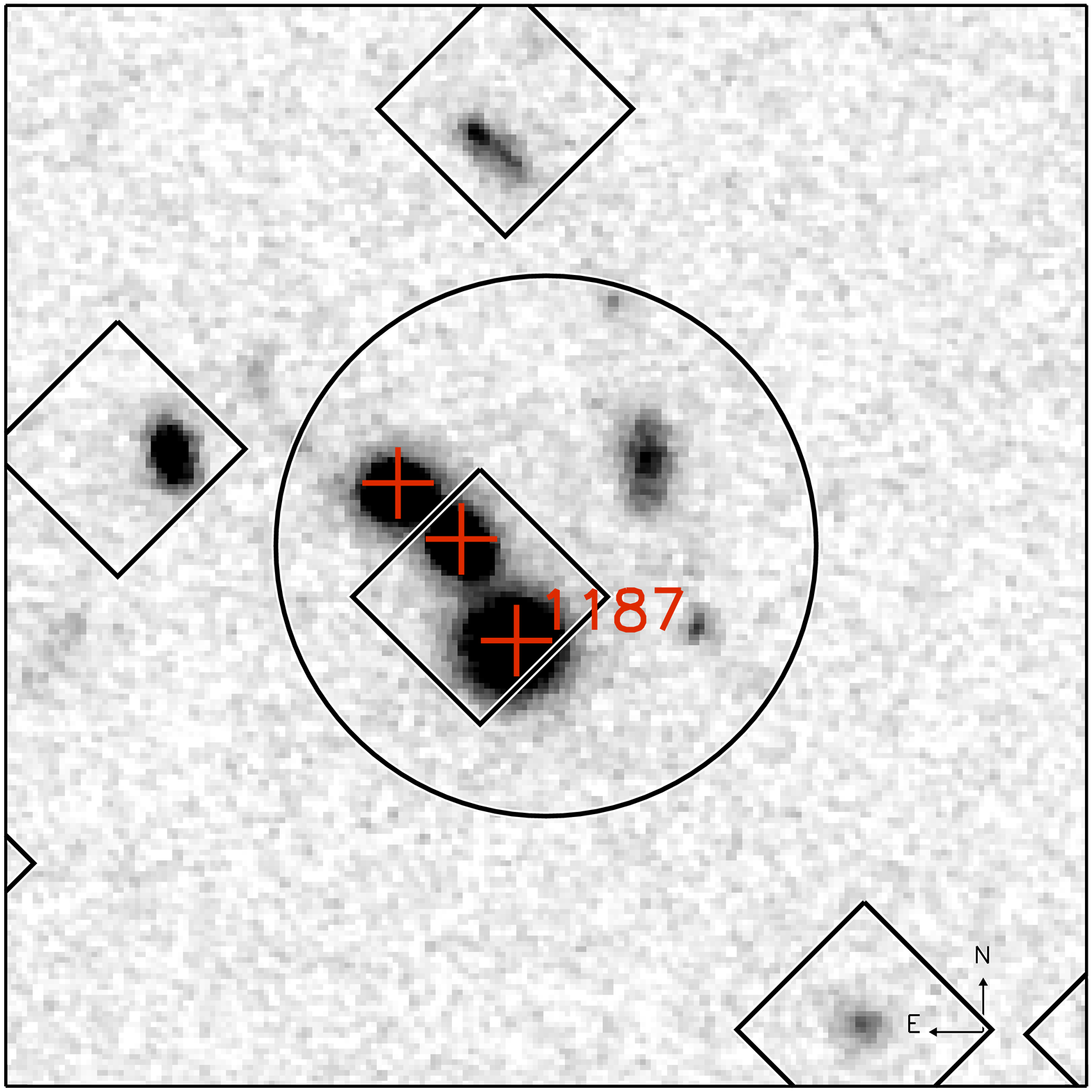}
\caption{Same as figure~\ref{appenfig:427_428}, for object \#1187.}\label{appenfig:1187}
\end{center}
\end{figure}

\begin{figure}
\begin{center}
\includegraphics[width=0.21\textwidth]{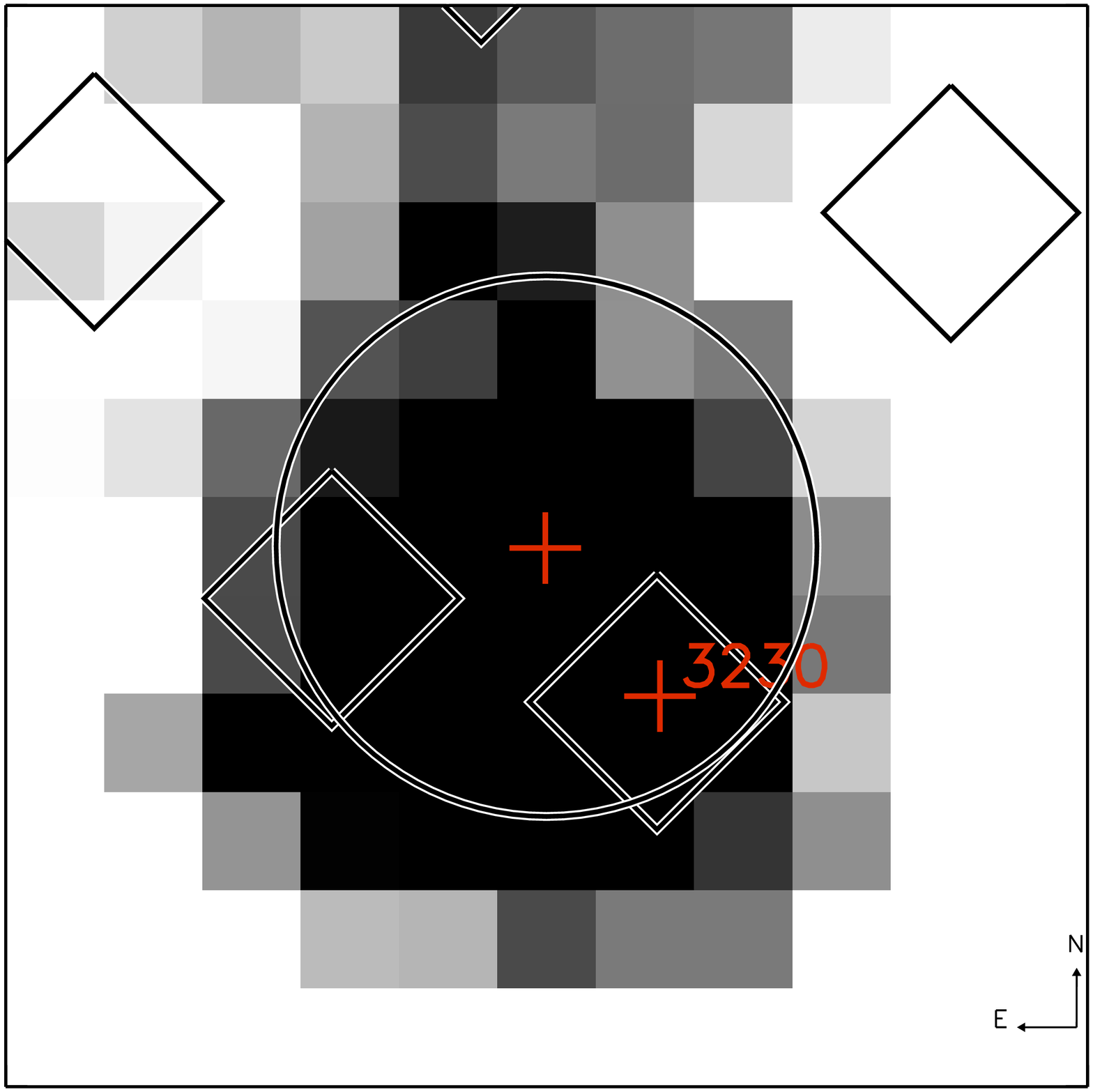}\includegraphics[width=0.21\textwidth]{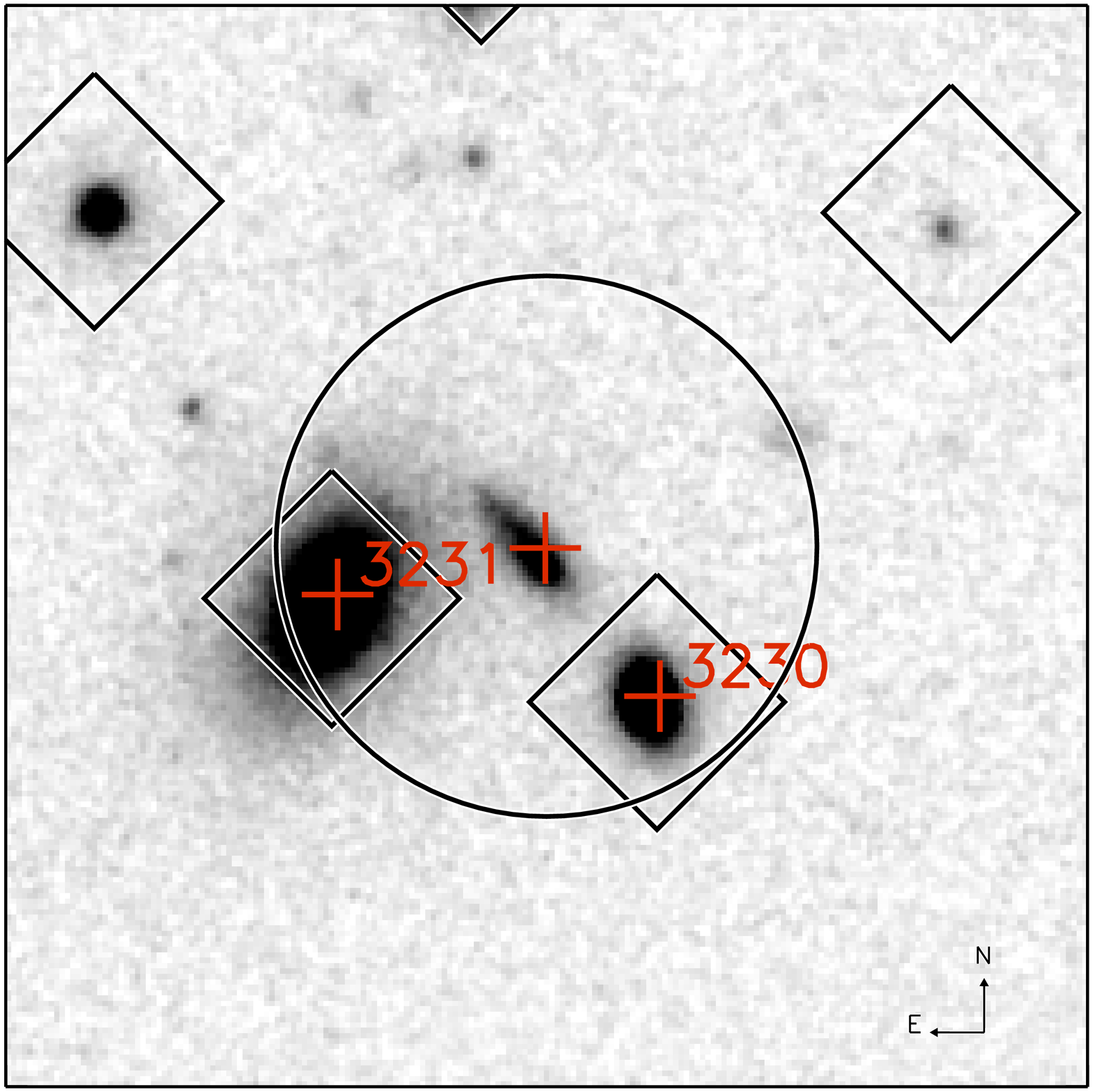}
\caption{Same as figure~\ref{appenfig:427_428}, for objects \#3230 and \#3231.}\label{appenfig:3230_3231}
\end{center}
\end{figure}

\begin{figure}
\begin{center}
\includegraphics[width=0.21\textwidth]{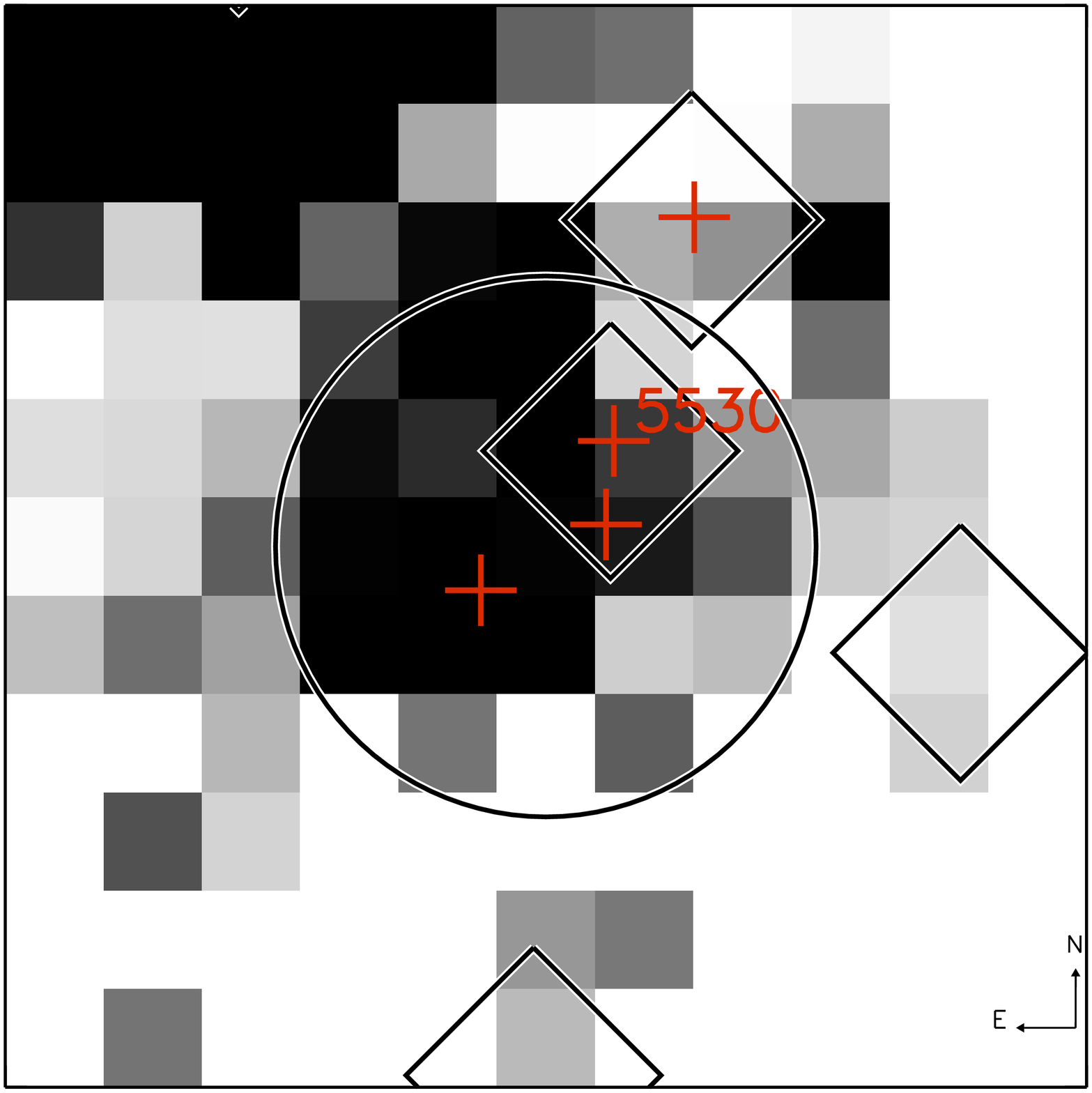}\includegraphics[width=0.21\textwidth]{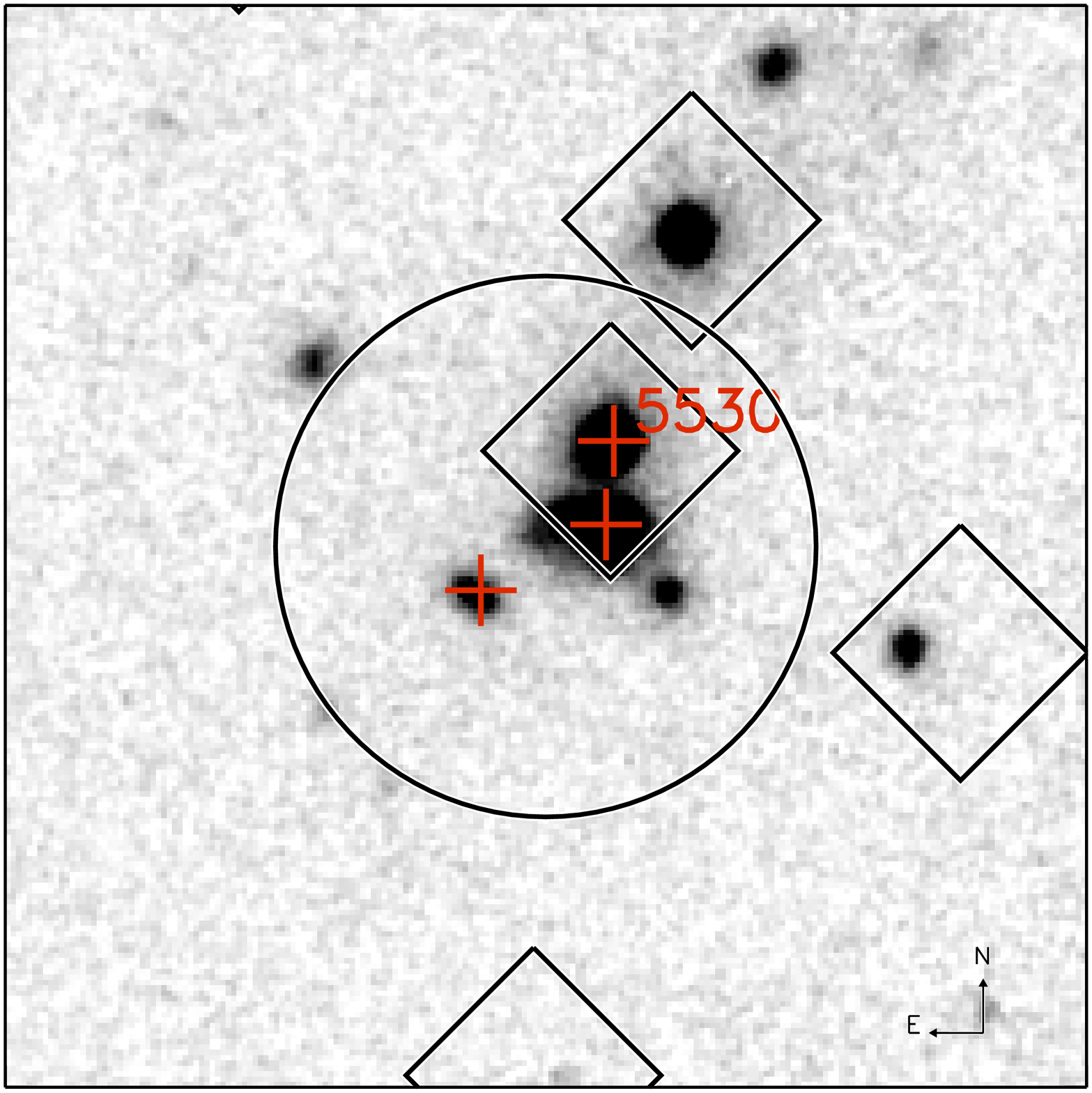}
\caption{Same as figure~\ref{appenfig:427_428}, for object \#5530.}\label{appenfig:5530}
\end{center}
\end{figure}

\begin{figure}
\begin{center}
\includegraphics[width=0.21\textwidth]{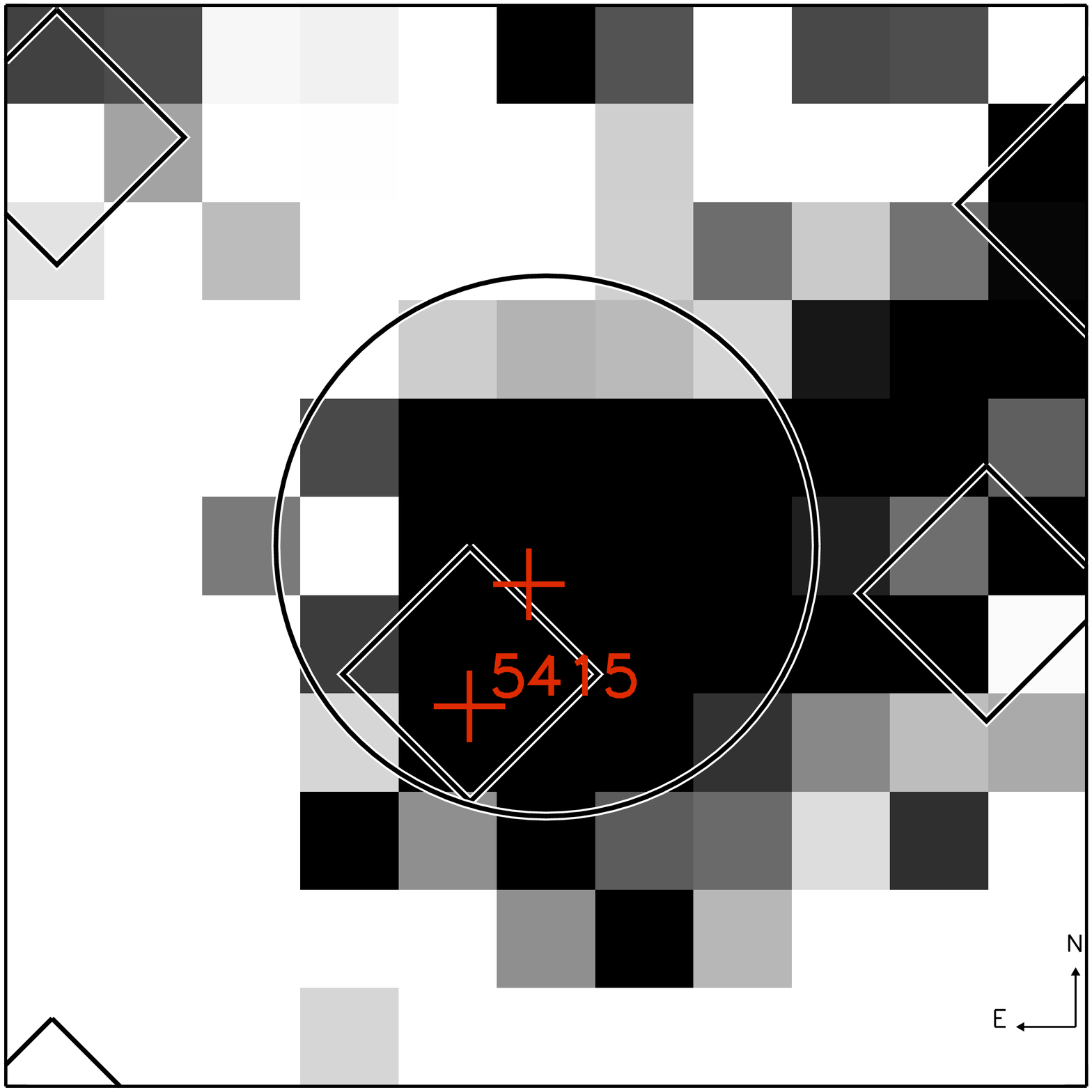}\includegraphics[width=0.21\textwidth]{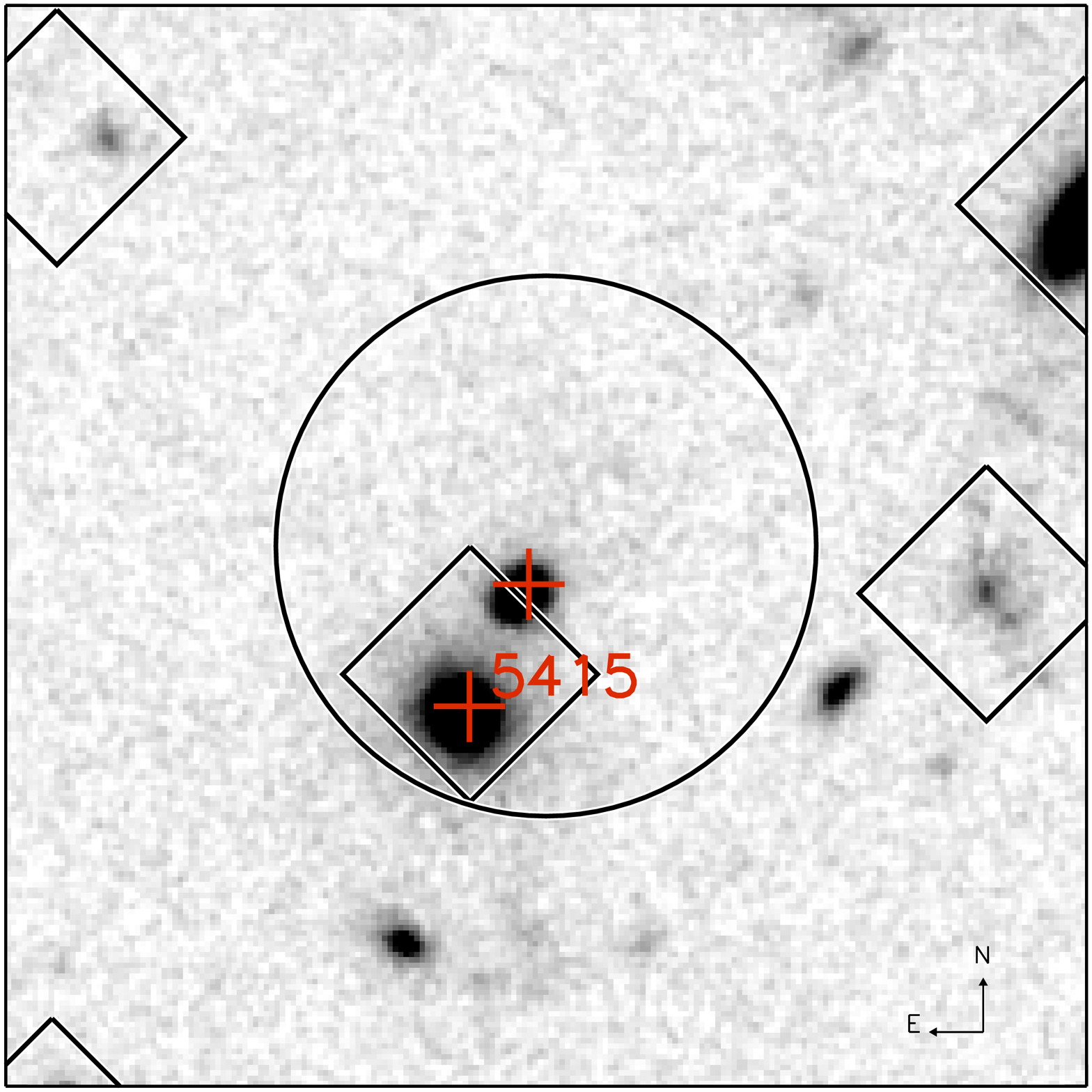}
\caption{Same as figure~\ref{appenfig:427_428}, for object \#5415.}\label{appenfig:5415}
\end{center}
\end{figure}

\begin{figure}
\begin{center}
\includegraphics[width=0.21\textwidth]{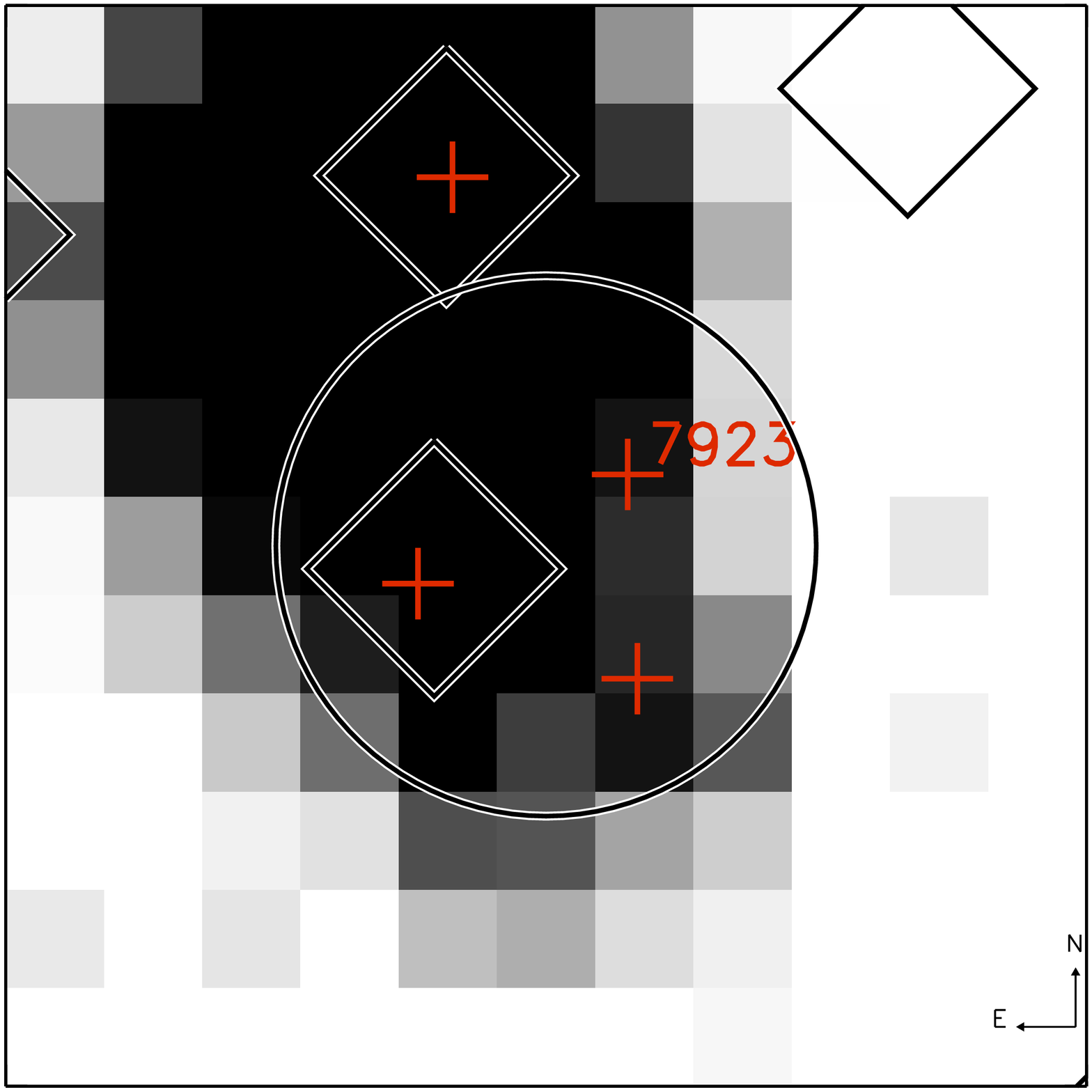}\includegraphics[width=0.21\textwidth]{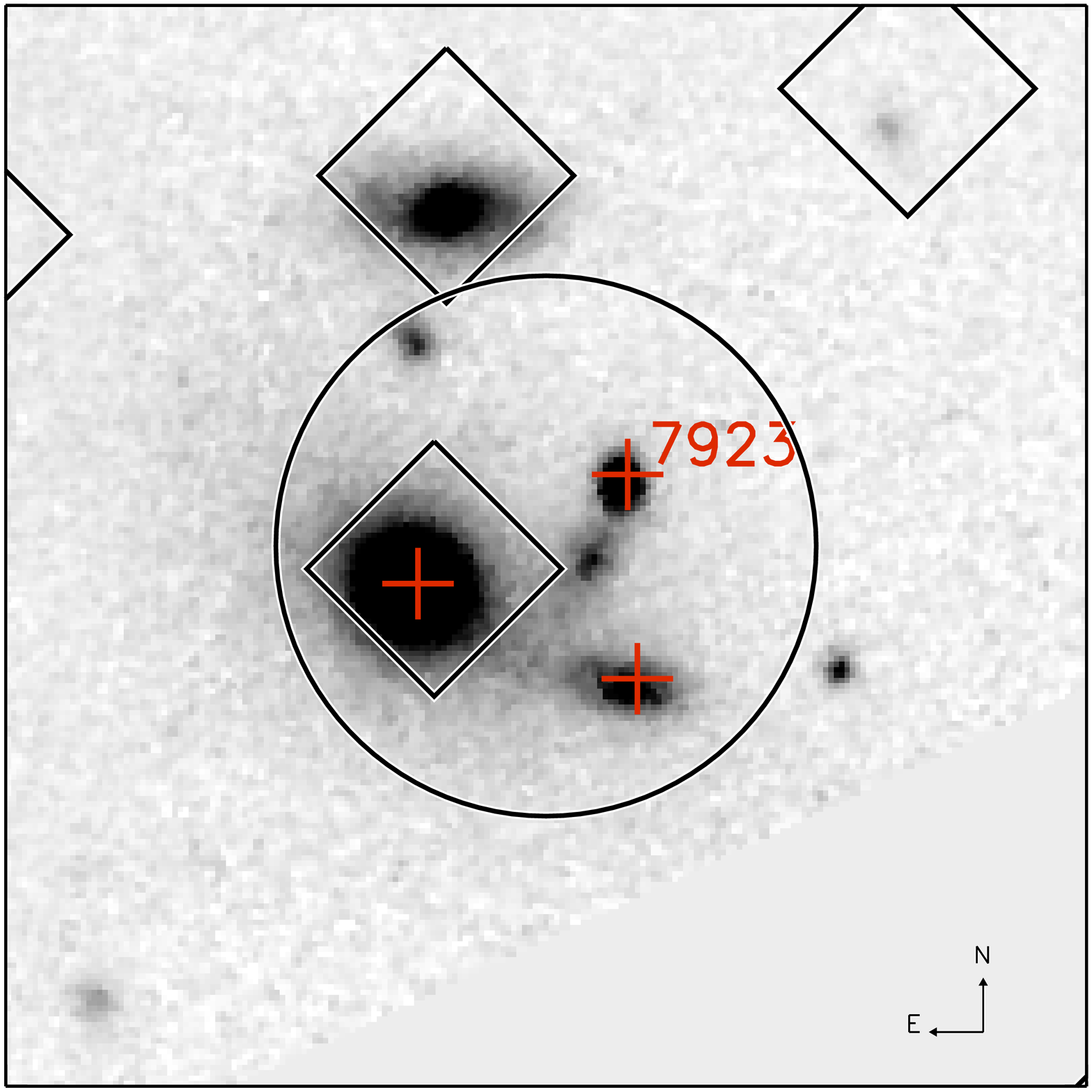}\\
\caption{Same as figure~\ref{appenfig:427_428}, for object \#7923.}\label{appenfig:7923}
\end{center}
\end{figure}

\section[]{Close pairs, groups and mergers}\label{app:B} 
Since many projected pairs are included in our sample, for each galaxy we checked the redshift of all the closest neighbors (dist$\leq 5$~arcsec), so to check if they are also gravitationally bound. As detailed below, some of the bound galaxy pairs in our sample are possibly included in the overdensity at $z\sim 1.61$. In the following we summarize the results for the most probable cases, including (i) pairs with at least one member with spectroscopic redshift and a companion with consistent photometric redshift, (ii) pairs with two photometric redshifts in agreement within $\Delta z\sim 0.1$, or -just in one case- (iii) pairs with visible tidal tails, suggesting interaction, although the difference in photometric redshifts is  $\Delta z> 0.1$.
\begin{itemize}

\item {\bf \#3231, \#3230} are two \mipsu\ galaxies with Early-Type morphology, at a projected distance of 3.62 arcsec, and spectroscopically identified at $z_{spec}=1.610$  and $z_{spec}=1.615$, respectively \citep[][i.e., K20 Survey]{2005A&A...437..883M}. There is a third bluer galaxy in the middle of this pair, with a photometric redshift of $z_{phot}$=2.09, which is the closest counterpart for the 24~$\mu$m/MIPS source. Recently, \citet{2013PASJ...65...17T} identified these galaxies, together with the X-ray source \#3258 in our sample (at $z=1.605$, and 23 arcsec far from \#3231), as likely members of an X-ray detected group of 8 galaxies, located $\sim$ 5' (2.5 Mpc) away from center of the highest-density region in the overdensity at $z\sim1.61$. Object \#3231 has been recognized by the authors as the brightest group galaxy (BGG). It is interesting to note that its size is comparable to that of local ETGs with similar mass.    

\item {\bf \#686 and \#687:} The \pacsd\ \#686 is classified as BLAGN/QSO-1 in the literature, and spectroscopically confirmed at $z=1.617$ \citep[][see also Table~1, and Section~\ref{sec:agn}]{2004ApJS..155..271S}. The quenched companion  \#687   has an early-type morphology, lies at a projected distance of d$\sim$ 1\farcs07 and has a ``likely'' spectroscopic redshift of  $z_{spec}=1.609$ \citep{2008A&A...478...83V}. 
Based on the criterion used in the previous section, these galaxies would be part of the highest density peak in the overdensity at $z=1.61$. However, if both redshifts are correct, these objects would have a  comoving radial distance of $\sim$ 12 Mpc from each other,  hence would not be interacting.

\item \citet{2009A&A...504..331K} identified a close triplet of  galaxies, including {\bf \#986} (z=1.609), indicated as the brightest confirmed galaxy member, {\bf \#1084} (z=1.614), and {\bf \#880} (z=1.612). All  three objects are  quenched/quenching ETGs, with the largest distance between them of 20\farcs0 (169 kpc) and with the largest difference in redshift corresponding to $\sim 575$ km\ s$^{-1}$. They lie about 1.5 arcmin (i.e., 760 kpc)  from the center of the highest density region described by Castellano et al. (2007). \citet{2009A&A...504..331K} concluded that the relatively small crossing time of a galaxy in this triplet ($\sim3 \times 10^8 $yr), indicates that they are possibly undergoing a merging. 

\item The \mipsu\ {\bf \#1187} galaxy ($z_{phot}=1.66$) has two  very close companions. While the closest object  at a projected distance of 1.27 arcsecond has $z_{spec}=0.86$ \citep{2005A&A...434...53V,2006A&A...454..423V,2008A&A...478...83V}, the second one (at a distance of 2.17 arcsecond) has  $z_{phot}=1.54$ (ID:GOODS-MUSIC-15228). This source seems to be a quenched galaxy (being also undetected at MIPS/24~$\mu$m), with red colors, and spheroidal morphology ($n=2.33$, \re$_{,circ}\sim1.2$ kpc). Moreover, also a bluer, fainter galaxy (\#1188 in the D07 catalog) at $z_{phot}=1.57$ lies at a distance of 4.27 arcsecond from \#1187. Hence, it may be that these three galaxies are gravitationally bound, and may even be part of the overdensity at $z=1.61$.

\item The highly obscured \pacsd\  galaxy {\bf \#5534} with  $z_{spec}=1.616$  has an asymmetric morphology  indicative  of a merging system. Moreover, it has a bluer neighbor at a distance of 3.02 arcsecond, with $z_{phot}$ =1.59 (ID:GOODS-MUSIC-6738). 

\item  Object {\bf\#720} ($z_{phot}=1.89$) shows a tidal feature toward the closest neighbor located at a projected distance of $\sim 2$ arcsecond,  which becomes even more evident when the derived 2D surface brightness best-fit model is subtracted to the real galaxy image (e.g., GALFIT residual image). The putative companion is a massive galaxy ($M_*\sim5\times 10^{11}$\msun) with visible spiral arms, with an insecure spectroscopic redshift $z_{spec}=2.49$ \citep[][]{2009A&A...494..443P}, in disagreement with the photometric one, designed as the best solution, at $z_{phot}=1.43$ (ID:GOODS-MUSIC-16094).   Difficult to say whether the two galaxies are a physical pair or not.

\item Objects {\bf \#427 and \#428} represent the most peculiar pair in our sample, since it seems to be a double lens system, as detailed in Section~\ref{sec:lens}. Although \#427 is spectroscopically confirmed at $z=1.427$ while \#428  has a photometric redshift $z_{phot}=1.59$, they may represent an interactive pair, as 
suggested by the {\it bridge} between them as discussed in Section~\ref{sec:lens}. 

\end{itemize}

\end{document}